\documentclass[aps,prb,twocolumn,groupedaddress,10pt]{revtex4-1}

\usepackage{graphicx}
\usepackage{hyperref}
\usepackage{hypcap}

\begin{document}


\title{Spin properties of 2D semiconductors probed by scanning tunneling microscopy}


\author{M. Morgenstern$^1$, A. Georgi$^1$, C. Stra\ss er$^2$, C. R.\ Ast$^2$, S. Becker$^1$, and M. Liebmann$^1$}
\affiliation{$^1$II. Physikalisches Institut B and JARA-FIT, RWTH Aachen University, D-52074 Aachen, Germany\\$^2$Max-Planck-Institute for Solid State Research, Heisenbergstr. 1, D-70569 Stuttgart, Germany}

%
%
\date{\today}

\begin{abstract}
The interrelation between spin and charge in semiconductors leads to interesting effects, e.g., the Rashba-type spin-orbit splitting or the
exchange enhancement. These properties are proposed to be used in applications such as spin transistors or spin qubits.
Probing them on the local scale with the ultimate spatial resolution of the
scanning tunneling microscope addresses their susceptibility to disorder directly. Here we review the results obtained on two-dimensional semiconductor systems (2DES).
We describe the preparation and characterization of an adequate 2DES which can be probed by scanning tunneling microscopy.
It is shown how the electron density and the disorder within the 2DES can be tuned and measured. The observed local density of states of weakly and strongly disordered
systems is discussed in detail. It is shown that the weakly disordered 2DES exhibits quantum Hall effect in magnetic field. The corresponding local density of states across a quantum Hall transition is mapped showing the development from localized states to extended states and back to localized states in real space. Decoupling the 2DES from screening electrons
of the bulk of the III-V semiconductor leads to a measurable exchange enhancement of up to 0.7 meV which depends on the local spin polarization of the 2DES. At stronger confinement potential, i.e. larger doping, the Rashba spin splitting with $\alpha$ as large as $7\cdot 10^{-11}$eVm is observed as a beating in the density of states in magnetic field. The Rashba spin splitting varies with position by about $\pm 50$ \%  being largest at potential hills.
\end{abstract}
%


\maketitle
\begin{widetext}
\tableofcontents
\end{widetext}
\section{Introduction}
\label{intro}

The proposal of the spin transistor by Datta and Das in 1990 \cite{datta:665} initiated a new field, which is called semiconductor spintronics. It has the goal to use the spin degree of freedom of the electron not only to store and read information, but also to process information.\cite{Wolf,Fabian} The initial idea of the spin transistor was to use the electric field of a gate electrode to induce an effective magnetic field $\underline{B}_{\rm eff}$ within the reference frame of moving electrons of a 2D electron system (Rashba effect).\cite{Rashba,BychkovRashba} This effective $B$-field induces precession of electron spins, if they are oriented perpendicular to $\underline{B}_{\rm eff}$ and if the electrons move perpendicular to $\underline{B}_{\rm eff}$. Using ferromagnetic source and drain electrodes, which are polarized both in the same direction perpendicular to $\underline{B}_{\rm eff}$, the transistor can be tuned on or off, if the precession angle between source and drain is $2\pi$ or $\pi$, respectively. This solely depends on the gate voltage and, thus, voltage pulses can be used for tuning.  Such a spin transistor might be faster and more energy efficient than standard field effect transistors.\\
Motivated by this idea,  various experimental techniques using electron spin resonance,\cite{Jantsch1,Jantsch2,Tyryshkin} time-resolved optics,\cite{Awschalom1,Awschalom2,Awschalom3,Awschalom4} and multi-terminal
spin transport \cite{Parkin,Parkin2,Ohno}
have been applied to understand the spin injection into semiconductors,\cite{Schmidt,Parkin,Appelbaum,Jonker}
the dynamics and possibilities of manipulation within semiconductors \cite{Grundler,Nitta,Schaepers,Schoenenberger} as well as the spin detection.\cite{Fiederling,Hanbicki,Heersche} While highly efficient spin injection \cite{Parkin, Jonker}, large spin relaxation lengths \cite{Kikkawa} as well as an efficient manipulation of spins by the Rashba effect \cite{Grundler,Nitta,Schaepers} have been demonstrated, the realization of a reliably operating spin transistor remains elusive.\cite{Meier,Schoenenberger}\\
Another branch of application in semiconductor spintronics is quantum computation based on spins in quantum dots.\cite{diVinc,QI} Using the spin degree of freedom of an electron as a qubit of information has the advantage that the spin is only marginally coupled to other degrees of freedom mostly via the spin-orbit interaction, the hyperfine interaction to the nuclear spins and the dipolar coupling to other electron spins. Thus, decoherence and relaxation of the qubit is weak, which might allow to implement adequate error correction schemes for information processing.\cite{Kouwenhoven2,Bennet} Using quantum dots as hosts for individual electron spins allows to transfer the immense knowledge on preparational and electrical control of the quantum dots gained in previous years.\cite{Kouwenhoven} However, adequately manipulating the spins on the local scale remains a challenge, since typical electron spin resonance schemes are never local on the nm-scale.\cite{vandersypen} Thus, researchers used singlet and triplet degrees of freedom of a pair of electron spins to perform the single qubit operations electrically by gates, thereby implementing a promising, scalable scheme.\cite{Yacoby} Coherence times up to 0.2 ms have been reached, which would lead to about 10$^5$ possible qubit operations within the coherence time.\cite{Bluhm} The fidelity of a two-qubit operation has been determined to be 70 \% so far and, thus, still has to be improved.\cite{Bluhm2} Although other approaches to solid-state based qubits like superconducting transmons \cite{Wallraff,Martinis,Schoelkopf} or NV centers  in diamond \cite{Hanson,Jelezko1,Jelezko2} are ahead with respect to fidelity or coherence times, it is still likely that the easier scalability of the spin qubits renders it the more favorable approach.\\

Avoiding disorder in electronic or spintronic devices is typically advantageous, although exceptions like the quantum Hall resistance standard \cite{KvK}
relying on a remaining amount of disorder are known. More importantly, disorder can never be avoided completely and, thus, the relation of the relevant effects to disorder
are important to be understood. Since disorder is a local property, getting knowledge on the local scale is the most straightforward approach. The highest spacial resolution is
provided by the scanning tunneling microscope (STM),\cite{Binnig} which in different spectroscopic modes is sensitive to the charge and  spin distribution of individual electronic states.\cite{Eigler,Maltezo,Bode,Loth,Heinrich,Heinrich2,Wiebe,Wiebe2} For certain structures, even the phase of the electron wave function has been reconstructed.\cite{Manoharan} A quasi-coherent manipulation of a
pair of wave functions within a quantum corral has also been realized by manipulating the local disorder via STM.\cite{Manoharan2} Thus, STM is an excellent tool to probe the interaction of relevant spin properties with disorder down to the atomic length scale.\\
{\it To this end, we will show that the exchange interaction between spins as well as the Rashba-type interaction of spins with electric fields can be detected in semiconductors on the nm-scale}. Therefore, we use a two-dimensional electron system (2DES) which is induced by surface doping of a low band gap III-V semiconductor.\cite{Aristov, FeInAs, NbInAs, PhysRevB.63.155315} This 2DES is located directly below the surface (depth: 10 nm) and, thus, can be probed by photoelectron spectroscopy\cite{Aristov} as well as by STM \cite{PhysRevLett.89.136806,PhysRevB.68.041402,Morgenstern2,Kanisawa}. This way, the parabolic dispersion of the 2DES can be measured directly. Moreover, the potential disorder can be determined with meV and nm accuracy using the so-called tip-induced quantum dot.\cite{Tipinduced} The 2DES exhibits the integer quantum Hall effect in magnetic field \cite{Masutomi,Tsuji} and the corresponding states across a quantum Hall transition have been detected by STM.\cite{hashimoto:256802}
Since the magnetic field allows to localize electrons, a sizable exchange interaction between the localized spins appears, which can be determined by careful inspection of the spin splitting within local Landau level fans.\cite{Becker2011} The spin splitting is roughly 0.7 meV larger, if the local filling factor is odd than if it is even.
Moreover, by preparing the 2DES such that it is confined within a large electric field ($2\cdot 10^7$ V/m), we can detect the Rashba-type spin splitting. This spin splitting is visible
as a beating pattern of the density of states (DOS) in magnetic field, which without Rashba-interaction exhibits regularly spaced Landau levels.\cite{Becker2010} The Rashba-type spin splitting can also be probed directly as a locally fluctuating energy distance between two spin levels appearing as peaks in the local density of states.\cite{Sherman}
Thus, we demonstrate that both, exchange interaction as well as Rashba-type interaction of electron spins can be probed down to the length scale of individual electronic states.\\
This review is organized as follows. Chapter \ref{ARPES} describes the characterization of the 2DES by photoemission. Chapter \ref{STS} is a short general introduction into scanning tunneling spectroscopy (STS).  Then, techniques to determine the potential disorder of the 2DES with high resolution are introduced (chapter \ref{disorder}). The resulting local density of states reacting onto different strength of the disorder differently is shown in chapter \ref{LDOS}.
After describing the observed local density of states within the quantum Hall regime (chapter \ref{QHE}), we finally show how the local exchange interaction can be directly extracted from the STS data of the 2DES
(chapter \ref{exchange2}) as well as how the Rashba effect is probed quantitatively on the local scale (chapter \ref{Rashba}).
\section{Probing the 2DES by photoemission}
\label{ARPES}
\begin{figure*}[htb]
\includegraphics[width= \textwidth]{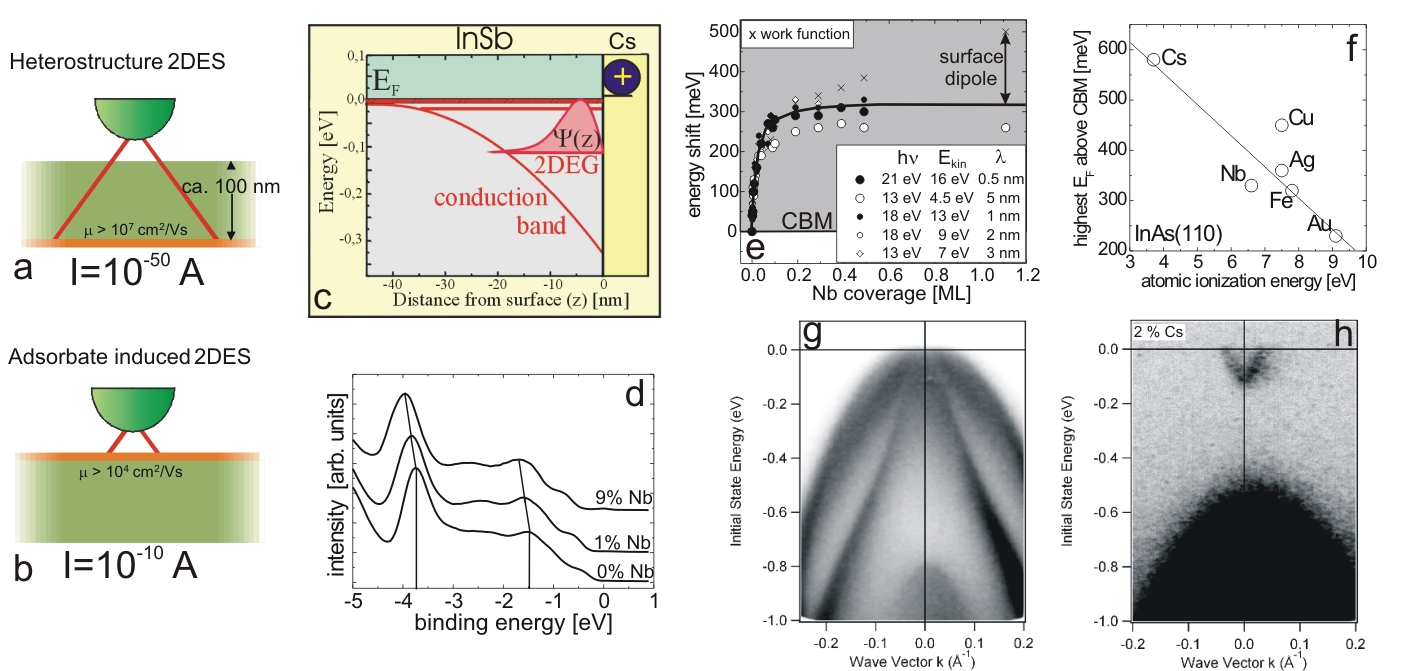}
\caption{(a) Sketch of buried 2DES as typically used in transport measurements with mobility $\mu$ and calculated tunneling current from the tip $I$ indicated; (b) same as (a), but for an adsorbate induced 2DES; (c) calculated band structure of InSb after depositing Cs atoms on top; Fermi level $E_{\rm F}$, charged Cs on the surface, subband energies (horizontal lines) and wave function $\Psi(z)$ of the first subband are shown; the valence band is omitted for clarity; (d) ARPES data of n-InAs(110) (energy distribution curves)
($N_D=1.1\cdot 10^{22}$ m$^{-3}$) with different Nb coverages on the surface as indicated; nearly vertical lines mark the shifting peaks, photon energy $h\nu=13$ eV, emission angle of photoelectrons $\Theta=0^\circ$; (e) energy shifts of different valence band peaks of n-InAs(110) as a function of Nb coverage, lower right inset shows photon energy $h\nu$, resulting kinetic energy $E_{\rm kin}$ of the photoelectrons and estimated escape depths $\lambda$ of the photoelectrons, the surface dipole determined by the differences between peak shift and workfunction shift is marked, notice that the peaks shift to lower binding energies with increasing coverage; (f) highest position of the Fermi level above the conduction band minimum at the surface for different adsorbates on InAs(110) as a function of ionization energy of the free adsorbate atom;\cite{NbInAs} (g) ARPES data of p-InSb(110) ($N_{\rm A}=1.5\cdot 10^{24}$ m$^{-3}$) directly after cleavage; (h) same data as (g) after adsorbing 2 \% Cs, $T=80$ K, $h\nu=21$ eV.  \label{Fig1}}
\end{figure*}
The general problem of studying 2DESs by surface science techniques is, that they are typically located about 100 nm below the surface. This depth is chosen intentionally in order to avoid scattering at surface defects and, indeed, high mobilities up to $10^4$ m$^2$/Vs of the corresponding 2DESs have been achieved at low temperature implying an electron mean free path close to mm. \cite{Pfeiffer,Umansky,Eisenstein} However, these 2DESs can not be tackled by surface science techniques as STM or angular resolved photoelectron spectroscopy (ARPES). If one estimates, e.g., the tunneling current from a tip above the surface into a GaAs 2DES covered by 100 nm of AlAs with the triangular barrier approximation,\cite{triang} one ends up with a tunneling current of the order of $10^{-50}$ A, which corresponds to about $10^{-13}$ electrons within  the anticipated age of the universe, well below the detection limit of available current amplifiers.
This impossibility is illustrated in Fig. \ref{Fig1}(a). One possibility to overcome the problem is to tunnel into the cleaved edge of the sample, which gives access to the cross section of the 2DES.\cite{Johnson,Koenraad} Indeed, subbands and confined states of the 2DES have been mapped by STS of cleaved heterostructures.\cite{Kanisawa2} Hovever, the distribution of electron wave functions within the 2DES plane, guiding, e.g., the localization phenomena of the 2DES, is not visible by this method. Another possibility is to induce a 2DES directly at the surface by surface doping (see Fig. \ref{Fig1}(b)).\cite{Baier,Aristov,Chen} This is possible for the low bandgap materials as InAs and InSb, which exhibit a relatively large electron affinity.\cite{Aristov} Basically, most adsorbates deposited on that surfaces exhibit a confined, occupied state above the conduction band edge of the substrate.
\cite{Baier,Aristov,Chen,FeInAs, NbInAs,CoInAs,PhysRevB.63.155315} Thus, the electron is moved from the adsorbate into the semiconductor leaving a positive charge behind, which induces confinement of electronic states in the conduction band. The resulting band bending in the near surface region together with the wave function of the confined state with lowest energy ($\Psi_1(z)$ of the first subband) is sketched in Fig. \ref{Fig1}(c). The band bending as well as the wave function can be calculated using so-called Poisson-Schr\"odinger solvers \cite{Poissonsolver}, which take the screening of the charged surface adsorbates by bulk dopants as well as by confined electrons self-consistently into account.
A reasonable approximation is given by assuming a triangular potential well in $z$ direction with infinite surface barrier leading to \cite{RevModPhys.54.437}
\begin{equation}
\label{Psi2DES}
\Psi(z)= (\frac{b^3}{2})^{0.5}\cdot z\cdot e^{-bz/2}
\end{equation}
with
\begin{eqnarray}
b=(\frac{48\cdot \pi m^*_z m_e \cdot e^2\cdot N}{4\pi \epsilon \epsilon_0 \hbar^2})^{1/3},\\ N=N_{\rm Dopant}+\frac{11}{32} N_{\rm 2DES}.
\end{eqnarray}

Here, $m^*_z$ is the effective mass of the conduction band in $z$-direction, $\epsilon$ the dielectric constant of the semiconductor,
$N_{\rm Dopant}$ the 2D density of charged dopants within the depletion or accumulation layer and $N_{\rm 2DES}$
the 2D density of confined electrons ($\epsilon_0= 8.8 \cdot 10^{-12}$ As/Vm, $\hbar=1.05 \cdot 10^{-34}$ Js, $m_e=9.1\cdot 10^{-31}$ kg). Typically, the center of mass of $\Psi(z)$, $z_0=3/b$, is about 10 nm below the surface and, more importantly, the 2D electrons spill out of the surface due to the finite barrier height at the surface. Thus, they can be detected by STM like surface states, e.g. on metals.\cite{Eigler}\\
A first characterization of the induced band bending and the resulting 2DES subbands can be performed using ARPES. Fig. \ref{Fig1}(d) shows a number of spectra obtained for different coverages of Nb on an n-type InAs(110) surface. Thereby, InAs(110) is used since other surfaces as (111) or (001) typically exhibit a band bending already after surface preparation without adsorbates,\cite{Olson} presumably because of the relatively large defect density. In contrast, the InAs(110) surface exhibits only step edges as natural defects with typical interstep distance of 5 $\mu$m.\cite{Meyer} Thus, it shows
flat band conditions after cleavage in ultrahigh vacuum. This is shown, e.g. in Fig. \ref{Fig1}(g), where the valence band of p-type InSb(110) as measured by ARPES exhibits heavy and light hole bands as well as a spin-orbit split band (maximum at $-0.8$ eV) and a Fermi level $E_{\rm F}$ that cuts the valence band, which evidences the absence of band bending.
Consequently, adsorbates can be used to tune the band bending. The subsequently deposited adsorbate density is determined by a quartz balance and is given with respect to the unit cell of InAs(110), i.e. 1 \% coverage corresponds to an adsorbate density of $3\cdot 10^{16}$ m$^{-2}$. The three spectra in Fig. \ref{Fig1}(d) all exhibit two peaks, which can be identified as bulk valence band states of InAs.\cite{NbInAs} They shift to lower binding energy with increasing
coverage indicating the adsorbate induced band shift in the near surface region. Fig. \ref{Fig1}(e) shows the binding energies of a number
of ARPES peaks observed at different photon energies as a function of Nb coverage. Generally, all peaks show similar energy shifts. However, the ones with the largest escape depths of the photoelectrons show an about 20 \% smaller peak shift in line with the idea that they probe the most shifted bands deeper into the bulk of the system. The fact that the measured work function shift is larger than the observed peak shifts, in particular, at large coverage implies an additional change of a surface dipole, which has been attributed to the lifting of the relaxation at the InAs(110) surface.\cite{NbInAs}
In the case of Nb on n-type InAs(110), the largest band bending is 320 meV. It is most likely limited
by the donor level of the adsorbate being $E_{\rm ads}=320$ meV above the conduction band minimum of InAs. If the band bending is large enough such that the donor level is shifted below the Fermi level of the substrate, no additional charging of the adsorbates occurs.
Thus, the process is self limiting. In turn, the adsorbate level can be determined via the largest band shift at the surface.\\
It has been found that this adsorbate level scales inversely with the ionization energy of the corresponding neutral atom as shown in Fig. \ref{Fig1}(f).\cite{NbInAs} Obviously surface band shifts up to 600 meV are possible using alkali atoms as adsorbates. This is much larger than the band gap of InAs being 430 meV at $T=0$ K or 360 meV at $T=300$ K.\cite{Huijser} Such a large band bending can even induce occupied confined states within a $p$-doped InAs crystal, i.e. a conducting 2DES within an inversion layer. Fig. \ref{Fig1}(g) and (h) show a direct proof of such an inversion layer using ARPES mapping of the $E(\underline{k})$ dispersion of p-type InSb(110) prior and after deposition of 2 \% of Cs. The valence band of InSb is visible in Fig. \ref{Fig1}(g) as described above. Adsorbing 2 \% of Cs moves the valence band down by about 450 meV and a single parabolic band curved upwards in the $E(\underline{k})$ dispersion appears at $E_{\rm F}$ (Fig. \ref{Fig1}(h)). This band is considerably weaker
in intensity than the valence bands, such that the valence band appears as a completely dark area, if the contrast is chosen such that the 2DES is visible.
This is due to the fact that ARPES probes only a few percent of the 2DES charge density, since the photoelectron escape length is a factor of 10-20 smaller than the extension of the 2DES. The fact that the energy distance between the origin of the upwards parabola of the 2DES to the valence band maximum is 350 meV, i.e. much larger than the known band gap of InSb (230 meV) \cite{Vurgaftman} directly reflects the confinement energy of the first subband of the 2DES being 120 meV. Higher subbands of this 2DES are not occupied and, thus, not visible in ARPES.\\
Importantly, a high level of control of the band structure properties of the 2DES
is obtained by ARPES. The 2DES properties can be designed by varying the dopant type (n- or p-type) and dopant density of the substrate as well as the adsorbate type and coverage.
\section{Scanning tunneling microscopy}
\label{STS}
Before describing the STS data of the 2DES, we will shortly summarize the most important features of STS.
In STS, a sharp metallic tip, mostly ending in a single atom, is positioned $3-8$ ${\rm \AA}$ above a
conducting surface. The surface is usually prepared in ultra high vacuum (UHV) in order to be atomically smooth. A voltage $V$ is applied between the tip and the conducting surface and the resulting tunneling current $I$ is measured. $I$ depends exponentially on the distance between surface and tip $\Delta z$ according to $I(z) \propto e^{-\alpha \Delta z}$. A good approximation for $\alpha$ is
$\alpha = \sqrt{(4m_{\rm e} \cdot (\Phi_s+\Phi_t -e|V|)}/\hbar)$ with Planck's constant $\hbar$, electron mass $m_{\rm e}$, electron charge $e$ and  work functions of tip and sample $\Phi_t$ and $\Phi_s$, respectively. A good estimate is $\alpha \simeq 2.1$/{{\rm \AA}}.\\
The tip is positioned with respect to the sample using piezoelectric elements. All three directions $x$, $y$, and $z$ can be changed with sub-pm precision.\cite{mashoff:053702} For STM, the tip is scanned in $x$ and $y$ direction and the tunneling current is kept constant by a feedback mechanism adjusting $z$. The resulting $z(x,y)$ is plotted and called constant-current image. It represents,
to first order, a contour of constant integrated local density of states  of the substrate $LDOS(x,y,z,E)$, where the integration has to be taken between the Fermi levels of sample and tip to be adjusted by $V$.\cite{Morgenstern3} The central position of the very last atom of the tip is given by $(x,y,z)$ and $E$ is the energy. Such images are often called topography of the sample indicating that corrugations of the atomic positions dominate the contour.\\
Differentiating $I$ with respect to $V$ (at low $V$ with respect to $\Phi_{s(t)}$) gives direct access to the $LDOS$ according to:
\begin{eqnarray}
dI/dV(x,y,z,V) \propto LDOS(x,y,z,E) =\\ \nonumber
\sum |\Psi_{\tilde{E}}(x,y,z)|^2\cdot \delta(E-\tilde{E}).
\end{eqnarray}
Thereby, $\Psi_{\tilde{E}}$ are the single-particle wave functions of the substrate at energy $\tilde{E}$ and $E=e\cdot V$.
Of course, this requires that the system is adequately described by independent single-particle wave functions. Moreover, an s-type
symmetry of the orbitals of the last atom of the tip is necessary, in principle. Chen has shown that the model remains largely correct even
if higher orbital momenta of the tip wave functions are contributing.\cite{Chen}
In real experiments, the $\delta$-function has to be replaced by an energy resolution function with approximate full width at half-maximum of $\delta E\approx\sqrt{(3.3\cdot k_BT)^2+(1.8\cdot eV_{\rm mod})^2}$. $T$ is the temperature ($k_{\rm B}$: Boltzmann's constant) and
$V_{\rm mod}$ is a modulation amplitude used to detect $dI/dV$ by lock-in technique. The resulting $dI/dV(x,y)$ recorded at constant $V$ and constant $I$ is often called $LDOS$-image implying that the image is proportional to
the LDOS distribution in the surface layer.
Strictly speaking, this is not correct since constant $I$ is not identical to constant height above the surface $z(x,y)\neq \rm const.$.
A local enhancement of density of states, e.g. around a dopant within a semiconductor, can locally lead to a larger tip-surface
distance and, thus, to a reduced differental conductivity. This can be cured by measuring the global distance dependence of the tunneling current $I(z)$ and dividing $dI/dV(x,y)$ by $I(z(x,y))$ resulting in:
\begin{equation}
LDOS(E,x,y)\propto \frac{dI/dV(x,y, V=E/e)}{I(z(x,y))}
\end{equation}
However, that requires that the decay of $I$ with respect to $z$ is spacially constant \cite{Ewelle}, which is a reasonable assumption for the LDOS of a 2DES, but could fail on the atomic scale \cite{Heinze}.\\
Another modification, which applies favorably for $dI/dV(V)$ {\it curves} at higher $V$ with respect to $\Phi_{s(t)}$, is
\begin{equation}
LDOS(E) \propto \frac{dI/dV(V=E/e)}{I(V)/V}
\end{equation}
This compensates for the effect that the transmission coefficient of electrons through the tunnel barrier also changes
with applied voltage, which is relevant if $|V| > 0.5$ V.\cite{Feenstra} Another problem with $dI/dV(V)$ curves is that they are sensitive to both, the $LDOS(E)$ of the sample as well as the $LDOS(E)$ of the tip. Thus, it is required to have a featureless $LDOS(E)$ of the tip. Typically, one repeats measurements with different tips on the same sample
and attributes repeatedly observed features to the sample or one changes the properties of the sample, e.g. by applying a $B$-field and measure them with the metallic tip being insensitive to the $B$-field.\cite{LLfluct}\\
Thus, STM can measure atomic structure with sub-pm resolution and electronic structure ($LDOS$) with sub-meV resolution.\cite{Wiebe3} The energy resolution makes STM complementary to the transmission electron microscope (TEM) which reveals atomically well defined structural information partly with chemical specificity,\cite{Meyer} but not the $LDOS$ down to the meV scale.
\section{Disorder potential}
\label{disorder}
\begin{figure*}[htb]
\includegraphics[width= \textwidth]{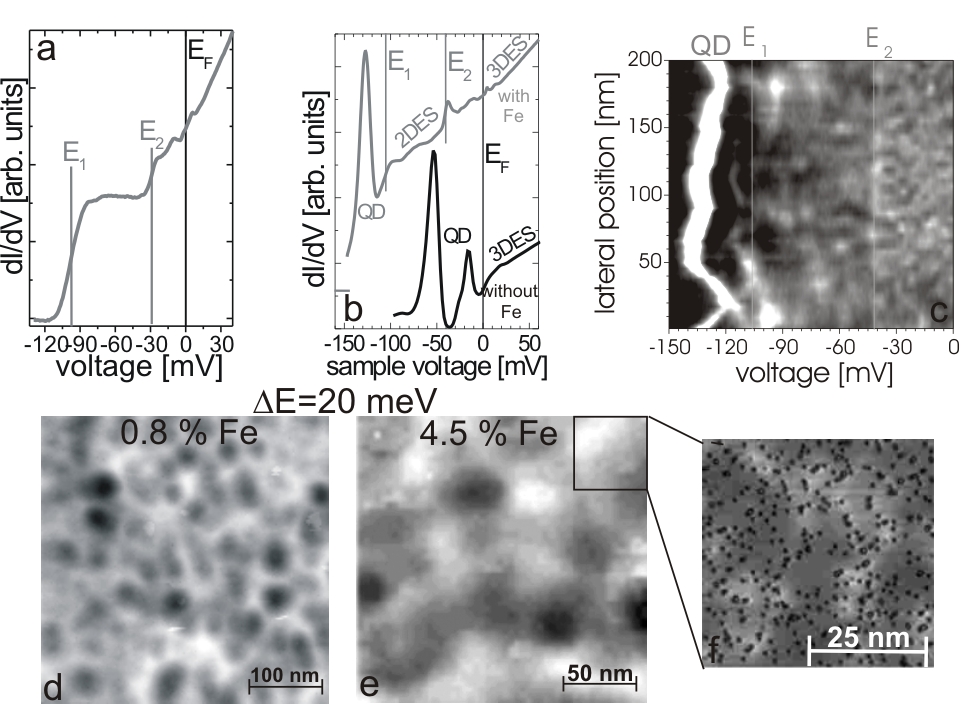}
\caption{(a) Spatially averaged $dI/dV$ curve of n-InAs(110) ($N_{\rm D}=1.1\cdot 10^{22}$ m$^{-3}$) after depositing 2.7 \% of Fe, Fermi level $E_{\rm F}$ and the subband energies $E_1$, $E_2$ as deduced from ARPES data are marked by vertical lines; (b) spatially averaged $dI/dV$ curves of the same n-InAs(110) as in (a) probed with a different tip prior
(lower curve) and after (upper curve) depositing 4.5 \% Fe, features belonging to the 2DES, the 3DES and the tip induced quantum dot (QD) are marked; (c) grey scale plot of $dI/dV$ as a function of voltage $V$ and position $x$, the subband energies $E_1$, $E_2$ and the feature belonging to the QD are marked, the latter is used to map the electrostatic potential of the 2DES; (d), (e) measured electrostatic potential landscape of the 2DES induced by 0.8 \% and 4.5 \% Fe, respectively, greysacle covers a range of 20 meV from black to white, note the larger disorder potential after depositing 0.8 \% Fe; (f) STM image of the area marked in (e) ($V=0.1$ V, $I=50$ pA), the black dots are the positions of Fe atoms; $T=6$ K.
\protect{\cite{PhysRevLett.89.136806}}
\label{Fig2}}
\end{figure*}
\subsection{Experimental determination}
One drawback of STS on semiconductors is the low charge density of semiconductors and the resulting
relatively large screening length $\lambda_{\rm s}$ of several nm, which can be calculated according to:
\begin{equation}
\lambda_{\rm s}= \sqrt{\frac{\pi\epsilon\cdot a_{\rm B}}{4 m^*}}\cdot(3\pi^2 N)^{-1/6}
\end{equation}
with the dielectric constant $\epsilon=14.6$ for InAs and $\epsilon=16.8$ for InSb, the Bohr radius $a_B=0.053$ nm, and $N$ being the 3D density of mobile charge
carriers.
Thus, electric fields from the tip can penetrate into the semiconductor leading to the so-called
tip-induced band bending.\cite{bandbend}\\
This band bending depends on the work function difference between tip and sample $\Delta \Phi$ and the applied voltage.
Typically, the work function of the tip varies by about 300-500 meV depending on details of the microtip.\cite{Tipinduced}
Thus, it is important to prepare the tip by trial and error such that $\Delta \Phi \simeq 0$ meV.\\
On the other hand, the tip-induced band bending can be used to determine the disorder potential of the 2DES. Therefore, one has to realize
that the band bending, if downwards towards the surface, leads to a conduction band region, which is confined in all three directions, i.e. only the
bands below the tip are pushed downwards by the electric field of the tip
leading to a quantum dot below the tip. This tip-induced quantum dot (TIQD) can be moved across the surface.\cite{Tipinduced,spinsplit}
In $z$-direction perpendicular to the surface, the confinement region of the quantum dot states is approximately the same as the confinement region of the 2DES, since screened by the same dopant density.
The corresponding probability amplitude is given by eq. \ref{Psi2DES}, however with slightly different $N_{\rm 2DES}$.
In ($x,y$)-direction parallel to the surface, the confinement is less, since the tip has a certain lateral extension, which adds to twice the screening length.
A good approximation is a Gaussian with a full width at half maximum (FWHM) of 50-100 nm. This width can be determined by the
energy distance of the confined states which appear as peaks in $dI/dV$ curves. The lowest-energy state of such a quantum
dot has a Gaussian lateral shape according to
\begin{equation}
|\Psi_1(x,y)|^2 \propto e^{-\frac{x^2+y^2}{2\sigma_{\rm 1}^2}}
\end{equation}
 with a FWHM $2\cdot \sigma_{\rm 1}$ about 4-5 times smaller than the FWHM of the QD, i.e. about 10 nm.\cite{Tipinduced}\\
 If the TIQD is absent, $dI/dV$ curves measure the unperturbed LDOS of the sample. An example of the spatially averaged
 $dI/dV$ curve of an adsorbate induced 2DES without TIQD is shown in Fig. \ref{Fig2}(a) exhibiting two steps according to the two
 low-energy subbands of the 2DES. Above the steps, the spatially averaged LDOS is flat as
 expected from the density of states of each subband of a 2DES
 \begin{equation}
 D_n(E) =\frac{m_e\cdot m^*}{\hbar^2\pi}.
 \end{equation}
The second subband  appears less intense, since it is buried deeper into the substrate.
Importantly, the 2DES leads to $dI/dV$ intensity in the region below $E_F$ although the InAs-sample is n-doped, i.e. the 2DES is within the region where
the bulk band gap of the semiconductor is expected.\\
In case of a TIQD, already 3DES before the surface doping by adsorption exhibits peaks below $E_F$, i.e.
in the band gap region of the 2DES. From the distance of the peaks of about $\Delta E= 40$ meV, one can calculate the
width of the quantum dot and the FWHM of the lowest state using the harmonic oscillator approximation
\begin{equation}
\Delta E = \hbar \omega \Rightarrow \sigma_1=\frac{\hbar}{\sqrt{m_e \cdot m^* \cdot \Delta E}}= 8.5 \,{\rm nm,}
\end{equation}
i.e. the lowest state has a FWHM of 17\,nm.\\
Moving this state across the surface exposes it to the potential disorder of the sample. Thus, the confinement potential of the QD is modified
by the 2D disorder potential of the 2DES $V(x,y)=\int \Psi_1(z)V(x,y,z)dz$ , where $V(x,y,z)$ is the bare 3D electrostatic potential of the sample.
Thus, within first order perturbation theory, the energy of the lowest state
is
\begin{equation}
E_1(x,y)=
<\Psi_1(\widehat{x}-x,\widehat{y}-y)|V(\widehat{x},\widehat{y})|\Psi_1(\widehat{x}-x,\widehat{y}-y)>_{\widehat{x},\widehat{y}}
\end{equation}
Consequently, an energy map of the lowest state is a map of the electrostatic potential of the 2DES, which is coarsened
by $2\sigma$. Notice that the 3D potential is already coarsened by $\simeq3/b \simeq 10$ nm
due to the extension of the 2DES in $z$ direction. Thus assuming an isotropic screening of point like charges as the origin of the
disorder, the lowest state maps the potential down to about the lowest relevant length scale.\\
Fig. \ref{Fig2}(c) shows a $dI/dV(V,x)$ map of an InAs-2DES as a greyscale plot. The subband energies $E_n$ determined from ARPES of the same system
are marked. At voltages above these energies, the 2DES shows wavy patterns along the lateral position $x$ which are related to the scattering of the electron waves (see chapter \ref{LDOS}).
Since more states contribute to the scattering above $E_2$, the apparent wave length gets smaller. Below $E_1$ a strong peak meandering in energy
as a function of $x$ is visible. This peak marked QD is the lowest state of the quantum dot (see also upper curve in Fig. \ref{Fig2}(b)). In the central $x$ region
at 70-130 nm even the second QD state is visible showing a similar meandering as the first one. If one plots the peak voltage of the low energy state as a function
of position, one gets the $E(x,y)$ maps shown in Fig. \ref{Fig2}(d) and (e) for two different 2DES, respectively. These images are called map of the potential disorder. They are both measured on n-InAs
with a donor density of $N_{\rm D}=1.1\cdot 10^{22}$ m$^{-3}$, but with different densities of Fe adsorbates as marked. The FWHM of the histograms of all potential values is 20 meV in
both cases, but surprisingly the potential disorder varies on a shorter length scale at lower adsorbate coverage.\\
Fig. \ref{Fig2}(e) features about 4 deep troughs and about 7 shallower troughs. The number of donors within the 2DES region of the imaged area $A=200 \times 200$ nm$^2$ amounts to
$N_{\rm D2D}=A\cdot N_{\rm D} \cdot 3/b =4$ implying that the deep troughs mark the donors of the substrate located within the 2DES. The shallower troughs are most likely substrate donors
lying below the 2DES, which are not completely screened. This is compatible with $\lambda_{\rm s}=20$ nm for the bulk of InAs being a factor of two larger than $3/b$, although $\lambda_{\rm s}$ is slightly shortened by the presence of the 2DES.\\
Thus, one concludes that the surface dopants are not relevant for the potential disorder. This gets even more obvious, if one compares the distribution of Fe atoms, which are visible as black dots in Fig. \ref{Fig2}(f). The potential fluctuations appear on a length scale covering several 100 adsorbate atoms. More precisely the 2D density of dopants within the 2DES is a factor of 1000 lower than the density of Fe atoms.
The potential troughs in Fig. \ref{Fig2}(d) are about $80/(400 \times 400)$ nm$^{-2}$, i.e., at least, a factor of two denser than in Fig. \ref{Fig2}(e). Thus, they are obviously influenced by the adsorbate layer. The likely reason for this different behavior will be discussed below.\\
Importantly, we can probe the disorder potential experimentally on the meV and nm scale.\cite{PhysRevLett.89.136806} Unfortunately, that requires the presence of a disturbing TIQD. Thus, the pristine
LDOS of a 2DES with exactly known potential disorder cannot be measured.\\
Nevertheless, we mapped the disturbed LDOS at different energies and compared it with the calculated LDOS resulting from a diagonalization
of the matrix
\begin{equation}
\label{mixing}
<\Psi_i(x,y)|\frac{\hbar^2k^2}{2m_e m^*}+V(x,y)|\Psi_j(x,x)>,
\end{equation}
where $\Psi_{i(j)}$ are plane waves, and we assumed periodic boundary conditions.
In order to get the LDOS, the resulting wave functions are averaged using a Gaussian energy distribution with FWHM$=\delta E $ . The length scale of LDOS fluctuations, the histogram of LDOS values (see. Fig. \ref{Fig5}(c)) and the Fourier transformation of the measured LDOS are well reproduced, but only few of the exact shapes of the LDOS($x,y$) are identical leading to a cross
correlation of 15 \%.\cite{PhysRevLett.89.136806} Possible reasons are the remaining influence of the TIQD, the coarsening of some of the details of the potential disorder and a localization length of the states which might be larger than the image size. Indeed, taking the measured mean free path of 300 nm (see below), an influence of the last argument
appears likely. \\
\begin{figure*}[htb]
\includegraphics[width= \textwidth]{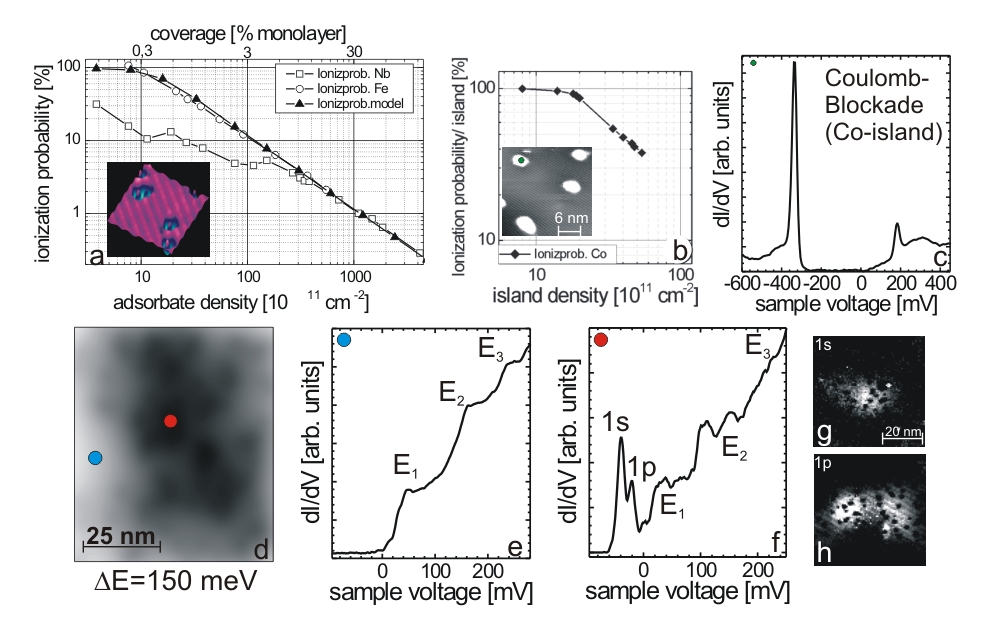}
\caption{(a) Open symbols: ionization probability per adsorbate atom for Fe and Nb deposited on n-InAs(110) ($N_{\rm D}=1.1\cdot 10^{22}$ m$^{-3}$) as deduced from the coverage dependence of the band shift measured by ARPES (see Fig. \ref{Fig1}e), full symbols: ionization probability calculated from a random distribution of monomers with an ionization level 300 meV above the conduction band minimum of InAs, inset: STM image of Fe adsorbates on InAs(110) ($I=200$ pA, $V=50$ mV); (b) Ionization probability as a function of Co island density on p-InAs(110) ($N_{\rm A}=4.6\cdot 10^{23}$ m$^{-3}$), inset: STM image of Co-islands on InAs(110) prepared at $T=120^\circ$ C ($I=50$ pA, $V=-0.86$ V); (c) $dI/dV$ spectrum obtained on a single Co island exhibiting a Coulomb gap ($V_{\rm stab}=0.7$ V, $I_{\rm stab}=0.7$ nA, $V_{\rm mod}=5$ mV); (d) simulated potential landscape of p-InAs(110) ($N_{\rm A}=4.6\cdot 10^{23}$ m$^{-3}$) covered with 15 \% Co atoms per InAs(110) unit cell, respectively, a Co island density of $N_{\rm island} =3.5\cdot 10^{16}$ m$^{-2}$, it is assumed that only the 50 \% of the islands with the lower Coulomb gap are charged;  (e), (f)
$dI/dV$ curves obtained from the areas marked in (d); ($V_{\rm stab}=0.7$ V, $I_{\rm stab}=0.7$ nA, $V_{\rm mod}=5$ mV); subband energies $E_1$, $E_2$, $E_3$ of the 2DES and the peaks of the strongly localized states (1s, 1p) are marked; (g), (h) $dI/dV$ images taken at the peak energies of the localized states marked in (f) ($V_{\rm stab}=0.43$ V, $I_{\rm stab}=0.7$ nA, $V_{\rm mod}=5$ mV);  $T=6$ K.
\protect{\cite{PhysRevB.68.041402,NbInAs}}
\label{Fig3}}
\end{figure*}
\subsection{Role of surface dopants}
Next, we will discuss the role of the adsorbates on the disorder. Therefore, it is essential to realize that only part of the adsorbates
are charged. The requirement for charging is that the electric potential produced by all other adsorbates at the position of a given one is lower than $E_{\rm ads}-E_{\rm BCBM}$.
Neglecting the screening in the adsorbate layer, one can easily simulate the charging probability of an adsorbate using
$E_{\rm ads}-E_{\rm BCBM}=320$ meV and $\lambda_s=10$ nm by depositing adsorbates subsequently onto random positions $\underline{x}_n$ on a 2D plane and
surrounding them by
\begin{equation}
V_n(\underline{x})=\frac{e^2}{4\pi\epsilon \epsilon_0 |\underline{x}-\underline{x}_n|}\cdot e^\frac{|\underline{x}-\underline{x}_n|}{\lambda_s}
\end{equation}
if
\begin{equation}
\sum_{m=1}^{n-1} V_m(\underline{x}_m) < E_{\rm_{ads}}-E_{\rm BCBM}
\end{equation}
and $V_n(\underline{x})= 0$ eV otherwise. One has to take into account that $\epsilon$ at the surface is reduced with respect to the bulk.\cite{NbInAs}
The result for a random choice of $\underline{x}_n$ is shown in Fig. \ref{Fig3}(a) as black triangles. For comparison the ionization probability for
Fe-adsorbates on n-InAs(110) \cite{FeInAs} is shown as open circles revealing reasonable agreement.
It has been calculated from the measured band bending by solving the Poisson-Schr\"odinger equation and dividing the sum of the resulting 2DES density $N_{\rm 2DES}$, which has also been determined experimentally,\cite{FeInAs} and the density of neutralized donors $N_{\rm Dopant}$ by the adsorbate density. Importantly, the ionization probability is 40-50 \% at 0.8 \% coverage
but only 8 \% at 4.5 \% coverage. Thus, the 2DES density changes only slightly by increasing the coverage to 4.5 \%, while most of the additional Fe atoms are uncharged. The remaining electrons in the Fe layer can screen the charged Fe-atoms effectively.  The charge density in the Fe layer of about $1.5\cdot 10^{17}$ m$^{-2}$ can probably even rearrange dynamically, since it is energetically coupled to the 2DES.  Given the very effective and density independent screening in 2D systems above a threshold density,\cite{Visscher} it is reasonable to assume that the
charge density in the Fe layer is effectively screened on the length scale of the average Fe-Fe distance being 2.5 nm at 4.5 \% coverage. Thus, the 2DES which is coarsening on the length scale of 10 nm due to its extension perpendicular to the surface basically ignores the potential disorder of the Fe layer provided that the adsorbate density is significantly larger than the density of transferred charge.\\
This simple picture breaks down, if the adsorbates are mobile on the surface and, thus, form clusters. Basically, each cluster is only charged once due to Coulomb blockade.
Thus, the ionization probability per adsorbate is reduced, if clusters are formed and the maximum band bending is achieved only at higher coverage. Moreover the ionization level
could be changed by hybridization. The squares in Fig. \ref{Fig3}(a) show the ionization probability for Nb which forms cluster consisting of 2-4 atoms depending on coverage
after room temperature deposition. The complex curve deviating strongly from the monomer model implies that both, single charging of multimers as well as hybridization in multimers is essential.\cite{NbInAs}\\
Importantly, already initial clustering of adsorbates leads to additional charge inhomogeneities. This is due the fact that cluster formation implies a reduced adsorbate density around the cluster and, thus, leads to a local potential hill due to reduced charging. This has been experimentally evidenced by Hashimoto {\it et al.}.\cite{hashimoto:256802}
Thus, it is important to avoid clustering and to deposit significantly more adsorbates than required for the charge transfer into the 2DES,
if one wants a high mobility 2DES. Measured values of the Hall mobility $\mu$ after Ag deposition on p-type InSb(110) ($N_A=1-2\cdot 10^{21}$ m$^{-3}$)
are up to $\mu=5$ m$^2$/Vs at a 2D carrier density $N_{\rm 2D}=1.5\cdot10^{15}$ m$^{-2}$.\cite{Masutomi}
This implies a mean free path $\lambda_{\rm MFP}$ of $\lambda_{\rm MFP}=\hbar/e \cdot \sqrt{2\pi N_{\rm 2D}}\cdot \mu =320$ nm.
Hall measurements at 40 \% Fe coverage or 1-20 \% Ag coverage on top of higher p-doped InAs(110) ($N_A=1.7 \cdot 10^{23}$ m$^{-3}$) revealed a lower mobility of $0.7$ m$^2$/Vs and $0.3$ m$^2$/Vs, respectively, again pointing to the leading importance of bulk doping for the disorder potential.\cite{Mochizuki,Tsuji}
In accordance with the above analysis, it was found that the  mobility decreased considerably below a Ag coverage of 1.5 \%.
Interestingly, spin-glass behavior in the Fe/p-InAs(110) system has been
deduced from hysteresis and a logarithmic time dependence of relaxation dynamics being present up to about 1000 s after a change of $B$ field. This effect has been correlated with the complex interplay of antiferromagnetic and ferromagnetic coupling in Fe-chains on InAs(110).\cite{Sacharow,Sacharow2}\\
Fig. \ref{Fig3}(c) shows the ionization probability for Co on p-InAs(110), which forms larger islands at room temperature as shown in the inset. The ionization probability
is plotted as a function of the island density and evidently gives a very similar curve as the one found for the Fe adsorbate density in Fig. \ref{Fig3}(a). Thus, indeed, also larger islands are only charged once. The reason is the Coulomb blockade which can be directly seen in $dI/dV(V)$ curves recorded on a particular Co island as a gap surrounded by two peaks
(see Fig. \ref{Fig3}(c)).\cite{PhysRevB.68.041402}. A histogram of the size of the Coulomb gaps has been performed for an island density of $3.5\cdot 10^{16}$ m$^{-2}$, where each island consists on average of 15 atoms leading to a an average gap size of 465 meV with a standard deviation of 150 meV. An estimate using a metallic sphere with the radius of the Co-clusters $R$ in a distance $d$ from a metallic plate results in a charging energy $E_{\rm Charge}=e^2/C$ with capacity $C= 4\pi \epsilon_0\times R\times(1+ R/(2R+2d))$, which is
additionally reduced by the image charge potential within the InAs of $\Delta E_{\rm charge}= \frac{\epsilon-1}{\epsilon+1}\cdot\frac{e^2}{8\pi\epsilon_0 R}$.\cite{Averin}
The calculated values are about a factor of two larger than the value of the measured Coulomb gap. This might be caused by the background charge\cite{Tinkham} as observed, e.g. for
Co clusters on MgO(0001),\cite{Pauly} or by an extension of the wave functions of the metal cluster into the InAs. A crude estimate of the 2D disorder potential can be made by assuming that all the clusters with the lower Coulomb gap up to the ionization density are charged and, are, thus, surrounded by a screened Coulomb potential. Adding charged acceptors randomly and averaging as usual over the $z$-extension of the 2DES leads to the 2D disorder potential shown in Fig. \ref{Fig3}(d). It fluctuates by about 150 meV on a length scale of
20 nm. Thus, the potential fluctuation is an order of magnitude stronger than in Fig. \ref{Fig2}(e). This allows to probe the 2DES at very different disorder strengths.

\section{Local density of states at different strength of disorder}
\label{LDOS}
\begin{figure*}[htb]
\includegraphics[width= \textwidth]{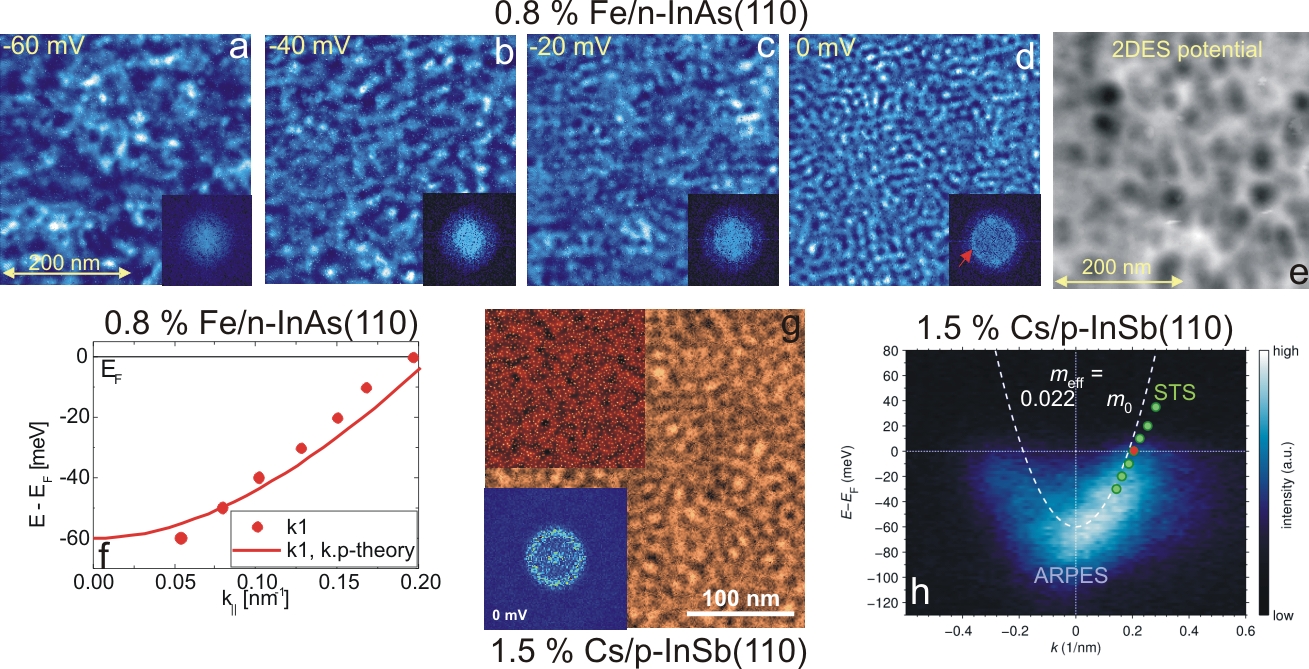}
\caption{(a)-(d) $dI/dV$ images at the voltages indicated of n-InAs(110) ($N_{\rm D}=1.1\cdot 10^{22}$ m$^{-3}$) covered with 0.8 \% Fe ($V_{\rm stab}=0.1$ V, $I_{\rm stab}=0.3$ nA, $V_{\rm mod}=1.8$ mV), insets: Fourier transformation of the real space data with preferential $|\underline{k}|$ vector marked in (d); (e) potential landscape of the same area determined by the lowest
tip induced quantum dot state as shown in Fig. \ref{Fig2}(c)-(e); (f) preferential $|\underline{k}|$ values as deduced from the Fourier transformations in comparison with unperturbed
InAs band structure as calculated within $k\cdot p$-approximation; (g) $dI/dV$ image at $V=0$ V of p-InSb(110) ($N_{\rm A}=1-2\cdot 10^{24}$ m$^{-3}$) covered with 1.5 \% Cs ($V_{\rm stab}=0.3$ V, $I_{\rm stab}=0.2$ nA, $V_{\rm mod}=1.0$ mV), lower inset: Fourier transformation of the real space data, upper inset: STM image of the same surface ($I=30$ pA, $V=0.3$ V) shown to scale and proving that the dark dots in the $dI/dV$ image are caused by the Cs atoms; (h) ARPES data of p-InSb(110) ($N_{\rm A}=1-2\cdot 10^{24}$ m$^{-3}$) covered with 1.5 \% Cs, $h\nu = 21$ eV; a parabolic approximation of the band of InSb with effective mass marked (dashed line) and the preferential $|\underline{k}|$-values from the Fourier transformation of STS data are included;  $T=6$ K (a-g); $T=80$ K (h).
\protect{\cite{PhysRevLett.89.136806,Becker2010}}
\label{Fig4}}
\end{figure*}
The two-dimensional electronic system (2DES) at the surface induced by adsorbates shows a nearly parabolic dispersion:
\begin{equation}
\label{Disp}
E(\underline{k})-E_n=\frac{\hbar \cdot|\underline{k}|^2}{2m_{\rm e}\cdot m^*}
\end{equation}
with $E(\underline{k})$ being the energy of the state, $E_n$ the subband energy corresponding to the confinement vertically to the surface, $\underline{k}$ the wave vector of the electron parallel to the surface,
$m_{\rm e}=9.1 \cdot 10^{-31}$ kg the bare electron mass, and $m^*$ the effective mass of the conduction band. Thus, the system is a good paradigm for free electrons, which also show parabolic dispersion and exhibit an extremely interesting phase diagram as a function of electron density, temperature and magnetic field. \cite{SRL}
At the energy of the bulk conduction band minimum $E_{\rm BCBM}$, the effective mass of InAs is $m^*_0=0.023$ and of InSb is $m^*_0=0.0135$.\cite{Vurgaftman} In fact, the bands are not perfectly parabolic, but exhibit a reduced curvature at higher energy, which can be approximated by an energy dependent
effective mass \cite{KaneInSb}
\begin{equation}
m^*(E)=m^*_0\cdot \frac{1+2\cdot(E-E_{\rm BCBM})}{E_{\rm Gap}}
\end{equation}
with the band gap being $E_{\rm Gap}=0.43$ eV for InAs and $E_{\rm Gap}=0.23$ eV for InSb at $T=0$ K.\cite{Vurgaftman}\\
For a 2DES, this leads to: \cite{PhysRevB.35.2460}
\begin{equation}
\label{M*}
m^*(E)=m^*_0\cdot \frac{1+2\cdot(1/3(E_n-E_{\rm BCBM})+E_{||})}{E_{\rm Gap}},
\end{equation}
i.e. the subband energy perpendicular to the surface $E_n$ counts differently than the in-plane energy $E_{||}$.
\\
The spin-orbit coupling within the valence band couples to the conduction band, which has preferentially a 5s-orbital character.\cite{Klijn} This can be described, e.g., by $k\cdot p$-theory.
As a consequence, the $g$-factor of the 2DES is strongly enhanced, negative, i.e. spin moments align preferentially antiparallel to $\underline{B}$, and energy dependent.
A good approximation is:\cite{Efros}
 \begin{equation}
 \label{g}
 \frac{g(E)}{g_0}=\frac{m^*_0}{m^*(E)},
 \end{equation}
with $g_0=-15$ for InAs and  $g_0=-51$ for InSb.
Moreover, the Rashba-type spin splitting within a potential gradient given, e.g. by the confinement of the 2DES in $z$ direction, is relatively strong as will be discussed in
chapter \ref{Rashba}.\\
One should keep in mind that the given approximations are only valid at energies $E-E_{\rm BCBM} < E_{\rm Gap}$. Importantly, the LDOS corresponding to this 2DES can be measured by STS, since only a small amount of the surface is covered by adsorbates. In between these adsorbates the 2DES of the substrate can be probed in $dI/dV$ images measured within the
bulk band gap of the semiconductor. The result at two different strength of the disorder potential are discussed next.\\
At strong disorder, some of the electronic wave functions of the 2DES get strongly localized, i.e. they are completely confined within a single valley of the potential disorder.
Approximating the potential trough marked by a red dot in Fig. \ref{Fig3}(d) as parabolic with an extension of $r=15$ nm at 75 meV, i.e.
\begin{equation}
V(x,y)=0.75 \frac{\rm meV}{\rm nm^2}(x^2+y^2),
\end{equation}
one gets single particle energies $E_n \simeq (n+0.5)\cdot 40$ meV, which is reasonably smaller than the height of the potential trough.
Thus, confinement of about the first two single particle levels is expected and, indeed, observed.\cite{PhysRevB.68.041402} Figure \ref{Fig3}(f) shows a $dI/dV$ curve exhibiting two sharp peaks
with an energy distance of 20 meV. A $dI/dV$ map at the corresponding energies, which are plotted here for another trough,
is shown in Fig. \ref{Fig3}(g) and (h) exhibiting an elliptical LDOS distribution for the lower peak and
a lobe-like distribution as known for p-states at the upper peak. Thus, the strongly localized states are similar to confined states within parabolic quantum dots.\cite{PhysRevB.68.041402} The quantitative discrepancy of the confinement energy is probably caused by the nonparabolicity of the potential trough.
\begin{figure*}[htb]
\includegraphics[width= \textwidth]{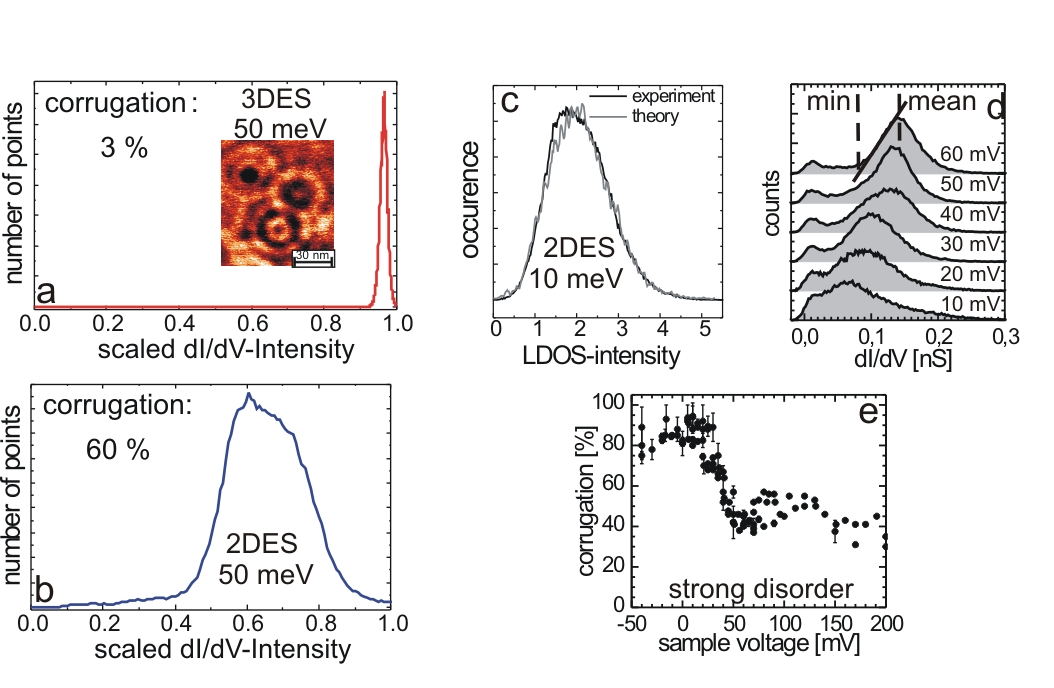}
\caption{(a) histogram of $dI/dV$ values obtained on  n-InAs(110) ($N_{\rm D}=1.1\cdot 10^{22}$ m$^{-3}$) after cleavage  with corrugation $C_{\rm LDOS}$ marked ($V=0.05$ V, $I_{}=0.4$ nA, $V_{\rm mod}=8.5$ mV), inset: part of the $dI/dV$ image used for evaluating the histogram; (b) same as (a) for
n-InAs(110) ($N_{\rm D}=1.1\cdot 10^{22}$ m$^{-3}$) covered with 2.7 \% Fe ($V=0.10$ V, $I_{}=0.3$ nA, $V_{\rm mod}=1.8$ mV); (c) histogram of $dI/dV$ values of
n-InAs(110) ($N_{\rm D}=1.1\cdot 10^{22}$ m$^{-3}$) covered with 0.8 \% Fe ($V=0.10$ V, $I_{}=0.3$ nA, $V_{\rm mod}=1.8$ mV) in comparison with the histogram of calculated LDOS
by diagonalizing the Hamiltonian with the experimentally determined disorder potential using periodic boundary conditions; the experimental curve is stretched by 5 \%;
(d) histogram of $dI/dV$ values obtained on  p-InAs(110) ($N_{\rm A}=4.6\cdot 10^{23}$ m$^{-3}$) covered with 15 \% Co at the voltages marked ($V_{\rm stab}=0.43$ V, $I_{\rm stab}=0.7$ nA, $V_{\rm mod}=5$ mV); $mean$ and $min$ as used for the determination of $C_{\rm LDOS}$ are marked; (e) corrugation $C_{\rm LDOS}$ as a function of energy deduced from the histograms shown in (d);  $T=6$ K.
\protect{\cite{PhysRevLett.89.136806,PhysRevB.68.041402}}
\label{Fig5}}
\end{figure*}
Notice that the black spots within the LDOS wave functions are caused by the presence of the Co islands, which are Coulomb blocked. \\
Next, we discuss the LDOS data obtained at lower disorder. Fig. \ref{Fig4}(a)-(d) shows the LDOS images recorded for the potential shown in Fig. \ref{Fig4}(e) exhibiting fluctuations of about 20 meV only. The total intensity in each image corresponds to 40 complete electronic states, but since the scattering length of individual states
is about the image size, more states contribute to the LDOS with part of its intensity distribution.\\
The LDOS images exhibit corrugations decreasing in length scale with increasing voltage. The corrugation patterns
are rather complicated and do not exhibit the simple circular ring structures found in the InAs 3DES \cite{Ewelle} or found for parabolic surface states
on the surface of metals.\cite{Crommie,Besenbacher}
Fourier transforms (FT's) of the LDOS (insets) reveal the distribution of contributing $k$-values.
At low voltages a circle is visible, which at higher voltages is confined by a ring. A plot of the $k$-values corresponding
to the rings  and, thus, dominating the spectrum is shown as a function of voltage in Fig. \ref{Fig4}(f). At low voltages where the ring is not apparent the outer diameter of the circle is taken. For comparison the $E(k)$-dispersion of InAs according to $k\cdot p$ theory (see eq. \ref{Disp} and \ref{M*}) is drawn which exhibits good agreement with the data.
However, we do not observe a ring only, which is due to the mixing of states by the disorder potential according to eq. \ref{mixing}.
The same conclusions can be drawn from Fig. \ref{Fig4}(g) and (h) which show the LDOS at the Fermi level of a p-InSb surface covered with 1.5 \% Cs. Again a regular wave pattern
not exhibiting simple ring like structures is visible and the Fourier transform exhibits a ring with additional intensity in the center. The resulting $k$-values of the ring are compared with ARPES results and a parabolic dispersion (Fig. \ref{Fig4}(h)), both resulting in good agreement. Thus, obviously the wave pattern belongs to the adsorbate induced 2DES.
Moreover, it is clearly visible that the adsorbates only marginally disturb the image quality of the 2DES appearing either as white spikes in Fig. \ref{Fig4}(a)-(d) or as black dots in Fig. \ref{Fig4}(g).\\

Interestingly, the corrugation of the LDOS pattern is distributed rather homogeneously across the sample.
The corrugation strength, defined by the ratio between spatially fluctuating $dI/dV$-intensity and total $dI/dV$-intensity according to
\begin{equation}
C_{\rm LDOS}=\frac{dI/dV_{\rm mean}-dI/dV_{\rm min}}{dI/dV_{\rm mean}}
\end{equation}
with $dI/dV_{\rm min}$ and $dI/dV_{\rm mean}$ determined as sketched in Fig. \ref{Fig5}(d),
is of the order of $50\pm 5$ \% as shown by the scaled histogram in Fig. \ref{Fig5}(b) and (c). This value is much larger than the corrugation strength in the 3DES, which is $3 \pm 0.5$ \% as shown in Fig. \ref{Fig4}(d).\cite{Ewelle}
Moreover, the corrugation can be reproduced by the single particle calculation according to eq. \ref{mixing} (see Fig. \ref{Fig5}(c)).
Both results reflect the much higher probability of multiple scattering of an electron within a 2DES which eventually leads to weak localization.\cite{Anderson}
Thus, many different scattering paths containing each many scattering events contribute to the LDOS leading to more intricate patterns. The fact,
that the probability for back scattering is enhanced in 2D, results in the stronger corrugation.\cite{Beenakker}
Since we map the LDOS and not individual states, we cannot determine the localization length as long as it is
larger than the average distance between localized states contributing to an LDOS image being about 65 nm in Fig. \ref{Fig4}(g).
However, the transition between strong localization and weak localization can be determined by the change of LDOS corrugation
This is demonstrated in Fig. \ref{Fig5}(d) showing histograms for different voltages of the Co/p-InAs(110) system.
The low $dI/dV$ peak at high voltages corresponds to the Co islands and, thus, not to the 2DES. Obviously, the 2DES peak starts
at $dI/dV= 0$ nS at low energy, but gets distinct from the peak of the Co islands at higher voltage. The corrugation strength $C_{\rm LDOS}$
drops from about 100 \% to about 50 \% at a sample voltage of 50 meV, where the electrons become free from local confinement, i.e. strong localization.
Although we do not have a quantitative understanding of this drop, it appears to be a strong indication of the transition
between strong and weak localization.\\
Obviously, disorder potential and LDOS of a 2DES can be mapped rather precisely. This is a crucial result, since the typically known band parameters,\cite{Vurgaftman} the potential
landscape, and the electron density completely determine the appearance of the LDOS. Thus, detailed comparison with theoretical models becomes possible.

\section{Localized and extended states in the Quantum Hall regime}
\label{QHE}
\begin{figure*}[htb]
\includegraphics[width= \textwidth]{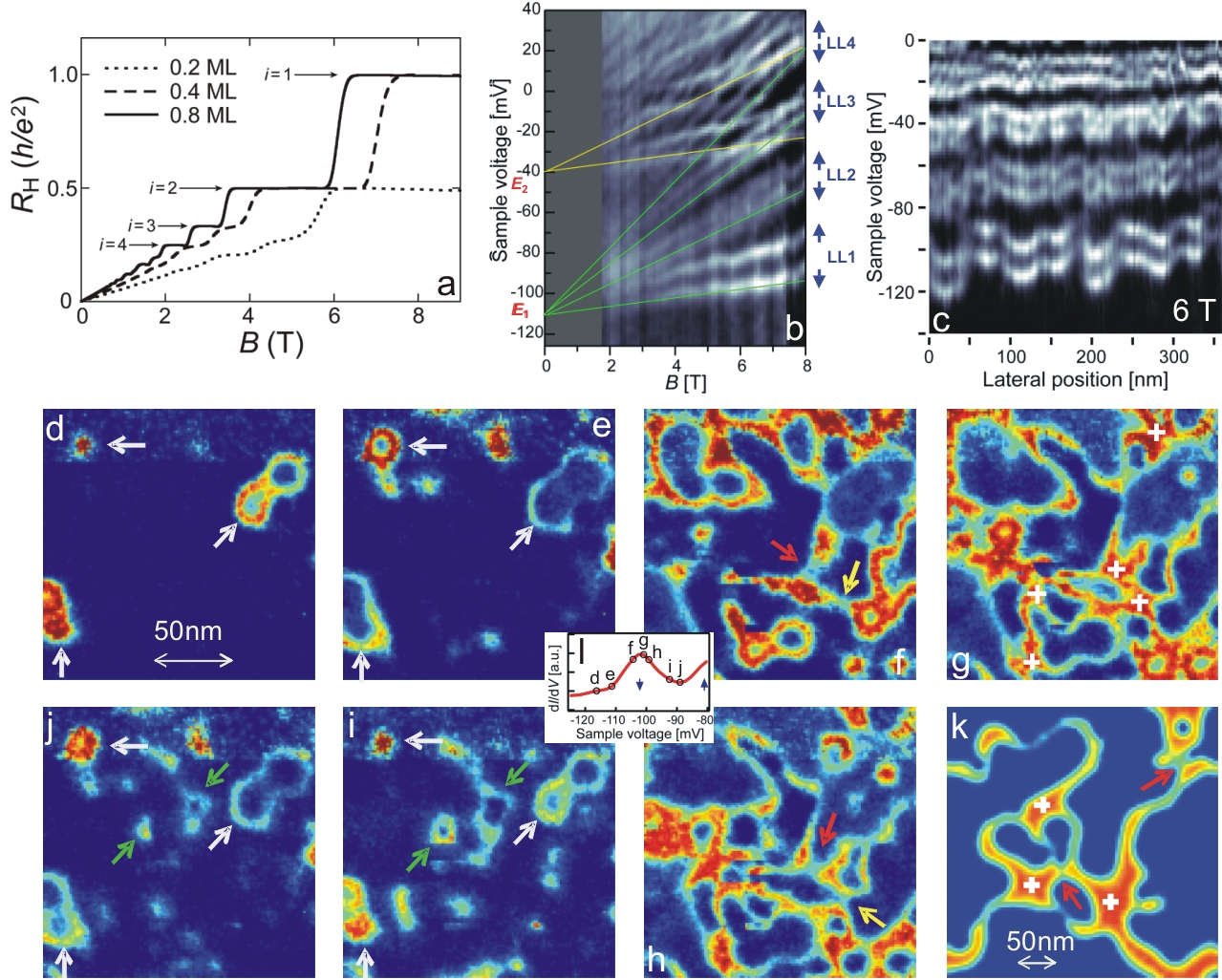}
\caption{(a) Hall resistance of p-InSb(110) ($N_{\rm A}=1-2\cdot 10^{21}$ m$^{-3}$) covered with Ag of the amount indicated, the resistance is measured by 4-probe lock-in technique (13 Hz), $T= 2$ K (courtesy of R. Masutomi, Tokyo) \protect{\cite{Masutomi}}; (b)
grayscale plot of $dI/dV$ intensity as a function of voltage and applied magnetic field $B$, green and yellow lines mark spin-down Landau levels of first ($E_1$) and second ($E_2$) subband, respectively, Landau level (LL) numbers and spin directions of the first subband are marked on the right ($V_{\rm stab}=0.15$ V, $I_{\rm stab}=0.1$ nA, $V_{\rm mod}=1.5$ mV); (c)  grayscale plot of $dI/dV$ intensity as a function of voltage and tip position with respect to the surface ($B=6$ T, $V_{\rm stab}=0.15$ V, $I_{\rm stab}=0.13$ nA, $V_{\rm mod}=1.3$ mV); (d)-(j) $dI/dV$ images at $B=12$ T and at the voltages indicated in (l): $V=-116,3$ mV, $-111.2$ mV, $-104.4$ mV, $-100.9$ mV, $-99.2$ mV, $-92.4$ mV, $-89.0$ mV, i.e. across the lowest spin-down Landau level, white arrows in (b), (e), (i), (j) mark drift states in potential valleys, green arrows in (h), (j) mark drift states on potential hills, red and yellow arrows in (f), (h) mark remaining intensity at saddle points; crosses in (g) mark saddle points, too ($V_{\rm stab}=0.15$ V, $I_{\rm stab}=0.1$ nA, $V_{\rm mod}=1.0$ mV);
(k) calculated LDOS at the center of LL0 $\downarrow$ at $B=12$ T, crosses and red arrows have the same meaning as in (f)-(h); (l) spatially averaged $dI/dV$ curve corresponding to the images in (d)-(j) as marked, arrows mark the spin direction of the lowest Landau level LL0; (b)-(j),(l): $T=0.3$ K.
\protect{\cite{hashimoto:256802}}
\label{Fig6}}
\end{figure*}
One of the most surprising effects of a 2DES is the quantum Hall effect (QHE)\cite{KvK},
which is often used to determine the quality of a 2DES.
Applying a magnetic field $\underline{B}$ perpendicular to the 2DES
and driving a current $I$ through the 2DES, one finds certain
$B$-field regions, where the voltage drop parallel to the current vanishes, i.e. $V_{xx}=0$ V, while
the voltage transversal to the current direction $V_{\rm Hall}$ exhibits plateaus exactly at
\begin{equation}
V_{\rm Hall}= \frac{h}{e^2\cdot i}\cdot I
\end{equation}
where $i$, the so called filling factor, is the integer closest to $h\cdot N_{\rm 2DES}/(eB)$. The value of $R_{\rm H}= V_{\rm Hall}/I$ is based on fundamental constants only and independent of the
details of the 2DES. It is used, e.g., as a resistance standard\cite{Jeckelmann} and discussed as an ingredient with respect to a mass standard.\cite{massstandard}\\
The Hall resistance of an adsorbate induced 2DES is shown in Fig. \ref{Fig6}(a) for different electron densities with the plateaus for $i=1-4$ marked.\\
The origin of the QHE is an interplay of Landau quantization and disorder.\cite{Prange, Ando2, Kramer} Without disorder the single particle energies of the 2DES are quantized
according to
\begin{equation}
\label{LL}
E_{m,n,s}=(n-0.5)\cdot\frac{\hbar eB}{m^*(E)m_{\rm e}}+s\cdot g(E)\mu_B B +E_m
\end{equation}
with $s=1/2,-1/2$ being the spin quantum number, $n$, $ m$ being positive integers and $\mu_B=5.8\cdot 10^{-5}$ eV/T being the Bohr magneton.
The first two terms are dubbed Landau quantization and spin quantization and $E_m$ is the subband energy of the 2DES discussed in chapter \ref{ARPES}.
The development of these energies with $B$ field can be probed by STS as shown in Fig. \ref{Fig6}(b). The measured $dI/dV$ intensity at a single point
is shown as a function of $V$ and $B$ exhibiting two fan-like ensembles of lines, which are labeled by subband energies $E_m$.
The different Landau levels (LL$n$) and spin levels ($\uparrow$, $\downarrow$) are visible and the linear energy dependence on $B$ field is
discernable. The corresponding single particle states without disorder are highly degenerate and can be described as rings with different radii for different LL$n$ encircling a single flux quantum
$\Psi_0= h/e$ each. This leads to a level degeneracy of $N_{n,m,s}=eB/h$ independent on the Landau level number or the spin direction.\\
However, these states are subject to random potential disorder within a semiconductor.
Thus, they change their energy as a function of position as demonstrated in Fig. \ref{Fig6}(c), where the pairs of Landau levels
belonging to different spin quantum numbers meander in energy as a function of position.
The fact that the meandering amplitude is larger for the lower lying Landau levels mimics the fact that the radius of the rings called cyclotron radius
$r_{{\rm c}n}=\sqrt{(2(n-1)+1)\cdot \hbar/(eB)}$ increases with the Landau level index.
Thus, the LDOS of higher Landau levels probes the potential on a rougher length scale.\cite{Florens1}
In 2D, the disorder potential has primarily a semiclassical effect:\cite{Prange} the electrons perform the fast
cyclotron rotation within the electrostatic disorder, which leads to additional drift motion along the equipotential lines of the disorder potential.\cite{Mirlin}
Basically, the electrons are accelerated and decelerated, if they move downhill or uphill within the potential disorder during their cyclotron rotation.
This results in
different radii of curvature of the electron path at lower and higher potential energy, directly implying a motion perpendicular to the gradient of the potential
as long as the gradient direction remains similar on the length scale of the cyclotron radius.
Quantum mechanically, so-called drift states meander along equipotential
lines with a width of about the cyclotron radius
$r_{{\rm c}n}$.\cite{Ando2,Driftstates}\\
If the potential energy of the state is low (high), the drift states are closed trajectories around potential minima (maxima), i.e.\ they are localized and represent insulating electron phases. Thus, whenever if the Fermi level is located at energies belonging to localized states, the longitudinal conductivity $\sigma_{xx}$ vanishes for $T\rightarrow 0$ K.
Since the longitudinal resistance $\rho_{xx}$ is:
\begin{equation}
\rho_{xx}=\frac{\sigma_{xx}}{\sigma_{xx}^2+\sigma_{xy}^2}
\end{equation}
and the Hall conductance is $\sigma_{xy}\neq 0$ S at $B>0$ T, $\rho_{xx}$ vanishes as well.\\
Only, in the center of a LL, the equipotential lines of adjacent potential valleys and hills
merge at the saddle points of the potential leading to an extended state. It can be shown that exactly one equipotential line
traverses the whole sample in the limit of infinitely large samples.\cite{Kramer} It is known that the state meandering along this line
is the quantum critical state of the integer QH transitions and responsible for the finite
longitudinal resistance between quantized values of the Hall conductance
\cite{Prange, Ando2, Kramer, Evers}. This quantum phase transition between localized states in the valleys via an extended state
towards localized states at hills of the potential disorder is repeated for each ($m,n,s$) level of eq. \ref{LL}.
This transition is universal, i.e. the energy dependent localization length $\xi(E)$ of the states, defined as
\begin{equation}
<|\Psi_i(\underline{x}-\underline{x}_{0,i})|^2)\cdot \delta(E-E_i)>_i \propto e^{-|\underline{x}|/\xi(E)}
\end{equation}
($\underline{x}_{0,i}$: center of mass of $|\Psi_i|^2$) is independent of details of the disorder being
\begin{equation}
\xi(E) \propto |E-E_{\rm crit}|^{\nu_{\rm c}}
\end{equation}
with $E_{\rm crit}$ being the energy of the extended state and $\nu_{\rm c} $ being the universal, critical exponent. The value of $\nu_{\rm c}$ is not known analytically,
but has been evaluated numerically, which results in $\nu_{\rm c} \simeq 2.4-2.6$.\cite{Slevin,Roemer,Chalker} Another universality is the multifractal spectrum of the critical state.\cite{Aoki,Evers2}
\\
The corresponding transition is shown for the LDOS of the lowest energy ($m,n,s$) level of an adsorbate induced 2DES in Fig. \ref{Fig6}(d)-(l).\cite{hashimoto:256802}
The spatially averaged d$I/$d$V$ curve is shown in Fig.\ \ref{Fig6}(l) exhibiting the single peak of LL1$\downarrow$. The energies of the LDOS images are marked.
In the low-energy tail of the peak (Fig.\ \ref{Fig6}(a)), the LDOS exhibits spatially isolated closed-loop
patterns with averaged full width at half maximum (FWHM) $\simeq$ 6.9 nm close to the cyclotron radius \textit{r}$_{\rm{c}1}$ = 7.4 nm.
These are localized drift states of the $n=1,m=1, S=-1/2$ level LL0$\downarrow$ aligning along
equipotential lines around a potential minimum. Accordingly, at slightly
higher energy [Fig.\ \ref{Fig6}(b)], the area encircled by the drift states increases
indicating that each drift state probes a longer equipotential line at higher
energy within the same valley. In contrast, the ring patterns at the
high-energy tail, marked by green arrows in Fig. \ \ref{Fig6}(i), (j), encircle an
area decreasing in size with increasing voltage. These states are attributed to
localized drift states around potential maxima. Notice that the structures
in Fig.\ \ref{Fig6}(d) and (e) appear nearly identical in Fig. \ref{Fig6}(i) and (j) as marked
by white arrows. The latter structures are the LL1$\uparrow$ states
localized around potential minima, which energetically overlap with the
high-energy LL1$\downarrow$ states localized around potential maxima. When the
voltage is close to the LL1$\downarrow$ center [Fig.\ \ref{Fig6}(f), (h)], adjacent drift states
coalesce and a dense network is observed directly at the LL center [Fig.\ \ref{Fig6}(g)]. This
is exactly the expected behavior of an extended drift state at
the QH transition as described above.\cite{Ando2,Kramer}  Fig.\ \ref{Fig6}(k)
shows the calculated LDOS around an extended state at $B = 12$ T in a 2DES of InSb with a random distribution of dopants with the experimentally
known densities $N_{\rm D}=9\cdot10^{21}$ m$^{-3}$ and $N_{\rm A}=5\cdot10^{21}$ m$^{-3}$. Each dopant as usual is surrounded by a screened Coulomb potential.
The Hartree approximation \cite{Sohrmann} is used for the diagonalization of the matrix. Good qualitative agreement with the measurement is achieved supporting the interpretation of the coalesced LDOS patterns as caused by the presence of an extended state, albeit the disorder in the calculations is obviously smoother, which could be traced back to a remaining influence of the Cs chains
that partially form on the surface.\cite{hashimoto:256802} The crosses within Fig. \ref{Fig6}(g) and (k) mark the saddle points of the potential appearing as extended LDOS areas due to the vanishing steepness of the potential at the saddle points. The same points are marked by arrows in Fig.\ref{Fig6}(f) and (h), where remaining intensity is found at the saddles.
It has been argued initially that this intensity is related to the quantum tunneling across the saddles which is decisive for the value of the critical exponent $\nu$,\cite{Kramer,Chalker} but numerical calculations revealed that this is not the case.\cite{Florens2} Instead, it reflects only the energy resolution of the experiment of 2.5 meV being significantly larger than the expected $\delta E \simeq 3.3\cdot k_{\rm B} T=0.1$ meV.\cite{hashimoto:256802} This might be related to the slow Coulomb glass dynamics in localized systems,\cite{Becker2011,Menashe} but further experiments are required to clarify this issue.\\
Similar transitions between localized and extended states have also been observed on the surface of highly oriented pyrolytic graphite (HOPG)\cite{Niimi} and on turbostratic graphene on top of graphite on SiC(000$\overline{1}$).\cite{Miller,Stroscio2,GrapheneReview} Recently, it was also investigated in higher Landau levels of the InSb 2DES, where an additional nodal structure of the LDOS perpendicular to the equipotential lines was revealed and explained.\cite{Hashi2}\\
Importantly, the LDOS across a universal quantum phase transition, the Quantum Hall transition, can be directly mapped
within the adsorbate induced 2DES. Improved theoretical tools using vortex
states within the framework of real-time Green's-functions allow, on the other hand, a direct calculation of LDOS properties for sufficiently smooth disorder.\cite{Florens1,Florens2,Florens3} Thus, a one-to-one
comparison of LDOS patterns pinpointing to the local signatures of the influence of particular interactions appears possible.\\
Of course, the quantum phase transition explains only the vanishing conductivity $\sigma_{xx}$ at most $B$ fields but not the quantization of the Hall conductance.
This can either be explained rather elegantly in terms of Chern numbers\cite{Kohmoto,Avron,Laughlin} or less abstract by the presence of the so-called edge states.\cite{Halperin}
The latter explanation is based on the fact that the drift states at the edge of the sample are meandering along the edge and, thereby, connect different leads. Moreover, they are chiral, i.e. they are allowing electron motion only in one direction and not in the opposite one, which corresponds to the drift direction explained above. If the Fermi level is within the localized states of the Landau level, the edge states at the Fermi level are the only ones which can carry the current. Since backscattering
is not possible within these states, no voltage drop can appear along the edge states.
Moreover, only a voltage imbalance between states on the left hand side of the sample carrying the
forward electrons and on the right hand side of the sample carrying the backward electrons can lead to an effective current $I$ through the sample.
It is straightforward to calculate that
the relation between effective current $I$ and voltage imbalance $V_{\rm Hall}$ is $I/V_{\rm Hall}= e^2/h$ for each pair of edge states. Moreover, each critical state below the Fermi level
requires one of these edge states crossing the Fermi level at each side due to the rising confinement potential at the edge.
These two facts together lead to
\begin{equation}
R_{\rm Hall}= \frac{h}{e^2\cdot i}
\end{equation}
with $i$ being the number of critical states below $E_{\rm F}$. So far the edge states have not been probed by STS, but by other scanning probe techniques as scanning
gate microscopy (SGM),\cite{McEuen} electrostatic force microscopy,\cite{Weiss} scanning capacitance microscopy (SCM),\cite{Ashoori} scanning near-field optical microscopy (SNOM),\cite{Ito} and a scanning single electron transistor (SSET).\cite{Yacoby2} However, all these techniques exhibit a much worse lateral resolution well above 100 nm.
Thus, it would be interesting to probe the edge states by STS, too. SSET has also been used to map properties within the fractional quantum Hall regime, which is guided
by electron-electron interactions. These experiments  showed rather directly the fractional excitation charge of $\Delta Q=e/3$.\cite{Ilani}

\section{Probing the exchange interaction}
\label{exchange2}
\begin{figure*}[htb]
\includegraphics[width= \textwidth]{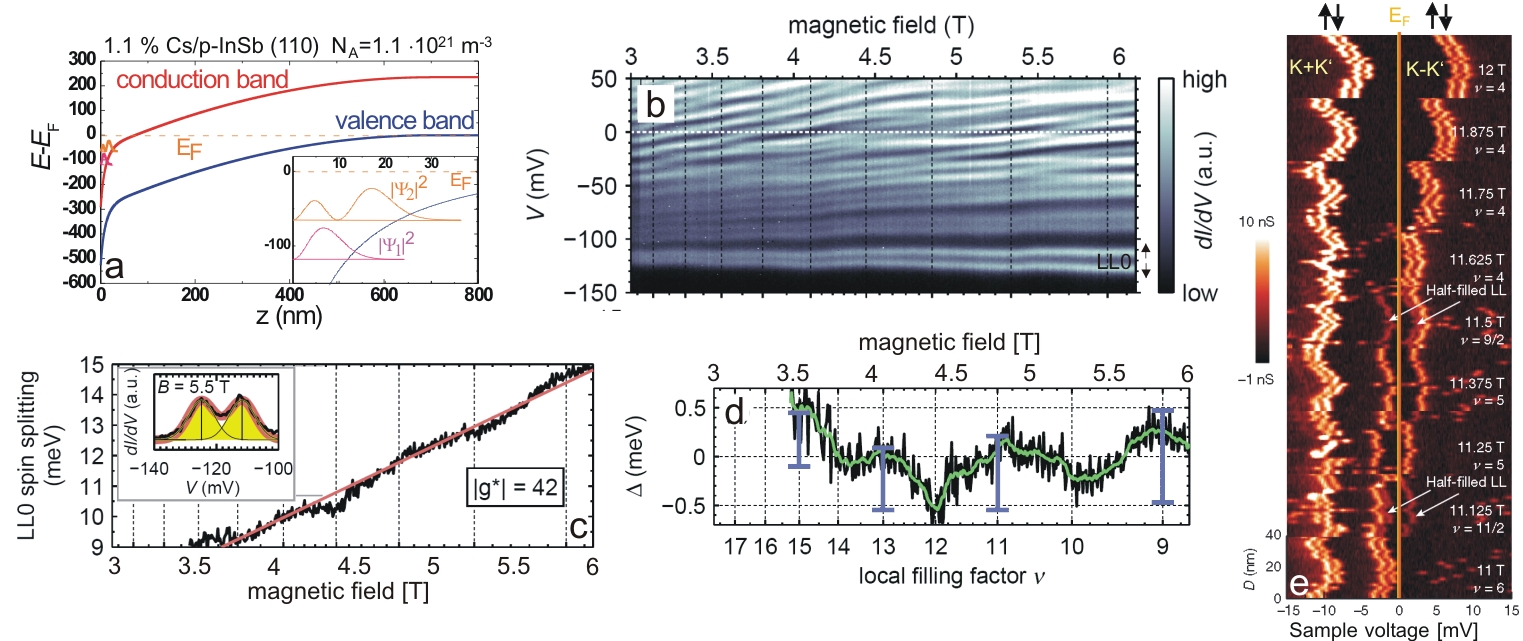}
\caption{(a) Band bending as calculated by a Poisson-Schr\"odinger solver for p-InSb(110) ($N_{\rm A}=1.1\cdot 10^{21}$ m$^{-3}$) covered with 1.1 \% Cs, valence band maximum (blue),
conduction band minimum (red), Fermi level $E_{\rm F}$ and $|\Psi_n(z)|^2$ of the first two subbands are shown in the direction perpendicular to the surface,
inset: magnification of the near surface area; (b) grayscale plot of $dI/dV$-intensity as a function of magnetic field $B$ and applied sample voltage $V$; Fermi level (dashed line) and two spin levels of LL0 are marked ($T=5$ K, $V_{\rm stab}=0.3$ V, $I_{\rm stab}=0.4$ nA, $V_{\rm mod}=1.6$ mV); (c) energy splitting between $\uparrow$ and $\downarrow$ level of the lowest Landau level LL0 (black line) in comparison with the splitting expected at $|g(E)|=42$ (red line), inset: $dI/dV$ curve (black) in the region of LL0 in comparison with two Gaussians colored yellow and leading to the transparent red $dI/dV$ curve, $B=5.5$ T; (d) difference between black and red curve from (c) (black line), smoothed curve is shown in green, scale bars mark the exchange enhancement at the corresponding filling factors calculated within the random phase approximation; (e) colorscale plots of $dI/dV$ intensity of graphene on SiC(000$\overline{1}$) as a function of voltage and tip position with respect to the surface for different magnetic fields $B$ and local filling factors $\nu$ as indicated, levels corresponding to different spins $\uparrow$ and $\downarrow$ and K/K' combinations are marked, ($T=0.01$ K, $V_{\rm stab}=0.25$ V, $I_{\rm stab}=0.2$ nA, $V_{\rm mod}=0.05$ mV) (courtesy of J. Stroscio, NIST Gaithersburg).
\protect{\cite{Becker2011,Stroscio1}}
\label{Fig7}}
\end{figure*}
One of the most simple type of electron-electron interaction is the exchange interaction. For a system of two particles, it reads
\begin{equation}
\label{exchange}
<\Psi_1(\underline{x}_1)\cdot \Psi_2(\underline{x}_2)|V_{\rm ee}(|\underline{x}_2-\underline{x}_1|)|\Psi_1(\underline{x}_2)\cdot \Psi_2(\underline{x}_1)>_{\underline{x}_1, \underline{x}_2}
\end{equation}
where $V_{\rm ee}(\Delta x)$ is the electron-electron interaction potential.\\
Due to symmetry of the many particle wave function, the exchange interaction is attractive for parallel spins and repulsive for antiparallel spins.
Thus, an electron within a spin-polarized 2DES exhibits a lower effective electron-electron repulsion energy, if parallel to the spin-polarization, than an electron of the same spin that is embedded
into a spin neutral 2DES. An electron with antiparallel spin to a spin-polarized 2DES exhibits an even stronger electron-electron repulsion.\\
Applying a magnetic field to a 2DES leads to an oscillating spin polarization. In case, the filling factor $\nu=h N/(eB)$ is odd, the spin polarization is maximum, i.e. only the half of the highest Landau level with the spin moments antiparallel to $\underline{B}$ is filled with electrons. This leads to a spin density
$N_s = s\cdot eB/h$ or to a magnetization $M \simeq \mu_B\cdot eB/\hbar\cdot b/3 = 1.4$ A/m$\cdot B[{\rm Tesla}]$. In case of an even $\nu$, there is no spin polarization, since the same amount of $\uparrow$ levels and $\downarrow$ levels are filled with electrons.\\
In the case of spin polarization, the lower energy spin levels of a Landau level will gain energy with respect to the ones in the higher energy spin level. Thus, the splitting of the two spin levels will be larger than $g(E)\mu_B B$ in the case of spin polarization, and
exactly $g(E)\mu_B B$ without spin polarization, i.e. at even $\nu$. This leads to an oscillatory effective $g_{\rm eff}(B)$-factor determining the spin splitting
$\Delta E_{\rm ss}(B)=g_{\rm eff}(B)\cdot \mu_B\cdot B$ as a function of $B$. The increased spin splitting at odd $\nu$ is called exchange enhancement.\cite{Janak,Ando_exchange}\\
Within the QH regime, exchange enhancement is a rather local effect, since the expression in eq. \ref{exchange} requires an overlap of different localized wave functions. Consequently, the exchange
enhancement depends on the local spin polarization mostly within a single valley or hill region of the potential disorder.
Exchange enhancement has, however, previously only been measured without lateral resolution, e.g. by capacitance spectroscopy.\cite{dial.nature}\\
Figure  \ref{Fig7}(a) shows the band structure of an adsorbate induced 2DES on p-InSb(110) at low doping. One observes that the band bending reaches about 600 nm
into the bulk of the sample. This implies an insulating region of more than 500 nm thickness between the 2DES at the front of the sample and the conducting bulk region of the InSb.
Consequently, the screening of the bulk electrons can be neglected on the scale of the electron-electron distance within the 2DES being only about 10 nm.
This increases the strength of $V_{\rm ee}$. Additionally, the insulating barrier cannot be penetrated by electrons, thus the current between tip and 2DES must flow through the
2DES, which had to be contacted on the side by a silver wire.\cite{Becker2011} Since localization reduces the conductivity of the 2DES in the QH regime, one has to be careful that the tunneling electrons, which tunnel from the 2DES to the tip, are replaced in between individual tunneling events, i.e. the current must be low enough such that local equilibrium is probed by
each tunneling event. Otherwise, the remaining local charge will give rise to an additional local
band bending and, thus, to a locally modified and time-dependent filling factor. Experimentally, we found that a tunneling resistance above 1 G$\Omega$ could fulfill this requirement up to $B= 7$ T and down to $T=5$ K, i.e. the band bending by this so-called spreading resistance remains significantly less than the distance between adjacent levels.\cite{Becker2011}\\
Figure \ref{Fig7}(b) shows $dI/dV(V)$ spectra of such a 2DES taken at a fixed position while ramping the magnetic field $B$. A Landau fan similar to Fig. \ref{Fig5}(b) is visible.
However, in contrast to Fig. \ref{Fig5}(b), the lines of conductance maxima are wavy and not straight.
The reason for waviness is the fixed electron density $N_{\rm 2DES}$ which has to adapt to the degeneracy of the Landau levels. Thus, the distance between $E_{\text{F}}$ and the lowest Landau level jumps each time, $E_{\rm F}$ has to move into the next lower Landau level.
Since $E_{\rm F}$ is fixed in the tunneling experiment a wavy movement of all the Landau levels with respect to $E_{\rm F}$ results.\\
More importantly, the distance between adjacent spin levels $\Delta E_{\text{SS}}$ deviates from $g(E)\mu_B B$.
To analyse this effect, barely visible in Fig. \ref{Fig7}(b), in more detail, we concentrate on the lowest LL around $-120~\text{mV}$, which gives the highest accuracy in determining $\Delta E_{\text{SS}}$. We adapted two Gaussians for all 386 spectra between $3.5$ T and $6.1~\text{T}$. The Gaussians, having equal width and height, were fitted using a nonlinear least squares method and a trust-region algorithm as implemented in Matlab.\footnote{\href{http://www.mathworks.com/help/toolbox/curvefit/}{MathWorks Curve Fitting Toolbox V2.1 User's Guide}} The fits are good as can be seen in the inset of Fig.~\ref{Fig7}(c) and by the large confidence value of $R^2=0.94$ ($0.97$ above $5~\text{T}$). The error for the resulting $\Delta E_{\text{SS}}$ is about $0.2~\text{meV}$. The resulting spin splitting $\Delta E_{\text{SS}}(B)$ as function of $B$ is shown in Fig.~\ref{Fig7}(c) in comparison to a straight line corresponding to ordinary Zeeman splitting of $|g(E)| \mu_{\text{B}} B$ with $|g(E)|=42$.\\
Figure~\ref{Fig7}(d), finally, shows the deviation $\Delta(B)$ from the straight line.
It oscillates around 0~meV with maxima (minima) around odd (even) filling factors as expected for the exchange enhancement.  Negative values of $\Delta(B)$ are probably caused by slight deviations from a spin splitting linear in $B$ due to increased spreading resistance with increasing $B$, which leads to superlinearity, and nonparabolicity of InSb leading to a smooth decrease of $g(E)$ with $B$, thus, supralinearity. However, both effects are, in first order, monotonic in $B$ and cannot explain the oscillations. One could imagine that the spreading resistance depends oscillatory on filling factor. But then, it would be largest at even filling factors, where $E_F$ has the largest energy distance to the next
critical state. This would lead to an oscillation of $\Delta(B)$ with maxima at even filling factor in contrast to the experimental observation. Thus, the oscillatory $B$ field dependence of the spin splitting is in accordance with an exchange enhancement of about 0.5-0.7 meV.\\
To substantiate this assignment, we calculated the expected exchange enhancement in the lowest Landau level using a random phase approximation (RPA). This approximation neglects the dielectric screening at other frequencies than the exciting one.\cite{Bohm,Gellmann}
This is well justified since the electron density in the 2DES $N_{\rm 2DES}$ is large compared to the scale set by the effective Bohr radius, i.e. $N_{\rm 2DES} \simeq 25\cdot (m^*/(\epsilon a_B))^2$,\cite{Ando_exchange,MacDonald} respectively, the $r_s$-parameter is much smaller than one.
We performed the calculation using $m^{*}=0.02$ and $g(E)=-42$ as deduced from the Landau level distance and spin level distance
visible in Fig. \ref{Fig7}(b).\\
The 2D bare Coulomb potential represented in Fourier space is given by
\begin{equation}
\label{Vq}
V(q) = F(q)\frac{2\pi e^2}{\varepsilon q}~
\end{equation}
with the form factor $F(q)$ accounting for the finite extension of the 2DES in $z$ direction:\cite{Ando_exchange}
\begin{eqnarray}
\label{Fq}
F(q) = \frac{3}{8x}+\frac{3}{8x^2}+\frac{1}{4x^3} \hspace{2mm}{\rm with}\\
~x=1+\frac{q}{\sqrt[3]{\frac{48\pi m^*m_{\rm e} e^2}{\varepsilon\hbar^2}\left(\frac{11}{32}N_{\text{2DES}}+N_{\text{Dopant}}\right)}}~.
\end{eqnarray}
with the dielectric constant of InSb $\varepsilon=16.8$, and $N_{\text{2DES}}=2.7 \times 10^{16}~\text{m}^{-2}$ as well as $N_{\text{Dopant}}=8 \times 10^{14}~\text{m}^{-2}$, the 2DES density and the density of ionized acceptors, respectively.\\
Restricting ourselves to the static response of the 2DES to perturbations, the dielectric screening of the potential requires to replace $V(q)$ by
\begin{equation}
V_{\rm RPA}(q) = \frac{V(q)}{1-V(q)\Pi_0(q)}~,
\end{equation}
with
\begin{equation}
\Pi_0(q) =\frac{1}{2\pi r_{{\rm c}1}^2}\sum_{n,m}P_{n,m}
\big(q^2r_{{\rm c}1}^2/2\big)
\sum_s \frac{f(E_{n,s}) - f(E_{m,s})}{E_{n,s}-E_{m,s}}\\
\end{equation}
Here, $f(E) = (1+e^{(E-E_{\rm F})/(k_{\rm B T})})^{-1}$ are Fermi functions evaluated at $T=5$ K, $E_{n,s}$ are the energies corresponding
to the $n$. Landau level and spin level $s$, and
\begin{equation}
P_{n,m}(x) = (-1)^{n+m}e^{-x}L_{m}^{n-m}(x) L_{n}^{m-n}(x)~,
\end{equation}
with $L_{n}^{m-n}(x)$ being associated Laguerre polynomials.
The total exchange energy for an electron in Landau level $n$ with spin $s$ then reads
\begin{widetext}
\begin{equation}
\Sigma_{n}^s =-k_{\rm B}T\int\frac{qdq}{2\pi}\sum_{m} P_{n,m}\left(q^2r_{{\rm c}1}^2/2\right)
\sum_{\Omega_l}\frac{V_{\rm RPA}(q)}{i\Omega_l-E_{m,s}+E_{\rm F}}~,
\end{equation}
\end{widetext}
where  $\Omega_l= (2l+1)\cdot \pi \cdot (k_{\rm B} T)$ (l: integer) are Matsubara frequencies.\\
The filling factor dependent exchange enhancement $EE_n(\nu)$ within the first Landau level ($n=1$) is then
\begin{equation}
EE_1(\nu) = \left(\Sigma^\downarrow_{n=1}-\Sigma^\uparrow_{n=1}\right)_\nu - \left(\Sigma^\downarrow_{n=1}-\Sigma^\uparrow_{n=1}\right)_{\nu-1}~.
\end{equation}
This leads numerically to, e.g., $EE_1(\nu=11)=0.76~\text{meV}$ and $EE_1(\nu=15)=0.55~\text{meV}$.
The calculated $EE_1(\nu)$ for $\nu=15,13,11,9$ are added as vertical bars in
Fig. \ref{Fig7}(d) showing very good agreement with the experimental data.
It has been checked that the static screening approximation used here
is sufficient to reproduce the results of more elaborate approaches \cite{MacDonald} on a quantitative level.
Moreover, the results for $EE_0$ barely change, if the system parameters $g(E)$ and $m^*$ (respectively $N$)
are varied within reasonable limits. For example, the change of $EE_0$ is less than 1~\% for $g(E)=-38$ and still below 10 \%, if $m^*$ is doubled to $m^*=0.04$.\cite{Becker2011}\\
The fact that magnitude and oscillation phase of $\Delta(B)$ compare favorably with a parameter free calculation of the exchange enhancement, is strong evidence
that this short-ranged electron-electron interaction effect is quantitatively probed by STS, despite the screening properties of the close-by tip.
Notice that the potential disorder and the second subband of the 2DES are not included within the calculations and, thus, are of minor importance
for the exchange interaction which happens on the length
scale $r_{\rm c}\simeq 10$ nm smaller than the correlation length of the disorder potential (30-50 nm).\\
The same 2DES also exhibited a Coulomb gap at $E_{\text{F}}$, which is a sign for the presence of long-ranged Coulomb repulsion.\cite{Pollak1970,Efros1975,Pikus1995}
Here, however, the screening by the tip had to be taken into account, in order to describe the observed Coulomb gap quantitatively,\cite{Becker2011} i.e., the tip is described as a screening gate
8.6 nm away from the plane of the 2DES. This stresses the need for less screening tips, as e.g.\ tips made from low-doped semiconductors.\cite{Floehr}\\
Exchange enhancement has also been observed on the system graphene on graphite on SiC(000$\overline{1}$).\cite{Song}
At very low temperature ($T=13$ mK), graphene shows spin splitting and valley splitting. This is demonstrated by the color plots in Fig. \ref{Fig7}(e),
where four lines belonging to a single Landau level meander in energy as a function of position.
Notice that the plots are rotated by $90^\circ$ with respect to Fig. \ref{Fig6}(c), i.e. position is along the vertical and  energy along the horizontal axis.
Moreover, plots recorded at different $B$, as marked on the right, are plotted above each other.
At $B=12$ T, one pair of lines separated by the Zeeman energy of 1.4 meV is observed above $E_{\rm F}$ and one pair of lines appears below the
Fermi level. Each pair corresponds to one of the sublattices of graphene at the measurement position, i.e. to a different, mutually orthogonal, combination of the two valleys
K and K' as marked.
All four lines belong to the $n=1$ Landau level. By decreasing the magnetic field, the degeneracy of each line $n_{\rm LL}$ decreases according to
$n_{\rm LL}=eB/h$ and the levels, which are above $E_{\rm F}$ at higher $B$ must cross $E_{\rm F}$. At the crossing point of the first spin level, the distance between the spin levels increases by about 4 meV and it decreases again, if both spin levels have crossed $E_{\rm F}$. This is an even clearer example of the exchange enhancement, which profits from the low bare $g$-factor $g=2$ of graphene  making the bare spin splitting smaller and of the smaller extension of the graphene electrons in $z$ direction enhancing their Coulomb interaction significantly
according to eq. \ref{Vq} and eq. \ref{Fq}. Careful inspection of Fig. \ref{Fig7}(e), reveals a weak third line (marked by arrows) at half-valued filling factor, which is interpreted as an interaction effect with the quantized 2D system of the underlying graphene layer.\cite{Song}\\
The experiments on exchange enhancement so far have not been used to probe local differences of the exchange enhancement at the same $B$, which is the obvious strength of STM, and, thus, outlines an interesting subject for future research. A prototype experiment has been done using the spin splitting of states within the TIQD, where a non-local correlation of the exchange enhancement with the disorder potential has been found.\cite{spinsplit}\\

\section{Probing the local Rashba effect}
\label{Rashba}
\begin{figure*}[htb]
\includegraphics[width= \textwidth]{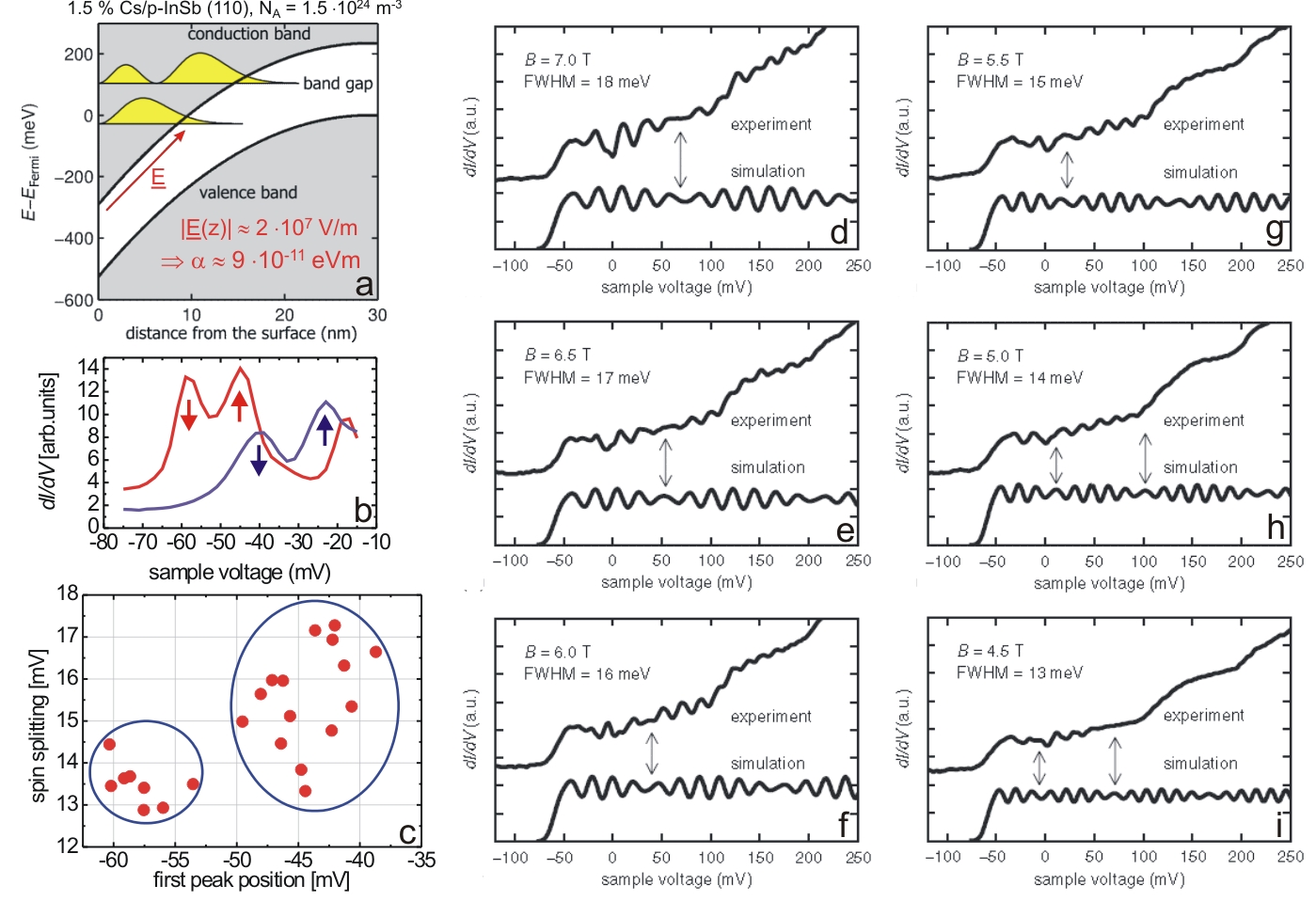}
\caption{(a) Band bending as calculated by a Poisson-Schr\"odinger solver for p-InSb(110) ($N_{\rm A}=1.5\cdot 10^{24}$ m$^{-3}$) covered with 1.5 \% Cs, conduction band, valence band,
and band gap are marked, $|\Psi_n(z)|^2$ of the first two subbands are shown in yellow and electric field$ \underline{E}$ is marked in red, field value and resulting Rashba parameter
$\alpha$ are given; (b) $dI/dV$ curves recorded at different positions of the sample, spin directions are marked ($B=7$ T, $V_{\rm stab}=0.3$ V, $I_{\rm stab}=0.2$ nA, $V_{\rm mod}=1.0$ mV); (c) relation between the peak voltage of the lower spin level and the determined spin splitting; rings mark two different areas where spin splitting is observed and indicate that it is only observed in hill and valley regions
of the potential; (d)-(i) spatially averaged $dI/dV$ curves (area for averaging: (300 nm)$^2$) at $B$-fields indicated ($V_{\rm stab}=0.3$ V, $I_{\rm stab}=0.2$ nA, $V_{\rm mod}=1.0$ mV) in comparison with calculated DOS using $m^*=0.035$, $\alpha=7\cdot 10^{-11}$ eVm, $g(E)=-21$, and a Gaussian level broadening with FWHM as marked, double arrows mark positions of nodes in DOS and $dI/dV$ curve; $T=5$ K.
\protect{\cite{Becker2010}}
\label{Fig8}}
\end{figure*}
As mentioned in the introduction, the exchange interaction between electrons is used within spin based qubits,\cite{Yacoby} which allows to manipulate the spin degree
of freedom electrically via the tunable overlap of wave functions in adjacent quantum dots. For propagating electrons, also the spin-orbit coupling, which eventually is a relativistic
effect originating from the Dirac equation, can be used to manipulate spins via electric fields.\cite{datta:665}
The Rashba term within the Schr\"odinger equation can be written as:\cite{BychkovRashba,Rashba}
\begin{equation}
H_{\rm Rashba} = \alpha \underline{\hat{E}}\cdot(\underline{k}\times {\underline{\bf \sigma}})
\end{equation}
with {\underline{k}} being the wave vector of the electron, $\underline{\bf \sigma}$ being the Pauli matrices and $\underline{\hat{E}}$ being the unit vector along the electric field
typically perpendicular to a surface or an interface. In first order, the Rashba-parameter $\alpha$ within III-V semiconductors depends on the strength of the electric field $\underline{E}$, the energy gap $E_{\rm Gap}$, the effective mass $m^*$ and the spin-orbit splitting $\Delta_{\rm SO}$ within the valence band, which is large for
crystals made of heavy atoms. A good approximation based on an 8-band $k\cdot p$ description of the band structure is:\cite{KaneInSb,AndradaeSilva1994}
\begin{widetext}
\begin{equation}
\alpha = \frac{\hbar^2}{2m^*m_e}\cdot\frac{\Delta_{\rm SO}}{E_{\rm Gap}}\cdot \frac{2E_{\rm Gap}+\Delta_{\rm SO}}{(E_{\rm Gap}+\Delta_{\rm SO})\cdot(3E_{\rm Gap}+2\Delta_{\rm SO})}\cdot e|\underline{E}|
\end{equation}
\end{widetext}
Since In and Sb are heavy 5p-atoms, they exhibit a large $\Delta_{\rm SO}=0.8$ eV at $T=4$ K. Moreover, as described in chapter \ref{ARPES} and \ref{LDOS}, band gap $E_{\rm Gap}=0.23$ eV and effective mass at the band edge $m_0^*=0.0135$ are quite low, such that a relatively large prefactor of $\alpha$ with respect to $\underline{E}$ results:
\begin{equation}
\label{alpha}
\alpha = 5.1 \cdot 10^{-18} \frac{\rm eV m^2}{\rm V}\cdot |\underline{E}|.
\end{equation}
Notice that the spin-orbit term $\Delta_{\rm SO}$ within binary, ternary or quaternary alloys of Bi, Sb, Se, and Te can even lead to a band inversion,
i.e. part of the p-type band becomes the conduction band while the part of the s-type band becomes a valence band.
This leads to the currently celebrated topological insulators,\cite{Fu, Moore, Murakami,Hsieh,Zhang} which exhibit, at least, one surface
state with spin chirality within the fundamental band gap, which, moreover, is protected by time-reversal symmetry.\\
In order to get a strong $\alpha$, one needs, in addition to the prefactor, a strong electric field.
An interesting question is, if charge neutrality prohibits the presence of electric fields within a 2DES.
Winkler has shown that the relevant property for $\alpha$ is the effective field within the valence band, which remains present due to different band parameters in valence and conduction band, and that this more correct description is except for prefactors close to one identical
to the quantitative description given above.\cite{Winkler2}
The strongest field is found at the surfaces of metals
($10^8 - 10^9$ V/m).
Consequently, the strongest $\alpha$ have been found for surface states on metals, e.g., for Bi(111), one has found $\alpha = 0.55\cdot 10^{-10}$ eVm,\cite{Hofmann} while for Bi alloys
even $\alpha = 3.05\cdot 10^{-10}$ eVm has been obtained.\cite{Ast} The latter value is partly attributed to additional interatomic electric fields.\cite{Ast}
However, within metals the electric field can barely be controlled by gates. Thus, spins can not be manipulated on ns time scales as required for information processing.
Most likely, InSb is the semiconductor with the
largest possible prefactor of $\alpha$ not being a topological insulator.\\
The first consequence of $H_{\rm SO}$ is a modification of the band structure. Assuming a parabolic dispersion of the kinetic energy as for the conduction band of
InAs and InSb, the origin of the parabola in $\underline{k}$ is offset from the $\Gamma$-point, which marks the center of the Brillouin zone, by
\begin{equation}
\label{dk}
|\Delta \underline{k}|=\frac{\alpha m^*m_e}{\hbar^2}.
\end{equation}
The $\underline{k}$-direction of offset is different for the states with spins pointing to the right with respect to $\underline{k}$ and for the spins pointing to the left with respect to
$\underline{k}$. Thus, spin degeneracy is lifted and the eigenstates of the spin are always oriented perpendicular to $\underline{k}$ and $\underline{E}$.
The resulting two parabolas for one $\underline{k}$ direction are rotationally symmetric leading to a circular trough in the $E(\underline{k})$ dispersion with radius $|\Delta \underline{k}|$. A cut through $\underline{k}$
space in $E(\underline{k})$ using the plane perpendicular to $\underline{E}$ and energies above the bottom of the trough exhibits two circles with chiral spin texture perpendicular to $\underline{k}$.
Opposite spin directions distinguish the two different circles. The difference between the radii of the circles is exactly the $|\Delta \underline{k}|$ of eq. \ref{dk}, i.e. independent of energy as
long as $m^*$ is independent of energy.\\
The Rashba-split $E(\underline{k})$ dispersion has been measured directly by ARPES,
but so far only for metals\cite{LaShell,Ast,Hofmann,Hoesch,Dedkov} or for metal/semiconductor interface states,\cite{gierz_silicon_2009,Hoepfner,Frantze,Ohtsubo,Yaji}
but not for intrinsic states of a semiconductor.
The reason is the low $m^*$, e.g. for InSb, that results in small $|\Delta \underline{k}| \simeq 1.3/{\rm V} \cdot |\underline{E}|$ between the two circles. This requires
excellent angular resolution in ARPES of the order of $0.1^\circ$.
Instead, the Rashba effect in III-V-semiconductors has been probed by the beating pattern of Shubnikov-de-Haas-oscillations\cite{Seiler,Nitta,Grundler} or by the analysis of weak antilocalization.\cite{Miller,Nitta2}
The beating can be explained by the two different circles in $\underline{k}$ space at the Fermi level, which are both not spin degenerate.
Since each circle with radius $|\underline{k}|_{{\rm F}, i}$ ($i=1,2$) encircles a density
of $n_i=|\underline{k}|_{{\rm F},i}^2/(4\pi) $ $\underline{k}$-points and the degeneracy of a spin polarized Landau level is $eB/h$, the number of filled Landau levels
(within the circle) is:
\begin{equation}
\nu_i= \frac{h|\underline{k}|_{{\rm F},i}^2}{4\pi eB}
\end{equation}
Each time, when $\nu_i$ is an integer, the Landau level highest in energy is completely filled and the conductivity exhibits a minimum. In contrast, a half-valued $\nu_i$, i.e. a Fermi level in the center of a Landau level leads to a maximum in conductivity.
If, both, $\nu_1$ and $\nu_2$ are integers the corresponding minimum is deep, while if $\nu_1$ is an integer and $\nu_2$ is half-valued,
the minimum disappears. From this beating of the oscillation in conductivity one can determine $|\Delta \underline{k}|$ and, thus, $\alpha$.\cite{Nitta}\\
The same effect leads to a beating of the density of states which can be probed by STS. A node of the beating will be obtained at an average $|\underline{k}|$-value of the two circles
$\overline{k}_n$, respectively an energy $\hat{E}_n= \hbar^2\overline{k}^2/(2m_em^*)$ with respect to the subband energy according to:
\begin{eqnarray}
 n+0.5&=&\nu_1-\nu_2\\
 &=&\frac{h(\overline{k}_{n}+|\Delta\underline{k}|/2)^2-(\overline{k}_{n}-|\Delta\underline{k}|/2)^2}{4\pi e B}\\
 &=&\frac{2h\overline{k}_{n}|\Delta\underline{k}|}{4\pi e B}\\
 \Rightarrow \overline{k}_{n} &=&\frac{ e B}{\hbar|\Delta\underline{k}|}\cdot (n+0.5) =\frac{\hbar e B}{\alpha m_e m^*}\cdot(n+0.5)\\
 \Rightarrow \hat{E}_n &=&\frac{\hbar^4e^2B^2}{2 \alpha^2 m^{*3}m_e^3}\cdot(n+0.5)
\end{eqnarray}
Using $\alpha=10^{-10}$ eVm, $B=6$ T, and $m^* =0.03$, one gets $\hat{E}_n =150$ meV $\cdot (n+0.5)$, i.e. the first node of the beating must be about
75 meV above the subband onset. Notice that the energy position of the beating node depends quadratically on
$B$-field and quadratically on the inverse of the Rashba parameter $\alpha$.\\
A more sophisticated description considering Zeeman splitting and Rashba spin splitting on equal footing
has been given by Rashba in his first publication on the effect:\cite{Rashba}
\begin{eqnarray}
\label{Rashbaeq}
E^{n,s}_{i}&=&E_{i} + \hbar\omega_{c} \left( n + 2s \left( \delta^2 + \gamma^2 n  \right)^{1/2} \right), \label{eq:Rashba}\\
\gamma &=& \alpha \left( 2 m^{*} /\hbar^{3} \omega_{c} \right)^{1/2}, \\
\delta &=&  \frac{1}{2} \left( 1 - \frac{m^{*} g(E)}{2 m^*_{0}} \right).
\end{eqnarray}
Here, $n = 0,1,2,\ldots$ is the Landau level index, $s = 1/2$ for $n = 0$, and $s = \pm 1/2$ for $n = 1,2,3,\ldots$ being the spin index, $i$ is the subband index and $\omega_{\rm c}= eB/(m^*m_{\rm e})$.\\
Fig. \ref{Fig8}(a) shows the Poisson-Schr\"odinger result for a 2DES, which is induced by $1.5\,\%$ Cs on the strongly $p$-doped InSb(110).
The strong doping results in a strong electric field ($\ge 10^7\,\text{V}/\text{m}$) within the 2DES and, thus, according to eq. \ref{alpha} to a large Rashba coefficient of
$\alpha \simeq 10^{-10} \,\text{eV}\text{m}$. A more sophisticated calculation taking the curvature of the bands and its overlap with
$\Psi_1(z)$ into account,\cite{Becker2011,AndradaeSilva1994} but still neglecting the penetration of the $\Psi_1(z)$ into vacuum,
leads to $\alpha \simeq 9\cdot 10^{-11} \,\text{eV}\text{m}$. This value is larger than the $\alpha$-values observed in InAs inversion-layers or heterostructures by transport measurements ($3\text{--}4 \cdot 10^{-11}\,\text{eV}\,\text{m}$).\cite{PhysRevB.61.15588} It should be noted that the calculated Rashba parameter is an upper estimate because of the ignored barrier penetration of the electronic wave functions. Furthermore, $\alpha$ is only the lowest order of an inversion asymmetry induced spin splitting and it is known that higher orders lead to a reduced effect.\cite{Winkler.PhysRevB.48.8918}\\
Figure~\ref{Fig8}(d)-(i) shows spatially averaged $dI/dV(V)$ spectra of the Cs covered InSb(110) at differnt $B$. The curves are averaged from 144 curves recorded on a regular grid covering an area of $(300\,\text{nm})^2$. The onset of the first subband at about $- 50$ mV and the onset of the second subband at about 150 mV can be identified as steps in the $dI/dV$ signal. The energies are in reasonable agreement with the result from the Poisson-Schr\"odinger equation. Obviously, only the first subband is occupied by electrons. The electron concentration of the 2DES is $N_{\text{2DES}}\simeq6.5 \times 10^{15}\,\text{m}^{-2}$. On top of the steps oscillations are visible due to Landau quantization, which feature, in addition, a changing amplitude.
The minima of the oscillation are marked by arrows. The distance of the peaks is in agreement with the expected distance of Landau levels. Thus, the spin splitting is not visible in the spatially averaged curves. This is different for single curves as shown for two examples in Fig. \ref{Fig8}(b), which also shows clearly that the energy shift by disorder of about 20 meV is too large to observe spin splitting in the averaged curve.
 Importantly, the beating of the spatially averaged Landau level intensity is quantitatively reproduced by the calculations described in eq. \ref{Rashbaeq} using
$\alpha = 7 \cdot 10^{-11}\,\text{eV}\text{m}$, $m^*=0.035$, $g(E)=-21$ and a disorder broadening of the levels as marked in Fig. \ref{Fig8}(d)-(i).
 Thereby, $m^*$ and $g(E)$ have been calculated as an average within the first subband up to the onset of the second subband from eq. \ref{M*} and eq. \ref{g}. The parameter FWHM is in excellent agreement with the strength of the disorder potential being about 25 meV peak-to-peak and shows the expected trend that it is coarsened by $r_{\rm c1}\propto 1/\sqrt{B}$.
 The remaining fitting parameter $\alpha$ is only slightly lower than the estimated one probably because of neglected barrier penetration and higher orders in $k\cdot p$-description as mentioned above. The nice correspondence between
the measured and the calculated node positions at different $B$ is strong evidence that the Rashba spin splitting has been detected by STM.\cite{Becker2010}\\
 Notice that the Rashba spin splitting is not directly visible in quasiparticle interference patterns probed by STS,
 as one would naively expect from the two Fermi circles. If only single scattering is considered theoretically, the scattering leads to a wave vector of the standing waves of exactly $\overline{k}$, the average of the radii of the two circles.\cite{Pettersen} Subtle changes of the quasiparticle interference appear,
if multiple scattering becomes relevant, e.g. within quantum corrals,\cite{Walls} and a very complex spin distribution within the standing wave should appear, if magnetic scatterers are used.\cite{Bluegel}
Indirect ways to probe the Rashba effect by STS used, e.g., the increased density of states at the band onset of the Bi/Ag(111) and Pb/Ag(111) surface alloys.\cite{Ast2}
This method reveals the Rashba parameter from the strength of the peak at the band onset, but it might be very sensitive to details of the tip density of states. Another method used the complex band structure of Bi(111), which exhibits 6 spin-split Fermi circles, and detects that quasiparticle interference requires spin conservation and, thus suppresses some of the naively expected features in the Fourier transform of STS images.\cite{Pascual} This method allows to prove spin textures in complex band structures and has also been used for topological insulators,\cite{Yazdani} but does not give access to the spin-orbit parameters.\\
Local differences of the Rashba spin splitting have not been reported so far, although they are of large relevance for spin relaxation processes.\cite{Sherman, Sherman2, sherman3, Sherman4}
Fig. \ref{Fig8}(b) shows that the spin splitting indeed varies with position. For a number of curves the spin splitting has been evaluated by the same type of fits as shown in the inset of Fig. \ref{Fig7}(c) and the resulting spin splitting is shown as a function of the first peak position in Fig. \ref{Fig8}(c). Obviously, the average spin splitting is larger at higher potential energies, i.e. at hills of the disorder potential. The difference in average values of the two circles is about 2 meV, although the spread is much larger than the uncertainty of the individual fits and, thus, depends on unknown details of the potential landscape. The stronger spin splitting at higher energy has to be contrasted by the reduced $g(E)$ with increasing energy.
Using $g(E)=-31$ as calculated for the subband onset from eq. \ref{M*} and eq. \ref{g} in combination with the result from the Poisson-Schr\"odinger equation and the average vaules of the spin splitting within the left circle  at around -57 mV, which is $\Delta E_{\rm SS}=13.5$ meV, and the right circle at around -43 meV, which is $\Delta E_{\rm SS}=15.6$ meV, we obtain
Rashba coefficients of $\alpha=5\cdot 10^{-11}$ eVm and $\alpha=9\cdot 10^{-11}$ eVm, respectively. The extrema of the spin splitting in Fig. \ref{Fig8}(c) require $\alpha$ values of
 $\alpha=3-11\cdot 10^{-11}$ eVm. Thus, the Rashba parameter fluctuates by more than $\pm 50$ \% within the potential disorder. A detailed understanding of this fluctuation
 requires further experiments, but a simple estimate offers a tentative explanation for the general effect. First, one has to understand that $\alpha$ is a local parameter, which depends on the local electric field only.\cite{Sherman}
Since the band shift at the surface is constant at high enough adsorbate density and the $z$-extension of the band bending is proportional to $\sqrt{N_{\rm Dopant}^{-1}}$ ($N_{\rm 2DES}$ is negligible at such large doping), the electric field and, thus, $\alpha$ is roughly proportional to $\sqrt{N_{\rm Dopant}}$. Thus the question is, how many dopants contribute to the local band bending
within the area of a single localized state. The length scale of the potential fluctuation is 20 nm, which corresponds to the lateral extension of localized states, while the extension of the electric field in $z$ direction is about 20 nm, too, leading to about $N_{\rm D}=10$ dopants in the electric field area of a single state. The standard deviation of $\pm \sqrt{N_D}=3.2$ leads to
 a fluctuation of $\alpha \propto 3.2 \pm 0.5$, i.e. a FWHM of the $\alpha$ distribution of 30 \%, which is only slightly lower than the experimental value of $\pm 25$ \%.
 Importantly, already a moderate doping leads to a strong fluctuation of $\alpha(x,y)$, which will lead to spin dephasing in spintronic devices independent of the well-known
 Dyakanov-Perel mechanism.\cite{Dyakonov}\\

\section{Conclusions}
\label{conclusion}
In conclusion, the adsorbate-induced 2DES on InAs(110) and InSb(110) have been used to detect the relevant spin parameters for semiconductor spintronics on the local scale.
In particular, the exchange splitting of electrons has been measured firstly. It depends on the local filling factor and is as large as 0.7 meV already at rather high
filling factors above $\nu=10$ for a state far away from the Fermi level. Future experiments will investigate the spatial dependence in more detail using systems with
lower filling factor, which naturally would lead also to lower distances to $E_{\rm F}$ and, thus, larger exchange energies. Using the conducting bulk of Fig. \ref{Fig7}(a) as an additional gate, would be the
best way to reduce the filling factor maybe even down to $\nu =1$ or $\nu=2$. Secondly, the Rashba parameter has been measured showing a rather large value of $\alpha=7\cdot 10^{-11}$ eVm on p-doped InSb(110) covered with Cs. Large fluctuations up to $\pm 50$ \% of the Rashba parameter have been observed within a disorder potential fluctuating by about 20 meV.
This is relevant for spintronic devices using the Rashba effect, since it limits the effective spin life-time implying a need for more detailed studies. Importantly all the STS measurements can be done on a 2DES where
the potential disorder can be measured independently using the tip-induced quantum dot. Also the other parameters like surface band shift and 2DES density can be probed by ARPES.
Thus, full control on the relevant parameters allows to tackle the influence of disorder in great detail.\\
Another interesting field with respect to semiconductor spintronics are the ferromagnetic semiconductors, where room-temperature ferromagnetism is still a matter of debate.\cite{Dietl,Volnianska,AndoK} Mostly, the III-V semiconductors doped with Mn have been studied so far by STS, showing, e.g., the anisotropic shape of the Mn acceptor wave function resembling a butterfly,\cite{Koenraad1} its persistence up to nearest neighbor Mn-Mn distances, where the local deviation from a simple overlap of butterflies is everywhere less than 30 \%,\cite{Koenraad2}
the additional mirror asymmetry of the wave function shape appearing at distance up to 8 monolayers from the (110) surface,\cite{MeierF,Koenraad3}
the exchange interaction between neighboring Mn on the surface,\cite{Yazdani2} and the critical properties of the wave functions
across the energy dependent metal-insulator transition at a Mn doping of 1.5 \%.\cite{Yazdani3} Here, the interplay of disorder and spin properties is again very relevant, most likely also for the ferromagnetic transition temperatures.\cite{Dederichs} This calls for detailed studies by STS also on other types of possible ferromagnetic materials with low conductivity.\\
Thus, STS probing the spin properties of semiconductors will sincerely remain an important and technologically relevant branch of research in the near future.

\begin{acknowledgments}
We want to thank Jan Klijn, Christian Meyer, Daniel Haude, Matthias Getzlaff, Roland Wiesendanger, Robert Johnson, Rudo R\"omer, Torge Mashoff, Lukas Plucinski, Renaud Brochier, Jens Wiebe, Andre Wachowiak, Mike Pezzotta, Katsushi Hashimoto, Marco Pratzer, Volker Meden, Christoph Karrasch, and Christoph Sohrmann for their contribution to the data which have been published previously, Joe Stroscio and Ryuichi Masutomi for contributing images as marked, and Serge Florens, Thierry Champel, Zhenya Sherman, and Roland Winkler for helpful discussions. We acknowledge financial support by the German Science Foundation (DFG) via Mo 858 11-1 and Mo 858 8-2.
\end{acknowledgments}

\bibliography{InSbBIB}

\begin{thebibliography}{100}%
\makeatletter
\providecommand \@ifxundefined [1]{%
 \ifx #1\undefined \expandafter \@firstoftwo
 \else \expandafter \@secondoftwo
\fi
}%
\providecommand \@ifnum [1]{%
 \ifnum #1\expandafter \@firstoftwo
 \else \expandafter \@secondoftwo
\fi
}%
\providecommand \enquote [1]{``#1''}%
\providecommand \bibnamefont  [1]{#1}%
\providecommand \bibfnamefont [1]{#1}%
\providecommand \citenamefont [1]{#1}%
\providecommand\href[0]{\@sanitize\@href}%
\providecommand\@href[1]{\endgroup\@@startlink{#1}\endgroup\@@href}%
\providecommand\@@href[1]{#1\@@endlink}%
\providecommand \@sanitize [0]{\begingroup\catcode`\&12\catcode`\#12\relax}%
\@ifxundefined \pdfoutput {\@firstoftwo}{%
 \@ifnum{\z@=\pdfoutput}{\@firstoftwo}{\@secondoftwo}%
}{%
 \providecommand\@@startlink[1]{\leavevmode\special{html:<a href="#1">}}%
 \providecommand\@@endlink[0]{\special{html:</a>}}%
}{%
 \providecommand\@@startlink[1]{%
  \leavevmode
  \pdfstartlink
   attr{/Border[0 0 1 ]/H/I/C[0 1 1]}%
   user{/Subtype/Link/A<</Type/Action/S/URI/URI(#1)>>}%
  \relax
 }%
 \providecommand\@@endlink[0]{\pdfendlink}%
}%
\providecommand \url  [0]{\begingroup\@sanitize \@url }%
\providecommand \@url [1]{\endgroup\@href {#1}{\urlprefix}}%
\providecommand \urlprefix [0]{URL }%
\providecommand \Eprint[0]{\href }%
\@ifxundefined \urlstyle {%
  \providecommand \doi [1]{doi:\discretionary{}{}{}#1}%
}{%
  \providecommand \doi [0]{doi:\discretionary{}{}{}\begingroup
  \urlstyle{rm}\Url }%
}%
\providecommand \doibase [0]{http://dx.doi.org/}%
\providecommand \Doi[1]{\href{\doibase#1}}%
\providecommand \bibAnnote [3]{%
  \BibitemShut{#1}%
  \begin{quotation}\noindent
    \textsc{Key:}\ #2\\\textsc{Annotation:}\ #3%
  \end{quotation}%
}%
\providecommand \bibAnnoteFile [2]{%
  \IfFileExists{#2}{\bibAnnote {#1} {#2} {\input{#2}}}{}%
}%
\providecommand \typeout [0]{\immediate \write \m@ne }%
\providecommand \selectlanguage [0]{\@gobble}%
\providecommand \bibinfo [0]{\@secondoftwo}%
\providecommand \bibfield [0]{\@secondoftwo}%
\providecommand \translation [1]{[#1]}%
\providecommand \BibitemOpen[0]{}%
\providecommand \bibitemStop [0]{}%
\providecommand \bibitemNoStop [0]{.\EOS\space}%
\providecommand \EOS [0]{\spacefactor3000\relax}%
\providecommand \BibitemShut [1]{\csname bibitem#1\endcsname}%
\bibitem{datta:665}%
  \BibitemOpen
  \bibfield{author}{%
  \bibinfo {author} {\bibfnamefont{S.}~\bibnamefont{Datta}}\ and\ \bibinfo
  {author} {\bibfnamefont{B.}~\bibnamefont{Das}},\ }%
  \bibfield{journal}{%
  \Doi{10.1063/1.102730}{\bibinfo {journal} {Appl. Phys. Lett.}}\ }%
  \textbf{\bibinfo {volume} {56}},\ \bibinfo {pages} {665} (\bibinfo {year}
  {1990})%
  \bibAnnoteFile{NoStop}{datta:665}%
\bibitem{Wolf}%
  \BibitemOpen
  \bibfield{author}{%
  \bibinfo {author} {\bibfnamefont{S.~A.}\ \bibnamefont{Wolf}}, \bibinfo
  {author} {\bibfnamefont{D.~D.}\ \bibnamefont{Awschalom}}, \bibinfo {author}
  {\bibfnamefont{R.~A.}\ \bibnamefont{Buhrman}}, \bibinfo {author}
  {\bibfnamefont{J.~M.}\ \bibnamefont{Daughton}}, \bibinfo {author}
  {\bibfnamefont{S.}~\bibnamefont{von Molnar}}, \bibinfo {author}
  {\bibfnamefont{M.~L.}\ \bibnamefont{Roukes}}, \bibinfo {author}
  {\bibfnamefont{A.~Y.}\ \bibnamefont{Chtchelkanova}},\ and\ \bibinfo {author}
  {\bibfnamefont{D.~M.}\ \bibnamefont{Treger}},\ }%
  \bibfield{journal}{%
  \bibinfo {journal} {Science}\ }%
  \textbf{\bibinfo {volume} {294}},\ \bibinfo {pages} {1488} (\bibinfo {year}
  {2001})%
  \bibAnnoteFile{NoStop}{Wolf}%
\bibitem{Fabian}%
  \BibitemOpen
  \bibfield{author}{%
  \bibinfo {author} {\bibfnamefont{I.}~\bibnamefont{Zutic}}, \bibinfo {author}
  {\bibfnamefont{J.}~\bibnamefont{Fabian}},\ and\ \bibinfo {author}
  {\bibfnamefont{S.}~\bibnamefont{DasSarma}},\ }%
  \bibfield{journal}{%
  \bibinfo {journal} {Rev. Mod. Phys.}\ }%
  \textbf{\bibinfo {volume} {76}},\ \bibinfo {pages} {323} (\bibinfo {year}
  {2004})%
  \bibAnnoteFile{NoStop}{Fabian}%
\bibitem{Rashba}%
  \BibitemOpen
  \bibfield{author}{%
  \bibinfo {author} {\bibfnamefont{E.~I.}\ \bibnamefont{Rashba}},\ }%
  \bibfield{journal}{%
  \bibinfo {journal} {Sov. Phys. Solid State}\ }%
  \textbf{\bibinfo {volume} {2}},\ \bibinfo {pages} {1109} (\bibinfo {year}
  {1960})%
  \bibAnnoteFile{NoStop}{Rashba}%
\bibitem{BychkovRashba}%
  \BibitemOpen
  \bibfield{author}{%
  \bibinfo {author} {\bibfnamefont{Y.~A.}\ \bibnamefont{Bychkov}}\ and\
  \bibinfo {author} {\bibfnamefont{E.~I.}\ \bibnamefont{Rashba}},\ }%
  \bibfield{journal}{%
  \Doi{10.1088/0022-3719/17/33/015}{\bibinfo {journal} {J. Phys. C}}\ }%
  \textbf{\bibinfo {volume} {17}},\ \bibinfo {pages} {6039} (\bibinfo {year}
  {1984})%
  \bibAnnoteFile{NoStop}{BychkovRashba}%
\bibitem{Jantsch1}%
  \BibitemOpen
  \bibfield{author}{%
  \bibinfo {author} {\bibfnamefont{Z.}~\bibnamefont{Wilamowski}}, \bibinfo
  {author} {\bibfnamefont{N.}~\bibnamefont{Sandersfeld}}, \bibinfo {author}
  {\bibfnamefont{W.}~\bibnamefont{Jantsch}}, \bibinfo {author}
  {\bibfnamefont{D.}~\bibnamefont{T\"obben}},\ and\ \bibinfo {author}
  {\bibfnamefont{F.}~\bibnamefont{Sch\"affler}},\ }%
  \bibfield{journal}{%
  \Doi{10.1103/PhysRevLett.87.026401}{\bibinfo {journal} {Phys. Rev. Lett.}}\
  }%
  \textbf{\bibinfo {volume} {87}},\ \bibinfo {pages} {026401} (\bibinfo {year}
  {2001})%
  \bibAnnoteFile{NoStop}{Jantsch1}%
\bibitem{Jantsch2}%
  \BibitemOpen
  \bibfield{author}{%
  \bibinfo {author} {\bibfnamefont{W.}~\bibnamefont{Jantsch}}, \bibinfo
  {author} {\bibfnamefont{Z.}~\bibnamefont{Wilamowski}}, \bibinfo {author}
  {\bibfnamefont{N.}~\bibnamefont{Sandersfeld}}, \bibinfo {author}
  {\bibfnamefont{M.}~\bibnamefont{Muhlberger}},\ and\ \bibinfo {author}
  {\bibfnamefont{F.}~\bibnamefont{Schaffler}},\ }%
  \bibfield{journal}{%
  \Doi{10.1016/S1386-9477(02)00179-0}{\bibinfo {journal} {Physica E}}\ }%
  \textbf{\bibinfo {volume} {13}},\ \bibinfo {pages} {504} (\bibinfo {year}
  {2002})%
  \bibAnnoteFile{NoStop}{Jantsch2}%
\bibitem{Tyryshkin}%
  \BibitemOpen
  \bibfield{author}{%
  \bibinfo {author} {\bibfnamefont{A.~M.}\ \bibnamefont{Tyryshkin}}, \bibinfo
  {author} {\bibfnamefont{S.}~\bibnamefont{Tojo}}, \bibinfo {author}
  {\bibfnamefont{J.~J.~L.}\ \bibnamefont{Morton}}, \bibinfo {author}
  {\bibfnamefont{H.}~\bibnamefont{Riemann}}, \bibinfo {author}
  {\bibfnamefont{N.~V.}\ \bibnamefont{Abrosimov}}, \bibinfo {author}
  {\bibfnamefont{P.}~\bibnamefont{Becker}}, \bibinfo {author}
  {\bibfnamefont{H.-J.}\ \bibnamefont{Pohl}}, \bibinfo {author}
  {\bibfnamefont{T.}~\bibnamefont{Schenkel}}, \bibinfo {author}
  {\bibfnamefont{M.~L.~W.}\ \bibnamefont{Thewalt}}, \bibinfo {author}
  {\bibfnamefont{K.~M.}\ \bibnamefont{Itoh}},\ and\ \bibinfo {author}
  {\bibfnamefont{S.~A.}\ \bibnamefont{Lyon}},\ }%
  \bibfield{journal}{%
  \bibinfo {journal} {Nature Mat.}\ }%
  \textbf{\bibinfo {volume} {11}},\ \bibinfo {pages} {143} (\bibinfo {year}
  {2012})%
  \bibAnnoteFile{NoStop}{Tyryshkin}%
\bibitem{Awschalom1}%
  \BibitemOpen
  \bibfield{author}{%
  \bibinfo {author} {\bibfnamefont{J.~M.}\ \bibnamefont{Kikkawa}}\ and\
  \bibinfo {author} {\bibfnamefont{D.~D.}\ \bibnamefont{Awschalom}},\ }%
  \bibfield{journal}{%
  \Doi{10.1103/PhysRevLett.80.4313}{\bibinfo {journal} {Phys. Rev. Lett.}}\ }%
  \textbf{\bibinfo {volume} {80}},\ \bibinfo {pages} {4313} (\bibinfo {year}
  {1998})%
  \bibAnnoteFile{NoStop}{Awschalom1}%
\bibitem{Awschalom2}%
  \BibitemOpen
  \bibfield{author}{%
  \bibinfo {author} {\bibfnamefont{S.~A.}\ \bibnamefont{Crooker}}, \bibinfo
  {author} {\bibfnamefont{D.~D.}\ \bibnamefont{Awschalom}}, \bibinfo {author}
  {\bibfnamefont{J.~J.}\ \bibnamefont{Baumberg}}, \bibinfo {author}
  {\bibfnamefont{F.}~\bibnamefont{Flack}},\ and\ \bibinfo {author}
  {\bibfnamefont{N.}~\bibnamefont{Samarth}},\ }%
  \bibfield{journal}{%
  \Doi{10.1103/PhysRevB.56.7574}{\bibinfo {journal} {Phys. Rev. B}}\ }%
  \textbf{\bibinfo {volume} {56}},\ \bibinfo {pages} {7574} (\bibinfo {year}
  {1997})%
  \bibAnnoteFile{NoStop}{Awschalom2}%
\bibitem{Awschalom3}%
  \BibitemOpen
  \bibfield{author}{%
  \bibinfo {author} {\bibfnamefont{J.~A.}\ \bibnamefont{Gupta}}, \bibinfo
  {author} {\bibfnamefont{D.~D.}\ \bibnamefont{Awschalom}}, \bibinfo {author}
  {\bibfnamefont{X.}~\bibnamefont{Peng}},\ and\ \bibinfo {author}
  {\bibfnamefont{A.~P.}\ \bibnamefont{Alivisatos}},\ }%
  \bibfield{journal}{%
  \bibinfo {journal} {Phys. Rev. B}\ }%
  \textbf{\bibinfo {volume} {402}},\ \bibinfo {pages} {790} (\bibinfo {year}
  {1999})%
  \bibAnnoteFile{NoStop}{Awschalom3}%
\bibitem{Awschalom4}%
  \BibitemOpen
  \bibfield{author}{%
  \bibinfo {author} {\bibfnamefont{Y.}~\bibnamefont{Ohno}}, \bibinfo {author}
  {\bibfnamefont{D.~K.}\ \bibnamefont{Young}}, \bibinfo {author}
  {\bibfnamefont{B.}~\bibnamefont{Beschoten}}, \bibinfo {author}
  {\bibfnamefont{F.}~\bibnamefont{Matsukura}}, \bibinfo {author}
  {\bibfnamefont{H.}~\bibnamefont{Ohno}},\ and\ \bibinfo {author}
  {\bibfnamefont{D.~D.}\ \bibnamefont{Awschalom}},\ }%
  \bibfield{journal}{%
  \Doi{10.1103/PhysRevB.59.R10421}{\bibinfo {journal} {Nature}}\ }%
  \textbf{\bibinfo {volume} {59}},\ \bibinfo {pages} {R10421} (\bibinfo {year}
  {1999})%
  \bibAnnoteFile{NoStop}{Awschalom4}%
\bibitem{Parkin}%
  \BibitemOpen
  \bibfield{author}{%
  \bibinfo {author} {\bibfnamefont{S.~S.~P.}\ \bibnamefont{Parkin}}, \bibinfo
  {author} {\bibfnamefont{C.}~\bibnamefont{Kaiser}}, \bibinfo {author}
  {\bibfnamefont{A.}~\bibnamefont{Panchula}}, \bibinfo {author}
  {\bibfnamefont{P.~M.}\ \bibnamefont{Rice}}, \bibinfo {author}
  {\bibfnamefont{B.}~\bibnamefont{Hughes}}, \bibinfo {author}
  {\bibfnamefont{M.}~\bibnamefont{Samant}},\ and\ \bibinfo {author}
  {\bibfnamefont{S.~H.}\ \bibnamefont{Yang}},\ }%
  \bibfield{journal}{%
  \bibinfo {journal} {Nature Mat.}\ }%
  \textbf{\bibinfo {volume} {3}},\ \bibinfo {pages} {862} (\bibinfo {year}
  {2004})%
  \bibAnnoteFile{NoStop}{Parkin}%
\bibitem{Parkin2}%
  \BibitemOpen
  \bibfield{author}{%
  \bibinfo {author} {\bibfnamefont{X.}~\bibnamefont{Jiang}}, \bibinfo {author}
  {\bibfnamefont{R.}~\bibnamefont{Wang}}, \bibinfo {author}
  {\bibfnamefont{R.~M.}\ \bibnamefont{Shelby}}, \bibinfo {author}
  {\bibfnamefont{R.~M.}\ \bibnamefont{Macfarlane}}, \bibinfo {author}
  {\bibfnamefont{S.~R.}\ \bibnamefont{Bank}}, \bibinfo {author}
  {\bibfnamefont{J.~S.}\ \bibnamefont{Harris}},\ and\ \bibinfo {author}
  {\bibfnamefont{S.~S.~P.}\ \bibnamefont{Parkin}},\ }%
  \bibfield{journal}{%
  \Doi{10.1103/PhysRevLett.94.056601}{\bibinfo {journal} {Phys. Rev. Lett.}}\
  }%
  \textbf{\bibinfo {volume} {94}},\ \bibinfo {pages} {056601} (\bibinfo {year}
  {2005})%
  \bibAnnoteFile{NoStop}{Parkin2}%
\bibitem{Ohno}%
  \BibitemOpen
  \bibfield{author}{%
  \bibinfo {author} {\bibfnamefont{S.}~\bibnamefont{Ikeda}}, \bibinfo {author}
  {\bibfnamefont{J.}~\bibnamefont{Hayakawa}}, \bibinfo {author}
  {\bibfnamefont{Y.}~\bibnamefont{Ashizawa}}, \bibinfo {author}
  {\bibfnamefont{Y.~M.}\ \bibnamefont{Lee}}, \bibinfo {author}
  {\bibfnamefont{K.}~\bibnamefont{Miura}}, \bibinfo {author}
  {\bibfnamefont{H.}~\bibnamefont{Hasegawa}}, \bibinfo {author}
  {\bibfnamefont{M.}~\bibnamefont{Tsunoda}}, \bibinfo {author}
  {\bibfnamefont{F.}~\bibnamefont{Matsukura}},\ and\ \bibinfo {author}
  {\bibfnamefont{H.}~\bibnamefont{Ohno}},\ }%
  \bibfield{journal}{%
  \bibinfo {journal} {Appl. Phys. Lett.},\ \bibinfo {pages} {082508}}%
   (\bibinfo {year} {2008})%
  \bibAnnoteFile{NoStop}{Ohno}%
\bibitem{Schmidt}%
  \BibitemOpen
  \bibfield{author}{%
  \bibinfo {author} {\bibfnamefont{G.}~\bibnamefont{Schmidt}}, \bibinfo
  {author} {\bibfnamefont{D.}~\bibnamefont{Ferrand}}, \bibinfo {author}
  {\bibfnamefont{L.~W.}\ \bibnamefont{Molenkamp}}, \bibinfo {author}
  {\bibfnamefont{A.~T.}\ \bibnamefont{Filip}},\ and\ \bibinfo {author}
  {\bibfnamefont{B.~J.}\ \bibnamefont{van Wees}},\ }%
  \bibfield{journal}{%
  \Doi{10.1103/PhysRevB.62.R4790}{\bibinfo {journal} {Phys. Rev. B}}\ }%
  \textbf{\bibinfo {volume} {62}},\ \bibinfo {pages} {R4790} (\bibinfo {year}
  {2000})%
  \bibAnnoteFile{NoStop}{Schmidt}%
\bibitem{Appelbaum}%
  \BibitemOpen
  \bibfield{author}{%
  \bibinfo {author} {\bibfnamefont{I.}~\bibnamefont{Appelbaum}}, \bibinfo
  {author} {\bibfnamefont{B.}~\bibnamefont{Huang}},\ and\ \bibinfo {author}
  {\bibfnamefont{D.~J.}\ \bibnamefont{Monsma}},\ }%
  \bibfield{journal}{%
  \bibinfo {journal} {Nature}\ }%
  \textbf{\bibinfo {volume} {447}},\ \bibinfo {pages} {295} (\bibinfo {year}
  {2007})%
  \bibAnnoteFile{NoStop}{Appelbaum}%
\bibitem{Jonker}%
  \BibitemOpen
  \bibfield{author}{%
  \bibinfo {author} {\bibfnamefont{B.~T.}\ \bibnamefont{Jonker}}, \bibinfo
  {author} {\bibfnamefont{Y.~D.}\ \bibnamefont{Park}}, \bibinfo {author}
  {\bibfnamefont{B.~R.}\ \bibnamefont{Bennett}}, \bibinfo {author}
  {\bibfnamefont{H.~D.}\ \bibnamefont{Cheong}}, \bibinfo {author}
  {\bibfnamefont{G.}~\bibnamefont{Kioseoglou}},\ and\ \bibinfo {author}
  {\bibfnamefont{A.}~\bibnamefont{Petrou}},\ }%
  \bibfield{journal}{%
  \Doi{10.1103/PhysRevB.62.8180}{\bibinfo {journal} {Phys. Rev. B}}\ }%
  \textbf{\bibinfo {volume} {62}},\ \bibinfo {pages} {8180} (\bibinfo {year}
  {2000})%
  \bibAnnoteFile{NoStop}{Jonker}%
\bibitem{Grundler}%
  \BibitemOpen
  \bibfield{author}{%
  \bibinfo {author} {\bibfnamefont{D.}~\bibnamefont{Grundler}},\ }%
  \bibfield{journal}{%
  \bibinfo {journal} {Phys. Rev. Lett.}\ }%
  \textbf{\bibinfo {volume} {84}},\ \bibinfo {pages} {6074} (\bibinfo {year}
  {2000})%
  \bibAnnoteFile{NoStop}{Grundler}%
\bibitem{Nitta}%
  \BibitemOpen
  \bibfield{author}{%
  \bibinfo {author} {\bibfnamefont{J.}~\bibnamefont{Nitta}}, \bibinfo {author}
  {\bibfnamefont{T.}~\bibnamefont{Akazaki}}, \bibinfo {author}
  {\bibfnamefont{H.}~\bibnamefont{Takayanagi}},\ and\ \bibinfo {author}
  {\bibfnamefont{T.}~\bibnamefont{Enoki}},\ }%
  \bibfield{journal}{%
  \bibinfo {journal} {Phys. Rev. Lett.}\ }%
  \textbf{\bibinfo {volume} {78}},\ \bibinfo {pages} {1337} (\bibinfo {year}
  {1997})%
  \bibAnnoteFile{NoStop}{Nitta}%
\bibitem{Schaepers}%
  \BibitemOpen
  \bibfield{author}{%
  \bibinfo {author} {\bibfnamefont{G.}~\bibnamefont{Engels}}, \bibinfo {author}
  {\bibfnamefont{J.}~\bibnamefont{Lange}}, \bibinfo {author}
  {\bibfnamefont{T.}~\bibnamefont{Sch\"apers}},\ and\ \bibinfo {author}
  {\bibfnamefont{H.}~\bibnamefont{L\"uth}},\ }%
  \bibfield{journal}{%
  \Doi{10.1103/PhysRevB.55.R1958}{\bibinfo {journal} {Phys. Rev. B}}\ }%
  \textbf{\bibinfo {volume} {55}},\ \bibinfo {pages} {R1958} (\bibinfo {year}
  {1997})%
  \bibAnnoteFile{NoStop}{Schaepers}%
\bibitem{Schoenenberger}%
  \BibitemOpen
  \bibfield{author}{%
  \bibinfo {author} {\bibfnamefont{S.}~\bibnamefont{Sahoo}}, \bibinfo {author}
  {\bibfnamefont{T.}~\bibnamefont{Kontos}}, \bibinfo {author}
  {\bibfnamefont{J.}~\bibnamefont{Furer}}, \bibinfo {author}
  {\bibfnamefont{C.}~\bibnamefont{Hoffmann}}, \bibinfo {author}
  {\bibfnamefont{M.}~\bibnamefont{Graber}}, \bibinfo {author}
  {\bibfnamefont{A.}~\bibnamefont{Cottet}},\ and\ \bibinfo {author}
  {\bibfnamefont{C.}~\bibnamefont{Schonenberger}},\ }%
  \bibfield{journal}{%
  \bibinfo {journal} {Nature Phys.}\ }%
  \textbf{\bibinfo {volume} {1}},\ \bibinfo {pages} {99} (\bibinfo {year}
  {2005})%
  \bibAnnoteFile{NoStop}{Schoenenberger}%
\bibitem{Fiederling}%
  \BibitemOpen
  \bibfield{author}{%
  \bibinfo {author} {\bibfnamefont{R.}~\bibnamefont{Fiederling}}, \bibinfo
  {author} {\bibfnamefont{M.}~\bibnamefont{Keim}}, \bibinfo {author}
  {\bibfnamefont{G.}~\bibnamefont{Reuscher}}, \bibinfo {author}
  {\bibfnamefont{W.}~\bibnamefont{Ossau}}, \bibinfo {author}
  {\bibfnamefont{G.}~\bibnamefont{Schmidt}}, \bibinfo {author}
  {\bibfnamefont{A.}~\bibnamefont{Waag}},\ and\ \bibinfo {author}
  {\bibfnamefont{L.}~\bibnamefont{Molenkamp}},\ }%
  \bibfield{journal}{%
  \bibinfo {journal} {Nature}\ }%
  \textbf{\bibinfo {volume} {402}},\ \bibinfo {pages} {787} (\bibinfo {year}
  {1999})%
  \bibAnnoteFile{NoStop}{Fiederling}%
\bibitem{Hanbicki}%
  \BibitemOpen
  \bibfield{author}{%
  \bibinfo {author} {\bibfnamefont{A.}~\bibnamefont{Hanbicki}}, \bibinfo
  {author} {\bibfnamefont{B.}~\bibnamefont{Jonker}}, \bibinfo {author}
  {\bibfnamefont{G.}~\bibnamefont{Itskos}}, \bibinfo {author}
  {\bibfnamefont{G.}~\bibnamefont{Kioseoglou}},\ and\ \bibinfo {author}
  {\bibfnamefont{A.}~\bibnamefont{Petrou}},\ }%
  \bibfield{journal}{%
  \bibinfo {journal} {Appl. Phys. Lett.}\ }%
  \textbf{\bibinfo {volume} {80}},\ \bibinfo {pages} {1240} (\bibinfo {year}
  {2002})%
  \bibAnnoteFile{NoStop}{Hanbicki}%
\bibitem{Heersche}%
  \BibitemOpen
  \bibfield{author}{%
  \bibinfo {author} {\bibfnamefont{F.}~\bibnamefont{Jedema}}, \bibinfo {author}
  {\bibfnamefont{H.}~\bibnamefont{Heersche}}, \bibinfo {author}
  {\bibfnamefont{A.}~\bibnamefont{Filip}}, \bibinfo {author}
  {\bibfnamefont{J.}~\bibnamefont{Baselmans}},\ and\ \bibinfo {author}
  {\bibfnamefont{B.}~\bibnamefont{van Wees}},\ }%
  \bibfield{journal}{%
  \bibinfo {journal} {Nature}\ }%
  \textbf{\bibinfo {volume} {416}},\ \bibinfo {pages} {713} (\bibinfo {year}
  {2002})%
  \bibAnnoteFile{NoStop}{Heersche}%
\bibitem{Kikkawa}%
  \BibitemOpen
  \bibfield{author}{%
  \bibinfo {author} {\bibfnamefont{J.}~\bibnamefont{Kikkawa}}, \bibinfo
  {author} {\bibfnamefont{I.}~\bibnamefont{Smorchkova}}, \bibinfo {author}
  {\bibfnamefont{N.}~\bibnamefont{Samarth}},\ and\ \bibinfo {author}
  {\bibfnamefont{D.}~\bibnamefont{Awschalom}},\ }%
  \bibfield{journal}{%
  \bibinfo {journal} {Science}\ }%
  \textbf{\bibinfo {volume} {277}},\ \bibinfo {pages} {1284} (\bibinfo {year}
  {1997})%
  \bibAnnoteFile{NoStop}{Kikkawa}%
\bibitem{Meier}%
  \BibitemOpen
  \bibfield{author}{%
  \bibinfo {author} {\bibfnamefont{T.}~\bibnamefont{Matsuyama}}, \bibinfo
  {author} {\bibfnamefont{C.-M.}\ \bibnamefont{Hu}}, \bibinfo {author}
  {\bibfnamefont{D.}~\bibnamefont{Grundler}}, \bibinfo {author}
  {\bibfnamefont{G.}~\bibnamefont{Meier}},\ and\ \bibinfo {author}
  {\bibfnamefont{U.}~\bibnamefont{Merkt}},\ }%
  \bibfield{journal}{%
  \Doi{10.1103/PhysRevB.65.155322}{\bibinfo {journal} {Phys. Rev. B}}\ }%
  \textbf{\bibinfo {volume} {65}},\ \bibinfo {pages} {155322} (\bibinfo {year}
  {2002})%
  \bibAnnoteFile{NoStop}{Meier}%
\bibitem{diVinc}%
  \BibitemOpen
  \bibfield{author}{%
  \bibinfo {author} {\bibfnamefont{D.}~\bibnamefont{Loss}}\ and\ \bibinfo
  {author} {\bibfnamefont{D.~P.}\ \bibnamefont{DiVincenzo}},\ }%
  \bibfield{journal}{%
  \Doi{10.1103/PhysRevA.57.120}{\bibinfo {journal} {Phys. Rev. A}}\ }%
  \textbf{\bibinfo {volume} {57}},\ \bibinfo {pages} {120} (\bibinfo {year}
  {1998})%
  \bibAnnoteFile{NoStop}{diVinc}%
\bibitem{QI}%
  \BibitemOpen
  \bibfield{author}{%
  \bibinfo {author} {\bibfnamefont{T.~D.}\ \bibnamefont{Ladd}}, \bibinfo
  {author} {\bibfnamefont{F.}~\bibnamefont{Jelezko}}, \bibinfo {author}
  {\bibfnamefont{R.}~\bibnamefont{Laflamme}}, \bibinfo {author}
  {\bibfnamefont{Y.}~\bibnamefont{Nakamura}}, \bibinfo {author}
  {\bibfnamefont{C.}~\bibnamefont{Monroe}},\ and\ \bibinfo {author}
  {\bibfnamefont{J.~L.}\ \bibnamefont{O'Brien}},\ }%
  \bibfield{journal}{%
  \bibinfo {journal} {Nature}\ }%
  \textbf{\bibinfo {volume} {464}},\ \bibinfo {pages} {45} (\bibinfo {year}
  {2010})%
  \bibAnnoteFile{NoStop}{QI}%
\bibitem{Kouwenhoven2}%
  \BibitemOpen
  \bibfield{author}{%
  \bibinfo {author} {\bibfnamefont{R.}~\bibnamefont{Hanson}}, \bibinfo {author}
  {\bibfnamefont{L.~P.}\ \bibnamefont{Kouwenhoven}}, \bibinfo {author}
  {\bibfnamefont{J.~R.}\ \bibnamefont{Petta}}, \bibinfo {author}
  {\bibfnamefont{S.}~\bibnamefont{Tarucha}},\ and\ \bibinfo {author}
  {\bibfnamefont{L.~M.~K.}\ \bibnamefont{Vandersypen}},\ }%
  \bibfield{journal}{%
  \Doi{10.1103/RevModPhys.79.1217}{\bibinfo {journal} {Rev. Mod. Phys.}}\ }%
  \textbf{\bibinfo {volume} {79}},\ \bibinfo {pages} {1217} (\bibinfo {year}
  {2007})%
  \bibAnnoteFile{NoStop}{Kouwenhoven2}%
\bibitem{Bennet}%
  \BibitemOpen
  \bibfield{author}{%
  \bibinfo {author} {\bibfnamefont{C.~H.}\ \bibnamefont{Bennett}}, \bibinfo
  {author} {\bibfnamefont{D.~P.}\ \bibnamefont{DiVincenzo}}, \bibinfo {author}
  {\bibfnamefont{J.~A.}\ \bibnamefont{Smolin}},\ and\ \bibinfo {author}
  {\bibfnamefont{W.~K.}\ \bibnamefont{Wootters}},\ }%
  \bibfield{journal}{%
  \Doi{10.1103/PhysRevA.54.3824}{\bibinfo {journal} {Phys. Rev. A}}\ }%
  \textbf{\bibinfo {volume} {54}},\ \bibinfo {pages} {3824} (\bibinfo {year}
  {1996})%
  \bibAnnoteFile{NoStop}{Bennet}%
\bibitem{Kouwenhoven}%
  \BibitemOpen
  \bibfield{author}{%
  \bibinfo {author} {\bibfnamefont{L.}~\bibnamefont{Kouwenhoven}}, \bibinfo
  {author} {\bibfnamefont{D.}~\bibnamefont{Austing}},\ and\ \bibinfo {author}
  {\bibfnamefont{S.}~\bibnamefont{Tarucha}},\ }%
  \bibfield{journal}{%
  \bibinfo {journal} {Rep. Prog. Phys.}\ }%
  \textbf{\bibinfo {volume} {64}},\ \bibinfo {pages} {701} (\bibinfo {year}
  {2001})%
  \bibAnnoteFile{NoStop}{Kouwenhoven}%
\bibitem{vandersypen}%
  \BibitemOpen
  \bibfield{author}{%
  \bibinfo {author} {\bibfnamefont{F.~H.~L.}\ \bibnamefont{Koppens}}, \bibinfo
  {author} {\bibfnamefont{C.}~\bibnamefont{Buizert}}, \bibinfo {author}
  {\bibfnamefont{K.~J.}\ \bibnamefont{Tielrooij}}, \bibinfo {author}
  {\bibfnamefont{I.~T.}\ \bibnamefont{Vink}}, \bibinfo {author}
  {\bibfnamefont{K.~C.}\ \bibnamefont{Nowack}}, \bibinfo {author}
  {\bibfnamefont{T.}~\bibnamefont{Meunier}}, \bibinfo {author}
  {\bibfnamefont{L.~P.}\ \bibnamefont{Kouwenhoven}},\ and\ \bibinfo {author}
  {\bibfnamefont{L.~M.~K.}\ \bibnamefont{Vandersypen}},\ }%
  \bibfield{journal}{%
  \bibinfo {journal} {Nature}\ }%
  \textbf{\bibinfo {volume} {442}},\ \bibinfo {pages} {766} (\bibinfo {year}
  {2006})%
  \bibAnnoteFile{NoStop}{vandersypen}%
\bibitem{Yacoby}%
  \BibitemOpen
  \bibfield{author}{%
  \bibinfo {author} {\bibfnamefont{J.}~\bibnamefont{Petta}}, \bibinfo {author}
  {\bibfnamefont{A.}~\bibnamefont{Johnson}}, \bibinfo {author}
  {\bibfnamefont{J.}~\bibnamefont{Taylor}}, \bibinfo {author}
  {\bibfnamefont{E.}~\bibnamefont{Laird}}, \bibinfo {author}
  {\bibfnamefont{A.}~\bibnamefont{Yacoby}}, \bibinfo {author}
  {\bibfnamefont{M.}~\bibnamefont{Lukin}}, \bibinfo {author}
  {\bibfnamefont{C.}~\bibnamefont{Marcus}}, \bibinfo {author}
  {\bibfnamefont{M.}~\bibnamefont{Hanson}},\ and\ \bibinfo {author}
  {\bibfnamefont{A.}~\bibnamefont{Gossard}},\ }%
  \bibfield{journal}{%
  \bibinfo {journal} {Science}\ }%
  \textbf{\bibinfo {volume} {309}},\ \bibinfo {pages} {2180} (\bibinfo {year}
  {2005})%
  \bibAnnoteFile{NoStop}{Yacoby}%
\bibitem{Bluhm}%
  \BibitemOpen
  \bibfield{author}{%
  \bibinfo {author} {\bibfnamefont{H.}~\bibnamefont{Bluhm}}, \bibinfo {author}
  {\bibfnamefont{S.}~\bibnamefont{Foletti}}, \bibinfo {author}
  {\bibfnamefont{I.}~\bibnamefont{Neder}}, \bibinfo {author}
  {\bibfnamefont{M.}~\bibnamefont{Rudner}}, \bibinfo {author}
  {\bibfnamefont{D.}~\bibnamefont{Mahalu}}, \bibinfo {author}
  {\bibfnamefont{V.}~\bibnamefont{Umansky}},\ and\ \bibinfo {author}
  {\bibfnamefont{A.}~\bibnamefont{Yacoby}},\ }%
  \bibfield{journal}{%
  \bibinfo {journal} {Nature Phys.}\ }%
  \textbf{\bibinfo {volume} {7}},\ \bibinfo {pages} {109} (\bibinfo {year}
  {2011})%
  \bibAnnoteFile{NoStop}{Bluhm}%
\bibitem{Bluhm2}%
  \BibitemOpen
  \bibfield{author}{%
  \bibinfo {author} {\bibfnamefont{M.~D.}\ \bibnamefont{Shulman}}, \bibinfo
  {author} {\bibfnamefont{O.~E.}\ \bibnamefont{Dial}}, \bibinfo {author}
  {\bibfnamefont{S.~P.}\ \bibnamefont{Harvey}}, \bibinfo {author}
  {\bibfnamefont{H.}~\bibnamefont{Bluhm}}, \bibinfo {author}
  {\bibfnamefont{V.}~\bibnamefont{Umansky}},\ and\ \bibinfo {author}
  {\bibfnamefont{A.}~\bibnamefont{Yacoby}},\ }%
  \bibfield{journal}{%
  \bibinfo {journal} {ArXiv},\ \bibinfo {pages} {1202.1828}}%
   (\bibinfo {year} {2012})%
  \bibAnnoteFile{NoStop}{Bluhm2}%
\bibitem{Wallraff}%
  \BibitemOpen
  \bibfield{author}{%
  \bibinfo {author} {\bibfnamefont{A.}~\bibnamefont{Fedorov}}, \bibinfo
  {author} {\bibfnamefont{L.}~\bibnamefont{Steffen}}, \bibinfo {author}
  {\bibfnamefont{M.}~\bibnamefont{Baur}}, \bibinfo {author}
  {\bibfnamefont{M.~P.}\ \bibnamefont{da~Silva}},\ and\ \bibinfo {author}
  {\bibfnamefont{A.}~\bibnamefont{Wallraff}},\ }%
  \bibfield{journal}{%
  \bibinfo {journal} {Nature}\ }%
  \textbf{\bibinfo {volume} {481}},\ \bibinfo {pages} {170} (\bibinfo {year}
  {2012})%
  \bibAnnoteFile{NoStop}{Wallraff}%
\bibitem{Martinis}%
  \BibitemOpen
  \bibfield{author}{%
  \bibinfo {author} {\bibfnamefont{M.}~\bibnamefont{Mariantoni}}, \bibinfo
  {author} {\bibfnamefont{H.}~\bibnamefont{Wang}}, \bibinfo {author}
  {\bibfnamefont{T.}~\bibnamefont{Yamamoto}}, \bibinfo {author}
  {\bibfnamefont{M.}~\bibnamefont{Neeley}}, \bibinfo {author}
  {\bibfnamefont{R.~C.}\ \bibnamefont{Bialczak}}, \bibinfo {author}
  {\bibfnamefont{Y.}~\bibnamefont{Chen}}, \bibinfo {author}
  {\bibfnamefont{M.}~\bibnamefont{Lenander}}, \bibinfo {author}
  {\bibfnamefont{E.}~\bibnamefont{Lucero}}, \bibinfo {author}
  {\bibfnamefont{A.~D.}\ \bibnamefont{O'Connell}}, \bibinfo {author}
  {\bibfnamefont{D.}~\bibnamefont{Sank}}, \bibinfo {author}
  {\bibfnamefont{M.}~\bibnamefont{Weides}}, \bibinfo {author}
  {\bibfnamefont{J.}~\bibnamefont{Wenner}}, \bibinfo {author}
  {\bibfnamefont{Y.}~\bibnamefont{Yin}}, \bibinfo {author}
  {\bibfnamefont{J.}~\bibnamefont{Zhao}}, \bibinfo {author}
  {\bibfnamefont{A.~N.}\ \bibnamefont{Korotkov}}, \bibinfo {author}
  {\bibfnamefont{A.~N.}\ \bibnamefont{Cleland}},\ and\ \bibinfo {author}
  {\bibfnamefont{J.~M.}\ \bibnamefont{Martinis}},\ }%
  \bibfield{journal}{%
  \bibinfo {journal} {Science}\ }%
  \textbf{\bibinfo {volume} {333}},\ \bibinfo {pages} {61} (\bibinfo {year}
  {2011})%
  \bibAnnoteFile{NoStop}{Martinis}%
\bibitem{Schoelkopf}%
  \BibitemOpen
  \bibfield{author}{%
  \bibinfo {author} {\bibfnamefont{M.~D.}\ \bibnamefont{Reed}}, \bibinfo
  {author} {\bibfnamefont{L.}~\bibnamefont{DiCarlo}}, \bibinfo {author}
  {\bibfnamefont{S.~E.}\ \bibnamefont{Nigg}}, \bibinfo {author}
  {\bibfnamefont{L.}~\bibnamefont{Sun}}, \bibinfo {author}
  {\bibfnamefont{L.}~\bibnamefont{Frunzio}}, \bibinfo {author}
  {\bibfnamefont{S.~M.}\ \bibnamefont{Girvin}},\ and\ \bibinfo {author}
  {\bibfnamefont{R.~J.}\ \bibnamefont{Schoelkopf}},\ }%
  \bibfield{journal}{%
  \bibinfo {journal} {Nature}\ }%
  \textbf{\bibinfo {volume} {482}},\ \bibinfo {pages} {382} (\bibinfo {year}
  {2012})%
  \bibAnnoteFile{NoStop}{Schoelkopf}%
\bibitem{Hanson}%
  \BibitemOpen
  \bibfield{author}{%
  \bibinfo {author} {\bibfnamefont{L.}~\bibnamefont{Robledo}}, \bibinfo
  {author} {\bibfnamefont{L.}~\bibnamefont{Childress}}, \bibinfo {author}
  {\bibfnamefont{H.}~\bibnamefont{Bernien}}, \bibinfo {author}
  {\bibfnamefont{B.}~\bibnamefont{Hensen}}, \bibinfo {author}
  {\bibfnamefont{P.~F.~A.}\ \bibnamefont{Alkemade}},\ and\ \bibinfo {author}
  {\bibfnamefont{R.}~\bibnamefont{Hanson}},\ }%
  \bibfield{journal}{%
  \bibinfo {journal} {Nature}\ }%
  \textbf{\bibinfo {volume} {477}},\ \bibinfo {pages} {574} (\bibinfo {year}
  {2011})%
  \bibAnnoteFile{NoStop}{Hanson}%
\bibitem{Jelezko1}%
  \BibitemOpen
  \bibfield{author}{%
  \bibinfo {author} {\bibfnamefont{G.}~\bibnamefont{Balasubramanian}}, \bibinfo
  {author} {\bibfnamefont{P.}~\bibnamefont{Neumann}}, \bibinfo {author}
  {\bibfnamefont{D.}~\bibnamefont{Twitchen}}, \bibinfo {author}
  {\bibfnamefont{M.}~\bibnamefont{Markham}}, \bibinfo {author}
  {\bibfnamefont{R.}~\bibnamefont{Kolesov}}, \bibinfo {author}
  {\bibfnamefont{N.}~\bibnamefont{Mizuochi}}, \bibinfo {author}
  {\bibfnamefont{J.}~\bibnamefont{Isoya}}, \bibinfo {author}
  {\bibfnamefont{J.}~\bibnamefont{Achard}}, \bibinfo {author}
  {\bibfnamefont{J.}~\bibnamefont{Beck}}, \bibinfo {author}
  {\bibfnamefont{J.}~\bibnamefont{Tissler}}, \bibinfo {author}
  {\bibfnamefont{V.}~\bibnamefont{Jacques}}, \bibinfo {author}
  {\bibfnamefont{P.~R.}\ \bibnamefont{Hemmer}}, \bibinfo {author}
  {\bibfnamefont{F.}~\bibnamefont{Jelezko}},\ and\ \bibinfo {author}
  {\bibfnamefont{J.}~\bibnamefont{Wrachtrup}},\ }%
  \bibfield{journal}{%
  \bibinfo {journal} {Nature Mat.}\ }%
  \textbf{\bibinfo {volume} {8}},\ \bibinfo {pages} {383} (\bibinfo {year}
  {2009})%
  \bibAnnoteFile{NoStop}{Jelezko1}%
\bibitem{Jelezko2}%
  \BibitemOpen
  \bibfield{author}{%
  \bibinfo {author} {\bibfnamefont{P.}~\bibnamefont{Neumann}}, \bibinfo
  {author} {\bibfnamefont{N.}~\bibnamefont{Mizuochi}}, \bibinfo {author}
  {\bibfnamefont{F.}~\bibnamefont{Rempp}}, \bibinfo {author}
  {\bibfnamefont{P.}~\bibnamefont{Hemmer}}, \bibinfo {author}
  {\bibfnamefont{H.}~\bibnamefont{Watanabe}}, \bibinfo {author}
  {\bibfnamefont{S.}~\bibnamefont{Yamasaki}}, \bibinfo {author}
  {\bibfnamefont{V.}~\bibnamefont{Jacques}}, \bibinfo {author}
  {\bibfnamefont{T.}~\bibnamefont{Gaebel}}, \bibinfo {author}
  {\bibfnamefont{F.}~\bibnamefont{Jelezko}},\ and\ \bibinfo {author}
  {\bibfnamefont{J.}~\bibnamefont{Wrachtrup}},\ }%
  \bibfield{journal}{%
  \bibinfo {journal} {Science}\ }%
  \textbf{\bibinfo {volume} {320}},\ \bibinfo {pages} {1326} (\bibinfo {year}
  {2008})%
  \bibAnnoteFile{NoStop}{Jelezko2}%
\bibitem{KvK}%
  \BibitemOpen
  \bibfield{author}{%
  \bibinfo {author} {\bibfnamefont{K.~v.}\ \bibnamefont{Klitzing}}, \bibinfo
  {author} {\bibfnamefont{G.}~\bibnamefont{Dorda}},\ and\ \bibinfo {author}
  {\bibfnamefont{M.}~\bibnamefont{Pepper}},\ }%
  \bibfield{journal}{%
  \Doi{10.1103/PhysRevLett.45.494}{\bibinfo {journal} {Phys. Rev. Lett.}}\ }%
  \textbf{\bibinfo {volume} {45}},\ \bibinfo {pages} {494} (\bibinfo {year}
  {1980})%
  \bibAnnoteFile{NoStop}{KvK}%
\bibitem{Binnig}%
  \BibitemOpen
  \bibfield{author}{%
  \bibinfo {author} {\bibfnamefont{G.}~\bibnamefont{Binnig}}\ and\ \bibinfo
  {author} {\bibfnamefont{H.}~\bibnamefont{Rohrer}},\ }%
  \bibfield{journal}{%
  \bibinfo {journal} {Helv. Phys. Acta}\ }%
  \textbf{\bibinfo {volume} {55}},\ \bibinfo {pages} {726} (\bibinfo {year}
  {1982})%
  \bibAnnoteFile{NoStop}{Binnig}%
\bibitem{Eigler}%
  \BibitemOpen
  \bibfield{author}{%
  \bibinfo {author} {\bibfnamefont{M.}~\bibnamefont{Crommie}}, \bibinfo
  {author} {\bibfnamefont{C.}~\bibnamefont{Lutz}},\ and\ \bibinfo {author}
  {\bibfnamefont{D.}~\bibnamefont{Eigler}},\ }%
  \bibfield{journal}{%
  \bibinfo {journal} {Science}\ }%
  \textbf{\bibinfo {volume} {262}},\ \bibinfo {pages} {218} (\bibinfo {year}
  {1993})%
  \bibAnnoteFile{NoStop}{Eigler}%
\bibitem{Maltezo}%
  \BibitemOpen
  \bibfield{author}{%
  \bibinfo {author} {\bibfnamefont{T.}~\bibnamefont{Maltezopoulos}}, \bibinfo
  {author} {\bibfnamefont{A.}~\bibnamefont{Bolz}}, \bibinfo {author}
  {\bibfnamefont{C.}~\bibnamefont{Meyer}}, \bibinfo {author}
  {\bibfnamefont{C.}~\bibnamefont{Heyn}}, \bibinfo {author}
  {\bibfnamefont{W.}~\bibnamefont{Hansen}}, \bibinfo {author}
  {\bibfnamefont{M.}~\bibnamefont{Morgenstern}},\ and\ \bibinfo {author}
  {\bibfnamefont{R.}~\bibnamefont{Wiesendanger}},\ }%
  \bibfield{journal}{%
  \Doi{10.1103/PhysRevLett.91.196804}{\bibinfo {journal} {Phys. Rev. Lett.}}\
  }%
  \textbf{\bibinfo {volume} {91}},\ \bibinfo {pages} {196804} (\bibinfo {year}
  {2003})%
  \bibAnnoteFile{NoStop}{Maltezo}%
\bibitem{Bode}%
  \BibitemOpen
  \bibfield{author}{%
  \bibinfo {author} {\bibfnamefont{M.}~\bibnamefont{Bode}},\ }%
  \bibfield{journal}{%
  \bibinfo {journal} {Rep. Prog. Phys.}\ }%
  \textbf{\bibinfo {volume} {66}},\ \bibinfo {pages} {523} (\bibinfo {year}
  {2003})%
  \bibAnnoteFile{NoStop}{Bode}%
\bibitem{Loth}%
  \BibitemOpen
  \bibfield{author}{%
  \bibinfo {author} {\bibfnamefont{S.}~\bibnamefont{Loth}}, \bibinfo {author}
  {\bibfnamefont{M.}~\bibnamefont{Etzkorn}}, \bibinfo {author}
  {\bibfnamefont{C.~P.}\ \bibnamefont{Lutz}}, \bibinfo {author}
  {\bibfnamefont{D.~M.}\ \bibnamefont{Eigler}},\ and\ \bibinfo {author}
  {\bibfnamefont{A.~J.}\ \bibnamefont{Heinrich}},\ }%
  \bibfield{journal}{%
  \bibinfo {journal} {Science}\ }%
  \textbf{\bibinfo {volume} {329}},\ \bibinfo {pages} {1628} (\bibinfo {year}
  {2010})%
  \bibAnnoteFile{NoStop}{Loth}%
\bibitem{Heinrich}%
  \BibitemOpen
  \bibfield{author}{%
  \bibinfo {author} {\bibfnamefont{A.}~\bibnamefont{Heinrich}}, \bibinfo
  {author} {\bibfnamefont{J.}~\bibnamefont{Gupta}}, \bibinfo {author}
  {\bibfnamefont{C.}~\bibnamefont{Lutz}},\ and\ \bibinfo {author}
  {\bibfnamefont{D.}~\bibnamefont{Eigler}},\ }%
  \bibfield{journal}{%
  \bibinfo {journal} {Science}\ }%
  \textbf{\bibinfo {volume} {306}},\ \bibinfo {pages} {466} (\bibinfo {year}
  {2004})%
  \bibAnnoteFile{NoStop}{Heinrich}%
\bibitem{Heinrich2}%
  \BibitemOpen
  \bibfield{author}{%
  \bibinfo {author} {\bibfnamefont{C.}~\bibnamefont{Hirjibehedin}}, \bibinfo
  {author} {\bibfnamefont{C.}~\bibnamefont{Lutz}},\ and\ \bibinfo {author}
  {\bibfnamefont{A.}~\bibnamefont{Heinrich}},\ }%
  \bibfield{journal}{%
  \bibinfo {journal} {Science}\ }%
  \textbf{\bibinfo {volume} {312}},\ \bibinfo {pages} {1021} (\bibinfo {year}
  {2006})%
  \bibAnnoteFile{NoStop}{Heinrich2}%
\bibitem{Wiebe}%
  \BibitemOpen
  \bibfield{author}{%
  \bibinfo {author} {\bibfnamefont{F.}~\bibnamefont{Meier}}, \bibinfo {author}
  {\bibfnamefont{L.}~\bibnamefont{Zhou}}, \bibinfo {author}
  {\bibfnamefont{J.}~\bibnamefont{Wiebe}},\ and\ \bibinfo {author}
  {\bibfnamefont{R.}~\bibnamefont{Wiesendanger}},\ }%
  \bibfield{journal}{%
  \bibinfo {journal} {Science}\ }%
  \textbf{\bibinfo {volume} {320}},\ \bibinfo {pages} {82} (\bibinfo {year}
  {2008})%
  \bibAnnoteFile{NoStop}{Wiebe}%
\bibitem{Wiebe2}%
  \BibitemOpen
  \bibfield{author}{%
  \bibinfo {author} {\bibfnamefont{A.~A.}\ \bibnamefont{Khajetoorians}},
  \bibinfo {author} {\bibfnamefont{B.}~\bibnamefont{Chilian}}, \bibinfo
  {author} {\bibfnamefont{J.}~\bibnamefont{Wiebe}}, \bibinfo {author}
  {\bibfnamefont{S.}~\bibnamefont{Schuwalow}}, \bibinfo {author}
  {\bibfnamefont{F.}~\bibnamefont{Lechermann}},\ and\ \bibinfo {author}
  {\bibfnamefont{R.}~\bibnamefont{Wiesendanger}},\ }%
  \bibfield{journal}{%
  \bibinfo {journal} {Nature}\ }%
  \textbf{\bibinfo {volume} {467}},\ \bibinfo {pages} {1084} (\bibinfo {year}
  {2010})%
  \bibAnnoteFile{NoStop}{Wiebe2}%
\bibitem{Manoharan}%
  \BibitemOpen
  \bibfield{author}{%
  \bibinfo {author} {\bibfnamefont{C.~R.}\ \bibnamefont{Moon}}, \bibinfo
  {author} {\bibfnamefont{L.~S.}\ \bibnamefont{Mattos}}, \bibinfo {author}
  {\bibfnamefont{B.~K.}\ \bibnamefont{Foster}}, \bibinfo {author}
  {\bibfnamefont{G.}~\bibnamefont{Zeltzer}}, \bibinfo {author}
  {\bibfnamefont{W.}~\bibnamefont{Ko}},\ and\ \bibinfo {author}
  {\bibfnamefont{H.~C.}\ \bibnamefont{Manoharan}},\ }%
  \bibfield{journal}{%
  \bibinfo {journal} {Science}\ }%
  \textbf{\bibinfo {volume} {319}},\ \bibinfo {pages} {782} (\bibinfo {year}
  {2008})%
  \bibAnnoteFile{NoStop}{Manoharan}%
\bibitem{Manoharan2}%
  \BibitemOpen
  \bibfield{author}{%
  \bibinfo {author} {\bibfnamefont{C.~R.}\ \bibnamefont{Moon}}, \bibinfo
  {author} {\bibfnamefont{C.~P.}\ \bibnamefont{Lutz}},\ and\ \bibinfo {author}
  {\bibfnamefont{H.~C.}\ \bibnamefont{Manoharan}},\ }%
  \bibfield{journal}{%
  \bibinfo {journal} {Nature Phys.}\ }%
  \textbf{\bibinfo {volume} {4}},\ \bibinfo {pages} {454} (\bibinfo {year}
  {2008})%
  \bibAnnoteFile{NoStop}{Manoharan2}%
\bibitem{Aristov}%
  \BibitemOpen
  \bibfield{author}{%
  \bibinfo {author} {\bibfnamefont{V.~Y.}\ \bibnamefont{Aristov}}, \bibinfo
  {author} {\bibfnamefont{G.}~\bibnamefont{LeLay}}, \bibinfo {author}
  {\bibfnamefont{P.}~\bibnamefont{Soukiassian}}, \bibinfo {author}
  {\bibfnamefont{K.}~\bibnamefont{Hricovini}}, \bibinfo {author}
  {\bibfnamefont{J.~E.}\ \bibnamefont{Bonnet}}, \bibinfo {author}
  {\bibfnamefont{J.}~\bibnamefont{Osvald}},\ and\ \bibinfo {author}
  {\bibfnamefont{O.}~\bibnamefont{Olsson}},\ }%
  \bibfield{journal}{%
  \bibinfo {journal} {Europhys. Lett.}\ }%
  \textbf{\bibinfo {volume} {26}},\ \bibinfo {pages} {359} (\bibinfo {year}
  {1994})%
  \bibAnnoteFile{NoStop}{Aristov}%
\bibitem{FeInAs}%
  \BibitemOpen
  \bibfield{author}{%
  \bibinfo {author} {\bibfnamefont{M.}~\bibnamefont{Morgenstern}}, \bibinfo
  {author} {\bibfnamefont{M.}~\bibnamefont{Getzlaff}}, \bibinfo {author}
  {\bibfnamefont{D.}~\bibnamefont{Haude}}, \bibinfo {author}
  {\bibfnamefont{R.}~\bibnamefont{Wiesendanger}},\ and\ \bibinfo {author}
  {\bibfnamefont{R.~L.}\ \bibnamefont{Johnson}},\ }%
  \bibfield{journal}{%
  \bibinfo {journal} {Phys. Rev. B}\ }%
  \textbf{\bibinfo {volume} {61}},\ \bibinfo {pages} {13805} (\bibinfo {year}
  {2000})%
  \bibAnnoteFile{NoStop}{FeInAs}%
\bibitem{NbInAs}%
  \BibitemOpen
  \bibfield{author}{%
  \bibinfo {author} {\bibfnamefont{M.}~\bibnamefont{Getzlaff}}, \bibinfo
  {author} {\bibfnamefont{M.}~\bibnamefont{Morgenstern}}, \bibinfo {author}
  {\bibfnamefont{C.}~\bibnamefont{Meyer}}, \bibinfo {author}
  {\bibfnamefont{R.}~\bibnamefont{Brochier}}, \bibinfo {author}
  {\bibfnamefont{R.~L.}\ \bibnamefont{Johnson}},\ and\ \bibinfo {author}
  {\bibfnamefont{R.}~\bibnamefont{Wiesendanger}},\ }%
  \bibfield{journal}{%
  \bibinfo {journal} {Phys. Rev. B}\ }%
  \textbf{\bibinfo {volume} {63}},\ \bibinfo {pages} {205305} (\bibinfo {year}
  {2001})%
  \bibAnnoteFile{NoStop}{NbInAs}%
\bibitem{PhysRevB.63.155315}%
  \BibitemOpen
  \bibfield{author}{%
  \bibinfo {author} {\bibfnamefont{M.~G.}\ \bibnamefont{Betti}}, \bibinfo
  {author} {\bibfnamefont{V.}~\bibnamefont{Corradini}}, \bibinfo {author}
  {\bibfnamefont{G.}~\bibnamefont{Bertoni}}, \bibinfo {author}
  {\bibfnamefont{P.}~\bibnamefont{Casarini}}, \bibinfo {author}
  {\bibfnamefont{C.}~\bibnamefont{Mariani}},\ and\ \bibinfo {author}
  {\bibfnamefont{A.}~\bibnamefont{Abramo}},\ }%
  \bibfield{journal}{%
  \Doi{10.1103/PhysRevB.63.155315}{\bibinfo {journal} {Phys. Rev. B}}\ }%
  \textbf{\bibinfo {volume} {63}},\ \bibinfo {pages} {155315} (\bibinfo {year}
  {2001})%
  \bibAnnoteFile{NoStop}{PhysRevB.63.155315}%
\bibitem{PhysRevLett.89.136806}%
  \BibitemOpen
  \bibfield{author}{%
  \bibinfo {author} {\bibfnamefont{M.}~\bibnamefont{Morgenstern}}, \bibinfo
  {author} {\bibfnamefont{J.}~\bibnamefont{Klijn}}, \bibinfo {author}
  {\bibfnamefont{C.}~\bibnamefont{Meyer}}, \bibinfo {author}
  {\bibfnamefont{M.}~\bibnamefont{Getzlaff}}, \bibinfo {author}
  {\bibfnamefont{R.}~\bibnamefont{Adelung}}, \bibinfo {author}
  {\bibfnamefont{R.~A.}\ \bibnamefont{R\"omer}}, \bibinfo {author}
  {\bibfnamefont{K.}~\bibnamefont{Rossnagel}}, \bibinfo {author}
  {\bibfnamefont{L.}~\bibnamefont{Kipp}}, \bibinfo {author}
  {\bibfnamefont{M.}~\bibnamefont{Skibowski}},\ and\ \bibinfo {author}
  {\bibfnamefont{R.}~\bibnamefont{Wiesendanger}},\ }%
  \bibfield{journal}{%
  \Doi{10.1103/PhysRevLett.89.136806}{\bibinfo {journal} {Phys. Rev. Lett.}}\
  }%
  \textbf{\bibinfo {volume} {89}},\ \bibinfo {pages} {136806} (\bibinfo {year}
  {2002})%
  \bibAnnoteFile{NoStop}{PhysRevLett.89.136806}%
\bibitem{PhysRevB.68.041402}%
  \BibitemOpen
  \bibfield{author}{%
  \bibinfo {author} {\bibfnamefont{J.}~\bibnamefont{Wiebe}}, \bibinfo {author}
  {\bibfnamefont{C.}~\bibnamefont{Meyer}}, \bibinfo {author}
  {\bibfnamefont{J.}~\bibnamefont{Klijn}}, \bibinfo {author}
  {\bibfnamefont{M.}~\bibnamefont{Morgenstern}},\ and\ \bibinfo {author}
  {\bibfnamefont{R.}~\bibnamefont{Wiesendanger}},\ }%
  \bibfield{journal}{%
  \Doi{10.1103/PhysRevB.68.041402}{\bibinfo {journal} {Phys. Rev. B}}\ }%
  \textbf{\bibinfo {volume} {68}},\ \bibinfo {pages} {041402} (\bibinfo {year}
  {2003})%
  \bibAnnoteFile{NoStop}{PhysRevB.68.041402}%
\bibitem{Morgenstern2}%
  \BibitemOpen
  \bibfield{author}{%
  \bibinfo {author} {\bibfnamefont{M.}~\bibnamefont{Morgenstern}}, \bibinfo
  {author} {\bibfnamefont{J.}~\bibnamefont{Klijn}}, \bibinfo {author}
  {\bibfnamefont{C.}~\bibnamefont{Meyer}},\ and\ \bibinfo {author}
  {\bibfnamefont{R.}~\bibnamefont{Wiesendanger}},\ }%
  \bibfield{journal}{%
  \bibinfo {journal} {Phys. Rev. Lett.}\ }%
  \textbf{\bibinfo {volume} {90}},\ \bibinfo {pages} {056804} (\bibinfo {year}
  {2003})%
  \bibAnnoteFile{NoStop}{Morgenstern2}%
\bibitem{Kanisawa}%
  \BibitemOpen
  \bibfield{author}{%
  \bibinfo {author} {\bibfnamefont{K.}~\bibnamefont{Kanisawa}}, \bibinfo
  {author} {\bibfnamefont{M.~J.}\ \bibnamefont{Butcher}}, \bibinfo {author}
  {\bibfnamefont{H.}~\bibnamefont{Yamaguchi}},\ and\ \bibinfo {author}
  {\bibfnamefont{Y.}~\bibnamefont{Hirayama}},\ }%
  \bibfield{journal}{%
  \bibinfo {journal} {Phys. Rev. Lett.}\ }%
  \textbf{\bibinfo {volume} {86}},\ \bibinfo {pages} {3384} (\bibinfo {year}
  {2001})%
  \bibAnnoteFile{NoStop}{Kanisawa}%
\bibitem{Tipinduced}%
  \BibitemOpen
  \bibfield{author}{%
  \bibinfo {author} {\bibfnamefont{R.}~\bibnamefont{Dombrowski}}, \bibinfo
  {author} {\bibfnamefont{C.}~\bibnamefont{Steinebach}}, \bibinfo {author}
  {\bibfnamefont{C.}~\bibnamefont{Wittneven}}, \bibinfo {author}
  {\bibfnamefont{M.}~\bibnamefont{Morgenstern}},\ and\ \bibinfo {author}
  {\bibfnamefont{R.}~\bibnamefont{Wiesendanger}},\ }%
  \bibfield{journal}{%
  \bibinfo {journal} {Phys. Rev. B}\ }%
  \textbf{\bibinfo {volume} {49}},\ \bibinfo {pages} {8043} (\bibinfo {year}
  {1999})%
  \bibAnnoteFile{NoStop}{Tipinduced}%
\bibitem{Masutomi}%
  \BibitemOpen
  \bibfield{author}{%
  \bibinfo {author} {\bibfnamefont{R.}~\bibnamefont{Masutomi}}, \bibinfo
  {author} {\bibfnamefont{M.}~\bibnamefont{Hio}}, \bibinfo {author}
  {\bibfnamefont{T.}~\bibnamefont{Mochizuki}},\ and\ \bibinfo {author}
  {\bibfnamefont{T.}~\bibnamefont{Okamoto}},\ }%
  \bibfield{journal}{%
  \bibinfo {journal} {Appl. Phys. Lett.}\ }%
  \textbf{\bibinfo {volume} {90}},\ \bibinfo {pages} {202104} (\bibinfo {year}
  {2007})%
  \bibAnnoteFile{NoStop}{Masutomi}%
\bibitem{Tsuji}%
  \BibitemOpen
  \bibfield{author}{%
  \bibinfo {author} {\bibfnamefont{Y.}~\bibnamefont{Tsuji}}, \bibinfo {author}
  {\bibfnamefont{T.}~\bibnamefont{Mochizuki}},\ and\ \bibinfo {author}
  {\bibfnamefont{T.}~\bibnamefont{Okamoto}},\ }%
  \bibfield{journal}{%
  \bibinfo {journal} {Appl. Phys. Lett.}\ }%
  \textbf{\bibinfo {volume} {87}},\ \bibinfo {pages} {062103} (\bibinfo {year}
  {2005})%
  \bibAnnoteFile{NoStop}{Tsuji}%
\bibitem{hashimoto:256802}%
  \BibitemOpen
  \bibfield{author}{%
  \bibinfo {author} {\bibfnamefont{K.}~\bibnamefont{Hashimoto}}, \bibinfo
  {author} {\bibfnamefont{C.}~\bibnamefont{Sohrmann}}, \bibinfo {author}
  {\bibfnamefont{J.}~\bibnamefont{Wiebe}}, \bibinfo {author}
  {\bibfnamefont{T.}~\bibnamefont{Inaoka}}, \bibinfo {author}
  {\bibfnamefont{F.}~\bibnamefont{Meier}}, \bibinfo {author}
  {\bibfnamefont{Y.}~\bibnamefont{Hirayama}}, \bibinfo {author}
  {\bibfnamefont{R.~A.}\ \bibnamefont{R\"omer}}, \bibinfo {author}
  {\bibfnamefont{R.}~\bibnamefont{Wiesendanger}},\ and\ \bibinfo {author}
  {\bibfnamefont{M.}~\bibnamefont{Morgenstern}},\ }%
  \bibfield{journal}{%
  \Doi{10.1103/PhysRevLett.101.256802}{\bibinfo {journal} {Phys. Rev. Lett.}}\
  }%
  \textbf{\bibinfo {volume} {101}},\ \bibinfo {eid} {256802} (\bibinfo {year}
  {2008})%
  \bibAnnoteFile{NoStop}{hashimoto:256802}%
\bibitem{Becker2011}%
  \BibitemOpen
  \bibfield{author}{%
  \bibinfo {author} {\bibfnamefont{S.}~\bibnamefont{Becker}}, \bibinfo {author}
  {\bibfnamefont{C.}~\bibnamefont{Karrasch}}, \bibinfo {author}
  {\bibfnamefont{T.}~\bibnamefont{Mashoff}}, \bibinfo {author}
  {\bibfnamefont{M.}~\bibnamefont{Pratzer}}, \bibinfo {author}
  {\bibfnamefont{M.}~\bibnamefont{Liebmann}}, \bibinfo {author}
  {\bibfnamefont{V.}~\bibnamefont{Meden}},\ and\ \bibinfo {author}
  {\bibfnamefont{M.}~\bibnamefont{Morgenstern}},\ }%
  \bibfield{journal}{%
  \bibinfo {journal} {Phys. Rev. Lett.}\ }%
  \textbf{\bibinfo {volume} {106}},\ \bibinfo {pages} {156805} (\bibinfo {year}
  {2011})%
  \bibAnnoteFile{NoStop}{Becker2011}%
\bibitem{Becker2010}%
  \BibitemOpen
  \bibfield{author}{%
  \bibinfo {author} {\bibfnamefont{S.}~\bibnamefont{Becker}}, \bibinfo {author}
  {\bibfnamefont{M.}~\bibnamefont{Liebmann}}, \bibinfo {author}
  {\bibfnamefont{T.}~\bibnamefont{Mashoff}}, \bibinfo {author}
  {\bibfnamefont{M.}~\bibnamefont{Pratzer}},\ and\ \bibinfo {author}
  {\bibfnamefont{M.}~\bibnamefont{Morgenstern}},\ }%
  \bibfield{journal}{%
  \Doi{10.1103/PhysRevB.81.155308}{\bibinfo {journal} {Phys. Rev. B}}\ }%
  \textbf{\bibinfo {volume} {81}},\ \bibinfo {pages} {155308} (\bibinfo {year}
  {2010})%
  \bibAnnoteFile{NoStop}{Becker2010}%
\bibitem{Sherman}%
  \BibitemOpen
  \bibfield{author}{%
  \bibinfo {author} {\bibfnamefont{M.~M.}\ \bibnamefont{Glazov}}, \bibinfo
  {author} {\bibfnamefont{E.~Y.}\ \bibnamefont{Sherman}},\ and\ \bibinfo
  {author} {\bibfnamefont{V.~K.}\ \bibnamefont{Dugaev}},\ }%
  \bibfield{journal}{%
  \bibinfo {journal} {Physica E}\ }%
  \textbf{\bibinfo {volume} {42}},\ \bibinfo {pages} {2157} (\bibinfo {year}
  {2010})%
  \bibAnnoteFile{NoStop}{Sherman}%
\bibitem{Pfeiffer}%
  \BibitemOpen
  \bibfield{author}{%
  \bibinfo {author} {\bibfnamefont{L.}~\bibnamefont{Pfeiffer}}, \bibinfo
  {author} {\bibfnamefont{K.}~\bibnamefont{West}}, \bibinfo {author}
  {\bibfnamefont{H.}~\bibnamefont{Stormer}},\ and\ \bibinfo {author}
  {\bibfnamefont{K.}~\bibnamefont{Baldwin}},\ }%
  \bibfield{journal}{%
  \bibinfo {journal} {Appl. Phys. Lett.}\ }%
  \textbf{\bibinfo {volume} {55}},\ \bibinfo {pages} {1888} (\bibinfo {year}
  {1989})%
  \bibAnnoteFile{NoStop}{Pfeiffer}%
\bibitem{Umansky}%
  \BibitemOpen
  \bibfield{author}{%
  \bibinfo {author} {\bibfnamefont{V.}~\bibnamefont{Umansky}}, \bibinfo
  {author} {\bibfnamefont{R.}~\bibnamefont{dePicciotto}},\ and\ \bibinfo
  {author} {\bibfnamefont{M.}~\bibnamefont{Heiblum}},\ }%
  \bibfield{journal}{%
  \bibinfo {journal} {Appl. Phys. Lett.}\ }%
  \textbf{\bibinfo {volume} {71}},\ \bibinfo {pages} {683} (\bibinfo {year}
  {1997})%
  \bibAnnoteFile{NoStop}{Umansky}%
\bibitem{Eisenstein}%
  \BibitemOpen
  \bibfield{author}{%
  \bibinfo {author} {\bibfnamefont{J.}~\bibnamefont{Xia}}, \bibinfo {author}
  {\bibfnamefont{V.}~\bibnamefont{Cvicek}}, \bibinfo {author}
  {\bibfnamefont{J.~P.}\ \bibnamefont{Eisenstein}}, \bibinfo {author}
  {\bibfnamefont{L.~N.}\ \bibnamefont{Pfeiffer}},\ and\ \bibinfo {author}
  {\bibfnamefont{K.~W.}\ \bibnamefont{West}},\ }%
  \bibfield{journal}{%
  \Doi{10.1103/PhysRevLett.105.176807}{\bibinfo {journal} {Phys. Rev. Lett.}}\
  }%
  \textbf{\bibinfo {volume} {105}},\ \bibinfo {pages} {176807} (\bibinfo {year}
  {2010})%
  \bibAnnoteFile{NoStop}{Eisenstein}%
\bibitem{triang}%
  \BibitemOpen
  \bibfield{author}{%
  \bibinfo {author} {\bibfnamefont{G.}~\bibnamefont{Vincent}}, \bibinfo
  {author} {\bibfnamefont{A.}~\bibnamefont{Chantre}},\ and\ \bibinfo {author}
  {\bibfnamefont{D.}~\bibnamefont{Bois}},\ }%
  \bibfield{journal}{%
  \bibinfo {journal} {J. Appl. Phys.}\ }%
  \textbf{\bibinfo {volume} {50}},\ \bibinfo {pages} {5484} (\bibinfo {year}
  {1979})%
  \bibAnnoteFile{NoStop}{triang}%
\bibitem{Johnson}%
  \BibitemOpen
  \bibfield{author}{%
  \bibinfo {author} {\bibfnamefont{M.}~\bibnamefont{Johnson}}, \bibinfo
  {author} {\bibfnamefont{O.}~\bibnamefont{Albrektsen}}, \bibinfo {author}
  {\bibfnamefont{R.}~\bibnamefont{Feenstra}},\ and\ \bibinfo {author}
  {\bibfnamefont{H.}~\bibnamefont{Salemink}},\ }%
  \bibfield{journal}{%
  \bibinfo {journal} {Appl. Phys. Lett.}\ }%
  \textbf{\bibinfo {volume} {63}},\ \bibinfo {pages} {2923} (\bibinfo {year}
  {1993})%
  \bibAnnoteFile{NoStop}{Johnson}%
\bibitem{Koenraad}%
  \BibitemOpen
  \bibfield{author}{%
  \bibinfo {author} {\bibfnamefont{D.}~\bibnamefont{Bruls}}, \bibinfo {author}
  {\bibfnamefont{J.}~\bibnamefont{Vugs}}, \bibinfo {author}
  {\bibfnamefont{P.}~\bibnamefont{Koenraad}}, \bibinfo {author}
  {\bibfnamefont{H.}~\bibnamefont{Salemink}}, \bibinfo {author}
  {\bibfnamefont{J.}~\bibnamefont{Wolter}}, \bibinfo {author}
  {\bibfnamefont{M.}~\bibnamefont{Hopkinson}}, \bibinfo {author}
  {\bibfnamefont{M.}~\bibnamefont{Skolnick}}, \bibinfo {author}
  {\bibfnamefont{F.}~\bibnamefont{Long}},\ and\ \bibinfo {author}
  {\bibfnamefont{S.}~\bibnamefont{Gill}},\ }%
  \bibfield{journal}{%
  \bibinfo {journal} {Appl. Phys. Lett.}\ }%
  \textbf{\bibinfo {volume} {81}},\ \bibinfo {pages} {1708} (\bibinfo {year}
  {2002})%
  \bibAnnoteFile{NoStop}{Koenraad}%
\bibitem{Kanisawa2}%
  \BibitemOpen
  \bibfield{author}{%
  \bibinfo {author} {\bibfnamefont{K.}~\bibnamefont{Suzuki}}, \bibinfo {author}
  {\bibfnamefont{K.}~\bibnamefont{Kanisawa}}, \bibinfo {author}
  {\bibfnamefont{C.}~\bibnamefont{Janer}}, \bibinfo {author}
  {\bibfnamefont{S.}~\bibnamefont{Perraud}}, \bibinfo {author}
  {\bibfnamefont{K.}~\bibnamefont{Takashina}}, \bibinfo {author}
  {\bibfnamefont{T.}~\bibnamefont{Fujisawa}},\ and\ \bibinfo {author}
  {\bibfnamefont{Y.}~\bibnamefont{Hirayama}},\ }%
  \bibfield{journal}{%
  \Doi{10.1103/PhysRevLett.98.136802}{\bibinfo {journal} {Phys. Rev. Lett.}}\
  }%
  \textbf{\bibinfo {volume} {98}},\ \bibinfo {pages} {136802} (\bibinfo {year}
  {2007})%
  \bibAnnoteFile{NoStop}{Kanisawa2}%
\bibitem{Baier}%
  \BibitemOpen
  \bibfield{author}{%
  \bibinfo {author} {\bibfnamefont{H.}~\bibnamefont{Baier}}, \bibinfo {author}
  {\bibfnamefont{L.}~\bibnamefont{Koenders}},\ and\ \bibinfo {author}
  {\bibfnamefont{W.}~\bibnamefont{Moench}},\ }%
  \bibfield{journal}{%
  \bibinfo {journal} {Sol. St. Com.}\ }%
  \textbf{\bibinfo {volume} {58}},\ \bibinfo {pages} {327} (\bibinfo {year}
  {1986})%
  \bibAnnoteFile{NoStop}{Baier}%
\bibitem{Chen}%
  \BibitemOpen
  \bibfield{author}{%
  \bibinfo {author} {\bibfnamefont{Y.}~\bibnamefont{Chen}}, \bibinfo {author}
  {\bibfnamefont{J.~C.}\ \bibnamefont{Hermanson}},\ and\ \bibinfo {author}
  {\bibfnamefont{G.~J.}\ \bibnamefont{Lapeyre}},\ }%
  \bibfield{journal}{%
  \Doi{10.1103/PhysRevB.39.12682}{\bibinfo {journal} {Phys. Rev. B}}\ }%
  \textbf{\bibinfo {volume} {39}},\ \bibinfo {pages} {12682} (\bibinfo {year}
  {1989})%
  \bibAnnoteFile{NoStop}{Chen}%
\bibitem{CoInAs}%
  \BibitemOpen
  \bibfield{author}{%
  \bibinfo {author} {\bibfnamefont{M.}~\bibnamefont{Morgenstern}}, \bibinfo
  {author} {\bibfnamefont{J.}~\bibnamefont{Wiebe}}, \bibinfo {author}
  {\bibfnamefont{A.}~\bibnamefont{Wachowiak}}, \bibinfo {author}
  {\bibfnamefont{M.}~\bibnamefont{Getzlaff}}, \bibinfo {author}
  {\bibfnamefont{J.}~\bibnamefont{Klijn}}, \bibinfo {author}
  {\bibfnamefont{L.}~\bibnamefont{Plucinski}}, \bibinfo {author}
  {\bibfnamefont{R.~L.}\ \bibnamefont{Johnson}},\ and\ \bibinfo {author}
  {\bibfnamefont{R.}~\bibnamefont{Wiesendanger}},\ }%
  \bibfield{journal}{%
  \Doi{10.1103/PhysRevB.65.155325}{\bibinfo {journal} {Phys. Rev. B}}\ }%
  \textbf{\bibinfo {volume} {65}},\ \bibinfo {pages} {155325} (\bibinfo {year}
  {2002})%
  \bibAnnoteFile{NoStop}{CoInAs}%
\bibitem{Poissonsolver}%
  \BibitemOpen
  \bibfield{author}{%
  \bibinfo {author} {\bibfnamefont{I.}~\bibnamefont{Tan}}, \bibinfo {author}
  {\bibfnamefont{G.}~\bibnamefont{Snider}}, \bibinfo {author}
  {\bibfnamefont{L.}~\bibnamefont{Chang}},\ and\ \bibinfo {author}
  {\bibfnamefont{E.}~\bibnamefont{Hu}},\ }%
  \bibfield{journal}{%
  \bibinfo {journal} {J. Appl. Phys.}\ }%
  \textbf{\bibinfo {volume} {68}},\ \bibinfo {pages} {4071} (\bibinfo {year}
  {1990})%
  \bibAnnoteFile{NoStop}{Poissonsolver}%
\bibitem{RevModPhys.54.437}%
  \BibitemOpen
  \bibfield{author}{%
  \bibinfo {author} {\bibfnamefont{T.}~\bibnamefont{Ando}}, \bibinfo {author}
  {\bibfnamefont{A.~B.}\ \bibnamefont{Fowler}},\ and\ \bibinfo {author}
  {\bibfnamefont{F.}~\bibnamefont{Stern}},\ }%
  \bibfield{journal}{%
  \Doi{10.1103/RevModPhys.54.437}{\bibinfo {journal} {Rev. Mod. Phys.}}\ }%
  \textbf{\bibinfo {volume} {54}},\ \bibinfo {pages} {437} (\bibinfo {year}
  {1982})%
  \bibAnnoteFile{NoStop}{RevModPhys.54.437}%
\bibitem{Olson}%
  \BibitemOpen
  \bibfield{author}{%
  \bibinfo {author} {\bibfnamefont{L.~O.}\ \bibnamefont{Olsson}}, \bibinfo
  {author} {\bibfnamefont{C.~B.~M.}\ \bibnamefont{Andersson}}, \bibinfo
  {author} {\bibfnamefont{M.~C.}\ \bibnamefont{H\aa{}kansson}}, \bibinfo
  {author} {\bibfnamefont{J.}~\bibnamefont{Kanski}}, \bibinfo {author}
  {\bibfnamefont{L.}~\bibnamefont{Ilver}},\ and\ \bibinfo {author}
  {\bibfnamefont{U.~O.}\ \bibnamefont{Karlsson}},\ }%
  \bibfield{journal}{%
  \Doi{10.1103/PhysRevLett.76.3626}{\bibinfo {journal} {Phys. Rev. Lett.}}\ }%
  \textbf{\bibinfo {volume} {76}},\ \bibinfo {pages} {3626} (\bibinfo {year}
  {1996})%
  \bibAnnoteFile{NoStop}{Olson}%
\bibitem{Meyer}%
  \BibitemOpen
  \bibfield{author}{%
  \bibinfo {author} {\bibfnamefont{C.}~\bibnamefont{Meyer}}, \bibinfo {author}
  {\bibfnamefont{J.}~\bibnamefont{Klijn}}, \bibinfo {author}
  {\bibfnamefont{M.}~\bibnamefont{Morgenstern}},\ and\ \bibinfo {author}
  {\bibfnamefont{R.}~\bibnamefont{Wiesendanger}},\ }%
  \bibfield{journal}{%
  \Doi{10.1103/PhysRevLett.91.076803}{\bibinfo {journal} {Phys. Rev. Lett.}}\
  }%
  \textbf{\bibinfo {volume} {91}},\ \bibinfo {pages} {076803} (\bibinfo {year}
  {2003})%
  \bibAnnoteFile{NoStop}{Meyer}%
\bibitem{Huijser}%
  \BibitemOpen
  \bibfield{author}{%
  \bibinfo {author} {\bibfnamefont{A.}~\bibnamefont{Huijser}}, \bibinfo
  {author} {\bibfnamefont{J.}~\bibnamefont{van Laar}},\ and\ \bibinfo {author}
  {\bibfnamefont{T.}~\bibnamefont{van Rooy}},\ }%
  \bibfield{journal}{%
  \bibinfo {journal} {Surf. Sci.}\ }%
  \textbf{\bibinfo {volume} {62}},\ \bibinfo {pages} {472 } (\bibinfo {year}
  {1977})%
  \bibAnnoteFile{NoStop}{Huijser}%
\bibitem{Vurgaftman}%
  \BibitemOpen
  \bibfield{author}{%
  \bibinfo {author} {\bibfnamefont{I.}~\bibnamefont{Vurgaftman}}, \bibinfo
  {author} {\bibfnamefont{J.~R.}\ \bibnamefont{Meyer}},\ and\ \bibinfo {author}
  {\bibfnamefont{L.~R.}\ \bibnamefont{Ram-Mohan}},\ }%
  \bibfield{journal}{%
  \Doi{10.1063/1.1368156}{\bibinfo {journal} {J. Appl. Phys.}}\ }%
  \textbf{\bibinfo {volume} {89}},\ \bibinfo {pages} {5815} (\bibinfo {year}
  {2001})%
  \bibAnnoteFile{NoStop}{Vurgaftman}%
\bibitem{mashoff:053702}%
  \BibitemOpen
  \bibfield{author}{%
  \bibinfo {author} {\bibfnamefont{T.}~\bibnamefont{Mashoff}}, \bibinfo
  {author} {\bibfnamefont{M.}~\bibnamefont{Pratzer}},\ and\ \bibinfo {author}
  {\bibfnamefont{M.}~\bibnamefont{Morgenstern}},\ }%
  \bibfield{journal}{%
  \Doi{10.1063/1.3127589}{\bibinfo {journal} {Rev. Sci. Instrum.}}\ }%
  \textbf{\bibinfo {volume} {80}},\ \bibinfo {eid} {053702} (\bibinfo {year}
  {2009})%
  \bibAnnoteFile{NoStop}{mashoff:053702}%
\bibitem{Morgenstern3}%
  \BibitemOpen
  \bibfield{author}{%
  \bibinfo {author} {\bibfnamefont{M.}~\bibnamefont{Morgenstern}}, \bibinfo
  {author} {\bibfnamefont{C.}~\bibnamefont{Wittneven}}, \bibinfo {author}
  {\bibfnamefont{R.}~\bibnamefont{Dombrowski}},\ and\ \bibinfo {author}
  {\bibfnamefont{R.}~\bibnamefont{Wiesendanger}},\ }%
  \bibfield{journal}{%
  \bibinfo {journal} {Phys. Rev. Lett.}\ }%
  \textbf{\bibinfo {volume} {84}},\ \bibinfo {pages} {5588} (\bibinfo {year}
  {2000})%
  \bibAnnoteFile{NoStop}{Morgenstern3}%
\bibitem{Ewelle}%
  \BibitemOpen
  \bibfield{author}{%
  \bibinfo {author} {\bibfnamefont{C.}~\bibnamefont{Wittneven}}, \bibinfo
  {author} {\bibfnamefont{R.}~\bibnamefont{Dombrowski}}, \bibinfo {author}
  {\bibfnamefont{M.}~\bibnamefont{Morgenstern}},\ and\ \bibinfo {author}
  {\bibfnamefont{R.}~\bibnamefont{Wiesendanger}},\ }%
  \bibfield{journal}{%
  \Doi{10.1103/PhysRevLett.81.5616}{\bibinfo {journal} {Phys. Rev. Lett.}}\ }%
  \textbf{\bibinfo {volume} {81}},\ \bibinfo {pages} {5616} (\bibinfo {year}
  {1998})%
  \bibAnnoteFile{NoStop}{Ewelle}%
\bibitem{Heinze}%
  \BibitemOpen
  \bibfield{author}{%
  \bibinfo {author} {\bibfnamefont{S.}~\bibnamefont{Heinze}}, \bibinfo {author}
  {\bibfnamefont{S.}~\bibnamefont{Bl\"ugel}}, \bibinfo {author}
  {\bibfnamefont{R.}~\bibnamefont{Pascal}}, \bibinfo {author}
  {\bibfnamefont{M.}~\bibnamefont{Bode}},\ and\ \bibinfo {author}
  {\bibfnamefont{R.}~\bibnamefont{Wiesendanger}},\ }%
  \bibfield{journal}{%
  \Doi{10.1103/PhysRevB.58.16432}{\bibinfo {journal} {Phys. Rev. B}}\ }%
  \textbf{\bibinfo {volume} {58}},\ \bibinfo {pages} {16432} (\bibinfo {year}
  {1998})%
  \bibAnnoteFile{NoStop}{Heinze}%
\bibitem{Feenstra}%
  \BibitemOpen
  \bibfield{author}{%
  \bibinfo {author} {\bibfnamefont{R.}~\bibnamefont{Feenstra}}, \bibinfo
  {author} {\bibfnamefont{J.}~\bibnamefont{Stroscio}},\ and\ \bibinfo {author}
  {\bibfnamefont{A.}~\bibnamefont{Fein}},\ }%
  \bibfield{journal}{%
  \bibinfo {journal} {Surf. Sci.}\ }%
  \textbf{\bibinfo {volume} {181}},\ \bibinfo {pages} {295} (\bibinfo {year}
  {1987})%
  \bibAnnoteFile{NoStop}{Feenstra}%
\bibitem{LLfluct}%
  \BibitemOpen
  \bibfield{author}{%
  \bibinfo {author} {\bibfnamefont{M.}~\bibnamefont{Morgenstern}}, \bibinfo
  {author} {\bibfnamefont{C.}~\bibnamefont{Wittneven}}, \bibinfo {author}
  {\bibfnamefont{R.}~\bibnamefont{Dombrowski}},\ and\ \bibinfo {author}
  {\bibfnamefont{R.}~\bibnamefont{Wiesendanger}},\ }%
  \bibfield{journal}{%
  \Doi{10.1103/PhysRevLett.84.5588}{\bibinfo {journal} {Phys. Rev. Lett.}}\ }%
  \textbf{\bibinfo {volume} {84}},\ \bibinfo {pages} {5588} (\bibinfo {year}
  {2000})%
  \bibAnnoteFile{NoStop}{LLfluct}%
\bibitem{Wiebe3}%
  \BibitemOpen
  \bibfield{author}{%
  \bibinfo {author} {\bibfnamefont{J.}~\bibnamefont{Wiebe}}, \bibinfo {author}
  {\bibfnamefont{A.}~\bibnamefont{Wachowiak}}, \bibinfo {author}
  {\bibfnamefont{F.}~\bibnamefont{Meier}}, \bibinfo {author}
  {\bibfnamefont{D.}~\bibnamefont{Haude}}, \bibinfo {author}
  {\bibfnamefont{T.}~\bibnamefont{Foster}}, \bibinfo {author}
  {\bibfnamefont{M.}~\bibnamefont{Morgenstern}},\ and\ \bibinfo {author}
  {\bibfnamefont{R.}~\bibnamefont{Wiesendanger}},\ }%
  \bibfield{journal}{%
  \bibinfo {journal} {Rev, Sci. Instr.}\ }%
  \textbf{\bibinfo {volume} {75}},\ \bibinfo {pages} {4871} (\bibinfo {year}
  {2004})%
  \bibAnnoteFile{NoStop}{Wiebe3}%
\bibitem{bandbend}%
  \BibitemOpen
  \bibfield{author}{%
  \bibinfo {author} {\bibfnamefont{R.}~\bibnamefont{Feenstra}}\ and\ \bibinfo
  {author} {\bibfnamefont{J.}~\bibnamefont{Stroscio}},\ }%
  \bibfield{journal}{%
  \bibinfo {journal} {J. Vac. Sci. Technol. B}\ }%
  \textbf{\bibinfo {volume} {5}},\ \bibinfo {pages} {923} (\bibinfo {year}
  {1987})%
  \bibAnnoteFile{NoStop}{bandbend}%
\bibitem{spinsplit}%
  \BibitemOpen
  \bibfield{author}{%
  \bibinfo {author} {\bibfnamefont{M.}~\bibnamefont{Morgenstern}}, \bibinfo
  {author} {\bibfnamefont{V.}~\bibnamefont{Gudmundsson}}, \bibinfo {author}
  {\bibfnamefont{R.}~\bibnamefont{Dombrowski}}, \bibinfo {author}
  {\bibfnamefont{C.}~\bibnamefont{Wittneven}},\ and\ \bibinfo {author}
  {\bibfnamefont{R.}~\bibnamefont{Wiesendanger}},\ }%
  \bibfield{journal}{%
  \Doi{10.1103/PhysRevB.63.201301}{\bibinfo {journal} {Phys. Rev. B}}\ }%
  \textbf{\bibinfo {volume} {63}},\ \bibinfo {pages} {201301} (\bibinfo {year}
  {2001})%
  \bibAnnoteFile{NoStop}{spinsplit}%
\bibitem{Visscher}%
  \BibitemOpen
  \bibfield{author}{%
  \bibinfo {author} {\bibfnamefont{P.~B.}\ \bibnamefont{Visscher}}\ and\
  \bibinfo {author} {\bibfnamefont{L.~M.}\ \bibnamefont{Falicov}},\ }%
  \bibfield{journal}{%
  \bibinfo {journal} {Phys. Rev. B}\ }%
  \textbf{\bibinfo {volume} {3}},\ \bibinfo {pages} {2541} (\bibinfo {year}
  {1971})%
  \bibAnnoteFile{NoStop}{Visscher}%
\bibitem{Mochizuki}%
  \BibitemOpen
  \bibfield{author}{%
  \bibinfo {author} {\bibfnamefont{T.}~\bibnamefont{Mochizuki}}, \bibinfo
  {author} {\bibfnamefont{R.}~\bibnamefont{Masutomi}},\ and\ \bibinfo {author}
  {\bibfnamefont{T.}~\bibnamefont{Okamoto}},\ }%
  \bibfield{journal}{%
  \Doi{10.1103/PhysRevLett.101.267204}{\bibinfo {journal} {Phys. Rev. Lett.}}\
  }%
  \textbf{\bibinfo {volume} {101}},\ \bibinfo {pages} {267204} (\bibinfo {year}
  {2008})%
  \bibAnnoteFile{NoStop}{Mochizuki}%
\bibitem{Sacharow}%
  \BibitemOpen
  \bibfield{author}{%
  \bibinfo {author} {\bibfnamefont{L.}~\bibnamefont{Sacharow}}, \bibinfo
  {author} {\bibfnamefont{M.}~\bibnamefont{Morgenstern}}, \bibinfo {author}
  {\bibfnamefont{G.}~\bibnamefont{Bihlmayer}},\ and\ \bibinfo {author}
  {\bibfnamefont{S.}~\bibnamefont{Bl\"ugel}},\ }%
  \bibfield{journal}{%
  \Doi{10.1103/PhysRevB.69.085317}{\bibinfo {journal} {Phys. Rev. B}}\ }%
  \textbf{\bibinfo {volume} {69}},\ \bibinfo {pages} {085317} (\bibinfo {year}
  {2004})%
  \bibAnnoteFile{NoStop}{Sacharow}%
\bibitem{Sacharow2}%
  \BibitemOpen
  \bibinfo {author} {\bibfnamefont{L.}~\bibnamefont{Sacharow}}, \bibinfo
  {author} {\bibfnamefont{R.}~\bibnamefont{Wiesendanger}}, \bibinfo {author}
  {\bibfnamefont{G.}~\bibnamefont{Bihlmeyer}}, \bibinfo {author}
  {\bibfnamefont{S.}~\bibnamefont{Bluegel}},\ and\ \bibinfo {author}
  {\bibfnamefont{M.}~\bibnamefont{Morgenstern}}%
  \bibAnnoteFile{NoStop}{Sacharow2}%
\bibitem{Averin}%
  \BibitemOpen
\bibfield{author}{%
    }%
  \bibfield{author}{%
  \bibinfo {author} {\bibfnamefont{D.}~\bibnamefont{Averin}}\ and\ \bibinfo
  {author} {\bibfnamefont{K.}~\bibnamefont{Likharev}},\ }%
  \bibfield{journal}{%
  \bibinfo {journal} {J. Low Temp. Phys.}\ }%
  \textbf{\bibinfo {volume} {62}},\ \bibinfo {pages} {345} (\bibinfo {year}
  {1986})%
  \bibAnnoteFile{NoStop}{Averin}%
\bibitem{Tinkham}%
  \BibitemOpen
  \bibfield{author}{%
  \bibinfo {author} {\bibfnamefont{A.~E.}\ \bibnamefont{Hanna}}\ and\ \bibinfo
  {author} {\bibfnamefont{M.}~\bibnamefont{Tinkham}},\ }%
  \bibfield{journal}{%
  \Doi{10.1103/PhysRevB.44.5919}{\bibinfo {journal} {Phys. Rev. B}}\ }%
  \textbf{\bibinfo {volume} {44}},\ \bibinfo {pages} {5919} (\bibinfo {year}
  {1991})%
  \bibAnnoteFile{NoStop}{Tinkham}%
\bibitem{Pauly}%
  \BibitemOpen
  \bibfield{author}{%
  \bibinfo {author} {\bibfnamefont{C.}~\bibnamefont{Pauly}}, \bibinfo {author}
  {\bibfnamefont{M.}~\bibnamefont{Grob}}, \bibinfo {author}
  {\bibfnamefont{M.}~\bibnamefont{Pezzotta}}, \bibinfo {author}
  {\bibfnamefont{M.}~\bibnamefont{Pratzer}},\ and\ \bibinfo {author}
  {\bibfnamefont{M.}~\bibnamefont{Morgenstern}},\ }%
  \bibfield{journal}{%
  \Doi{10.1103/PhysRevB.81.125446}{\bibinfo {journal} {Phys. Rev. B}}\ }%
  \textbf{\bibinfo {volume} {81}},\ \bibinfo {pages} {125446} (\bibinfo {year}
  {2010})%
  \bibAnnoteFile{NoStop}{Pauly}%
\bibitem{SRL}%
  \BibitemOpen
  \bibfield{author}{%
  \bibinfo {author} {\bibfnamefont{M.}~\bibnamefont{Morgenstern}},\ }%
  \bibfield{journal}{%
  \bibinfo {journal} {Surf. Rev. Lett.}\ }%
  \textbf{\bibinfo {volume} {10}},\ \bibinfo {pages} {933} (\bibinfo {year}
  {2003})%
  \bibAnnoteFile{NoStop}{SRL}%
\bibitem{KaneInSb}%
  \BibitemOpen
  \bibfield{author}{%
  \bibinfo {author} {\bibfnamefont{E.~O.}\ \bibnamefont{Kane}},\ }%
  \bibfield{journal}{%
  \Doi{10.1016/0022-3697(57)90013-6}{\bibinfo {journal} {J. Phys. Chem.
  Solids}}\ }%
  \textbf{\bibinfo {volume} {1}},\ \bibinfo {pages} {249} (\bibinfo {year}
  {1957})%
  \bibAnnoteFile{NoStop}{KaneInSb}%
\bibitem{PhysRevB.35.2460}%
  \BibitemOpen
  \bibfield{author}{%
  \bibinfo {author} {\bibfnamefont{U.}~\bibnamefont{Merkt}}\ and\ \bibinfo
  {author} {\bibfnamefont{S.}~\bibnamefont{Oelting}},\ }%
  \bibfield{journal}{%
  \Doi{10.1103/PhysRevB.35.2460}{\bibinfo {journal} {Phys. Rev. B}}\ }%
  \textbf{\bibinfo {volume} {35}},\ \bibinfo {pages} {2460} (\bibinfo {year}
  {1987})%
  \bibAnnoteFile{NoStop}{PhysRevB.35.2460}%
\bibitem{Klijn}%
  \BibitemOpen
  \bibfield{author}{%
  \bibinfo {author} {\bibfnamefont{J.}~\bibnamefont{Klijn}}, \bibinfo {author}
  {\bibfnamefont{L.}~\bibnamefont{Sacharow}}, \bibinfo {author}
  {\bibfnamefont{C.}~\bibnamefont{Meyer}}, \bibinfo {author}
  {\bibfnamefont{S.}~\bibnamefont{Bl\"ugel}}, \bibinfo {author}
  {\bibfnamefont{M.}~\bibnamefont{Morgenstern}},\ and\ \bibinfo {author}
  {\bibfnamefont{R.}~\bibnamefont{Wiesendanger}},\ }%
  \bibfield{journal}{%
  \bibinfo {journal} {Phys. Rev. B}\ }%
  \textbf{\bibinfo {volume} {68}},\ \bibinfo {pages} {205327} (\bibinfo {month}
  {Nov}\ \bibinfo {year} {2003})%
  \bibAnnoteFile{NoStop}{Klijn}%
\bibitem{Efros}%
  \BibitemOpen
  \bibinfo {author} {\bibfnamefont{L.}~\bibnamefont{Kanskaya}}, \bibinfo
  {author} {\bibfnamefont{S.}~\bibnamefont{Kokhanovskii}}, \bibinfo {author}
  {\bibfnamefont{R.}~\bibnamefont{Seysan}},\ and\ \bibinfo {author}
  {\bibfnamefont{A.}~\bibnamefont{Efros}}%
  \bibAnnoteFile{NoStop}{Efros}%
\bibitem{Crommie}%
  \BibitemOpen
\bibfield{author}{%
    }%
  \bibfield{author}{%
  \bibinfo {author} {\bibfnamefont{M.}~\bibnamefont{Crommie}}, \bibinfo
  {author} {\bibfnamefont{C.}~\bibnamefont{Lutz}},\ and\ \bibinfo {author}
  {\bibfnamefont{D.}~\bibnamefont{Eigler}},\ }%
  \bibfield{journal}{%
  \bibinfo {journal} {Nature}\ }%
  \textbf{\bibinfo {volume} {363}},\ \bibinfo {pages} {524} (\bibinfo {year}
  {1993})%
  \bibAnnoteFile{NoStop}{Crommie}%
\bibitem{Besenbacher}%
  \BibitemOpen
  \bibfield{author}{%
  \bibinfo {author} {\bibfnamefont{L.}~\bibnamefont{Petersen}}, \bibinfo
  {author} {\bibfnamefont{P.~T.}\ \bibnamefont{Sprunger}}, \bibinfo {author}
  {\bibfnamefont{P.}~\bibnamefont{Hofmann}}, \bibinfo {author}
  {\bibfnamefont{E.}~\bibnamefont{L\ae{}gsgaard}}, \bibinfo {author}
  {\bibfnamefont{B.~G.}\ \bibnamefont{Briner}}, \bibinfo {author}
  {\bibfnamefont{M.}~\bibnamefont{Doering}}, \bibinfo {author}
  {\bibfnamefont{H.-P.}\ \bibnamefont{Rust}}, \bibinfo {author}
  {\bibfnamefont{A.~M.}\ \bibnamefont{Bradshaw}}, \bibinfo {author}
  {\bibfnamefont{F.}~\bibnamefont{Besenbacher}},\ and\ \bibinfo {author}
  {\bibfnamefont{E.~W.}\ \bibnamefont{Plummer}},\ }%
  \bibfield{journal}{%
  \Doi{10.1103/PhysRevB.57.R6858}{\bibinfo {journal} {Phys. Rev. B}}\ }%
  \textbf{\bibinfo {volume} {57}},\ \bibinfo {pages} {R6858} (\bibinfo {year}
  {1998})%
  \bibAnnoteFile{NoStop}{Besenbacher}%
\bibitem{Anderson}%
  \BibitemOpen
  \bibfield{author}{%
  \bibinfo {author} {\bibfnamefont{E.}~\bibnamefont{Abrahams}}, \bibinfo
  {author} {\bibfnamefont{P.~W.}\ \bibnamefont{Anderson}}, \bibinfo {author}
  {\bibfnamefont{D.}~\bibnamefont{Licciardello}},\ and\ \bibinfo {author}
  {\bibfnamefont{T.~V.}\ \bibnamefont{Ramakrishnan}},\ }%
  \bibfield{journal}{%
  \bibinfo {journal} {Phys. Rev. Lett.}\ }%
  \textbf{\bibinfo {volume} {42}},\ \bibinfo {pages} {673} (\bibinfo {year}
  {1979})%
  \bibAnnoteFile{NoStop}{Anderson}%
\bibitem{Beenakker}%
  \BibitemOpen
  \bibfield{author}{%
  \bibinfo {author} {\bibfnamefont{C.}~\bibnamefont{Beenakker}}\ and\ \bibinfo
  {author} {\bibfnamefont{H.}~\bibnamefont{VanHouten}},\ }%
  \bibfield{journal}{%
  \bibinfo {journal} {Solid State Phys.}\ }%
  \textbf{\bibinfo {volume} {44}},\ \bibinfo {pages} {1} (\bibinfo {year}
  {1991})%
  \bibAnnoteFile{NoStop}{Beenakker}%
\bibitem{Jeckelmann}%
  \BibitemOpen
  \bibfield{author}{%
  \bibinfo {author} {\bibfnamefont{B.}~\bibnamefont{Jeckelmann}}\ and\ \bibinfo
  {author} {\bibfnamefont{B.}~\bibnamefont{Jeanneret}},\ }%
  \bibfield{journal}{%
  \bibinfo {journal} {Rep. Prog. Phys.}\ }%
  \textbf{\bibinfo {volume} {64}},\ \bibinfo {pages} {1603} (\bibinfo {year}
  {2001})%
  \bibAnnoteFile{NoStop}{Jeckelmann}%
\bibitem{massstandard}%
  \BibitemOpen
  \bibfield{author}{%
  \bibinfo {author} {\bibfnamefont{F.}~\bibnamefont{Piquemal}}, \bibinfo
  {author} {\bibfnamefont{A.}~\bibnamefont{Bounouh}}, \bibinfo {author}
  {\bibfnamefont{L.}~\bibnamefont{Devoille}}, \bibinfo {author}
  {\bibfnamefont{N.}~\bibnamefont{Feltin}}, \bibinfo {author}
  {\bibfnamefont{O.}~\bibnamefont{Thevenot}},\ and\ \bibinfo {author}
  {\bibfnamefont{G.}~\bibnamefont{Trapon}},\ }%
  \bibfield{journal}{%
  \bibinfo {journal} {Comptes Rendus Phys.}\ }%
  \textbf{\bibinfo {volume} {5}},\ \bibinfo {pages} {857} (\bibinfo {year}
  {2004})%
  \bibAnnoteFile{NoStop}{massstandard}%
\bibitem{Prange}%
  \BibitemOpen
  \bibfield{author}{%
  \bibinfo {author} {\bibfnamefont{R.}~\bibnamefont{Joynt}}\ and\ \bibinfo
  {author} {\bibfnamefont{R.~E.}\ \bibnamefont{Prange}},\ }%
  \bibfield{journal}{%
  \bibinfo {journal} {Phys. Rev. B}\ }%
  \textbf{\bibinfo {volume} {29}},\ \bibinfo {pages} {3303} (\bibinfo {year}
  {1984})%
  \bibAnnoteFile{NoStop}{Prange}%
\bibitem{Ando2}%
  \BibitemOpen
  \bibfield{author}{%
  \bibinfo {author} {\bibfnamefont{T.}~\bibnamefont{Ando}},\ }%
  \bibfield{journal}{%
  \bibinfo {journal} {J. Phys. Soc. Jpn.}\ }%
  \textbf{\bibinfo {volume} {53}},\ \bibinfo {pages} {3101} (\bibinfo {year}
  {1984})%
  \bibAnnoteFile{NoStop}{Ando2}%
\bibitem{Kramer}%
  \BibitemOpen
  \bibfield{author}{%
  \bibinfo {author} {\bibfnamefont{B.}~\bibnamefont{Kramer}}, \bibinfo {author}
  {\bibfnamefont{T.}~\bibnamefont{Ohtsuki}},\ and\ \bibinfo {author}
  {\bibfnamefont{S.}~\bibnamefont{Kettemann}},\ }%
  \bibfield{journal}{%
  \bibinfo {journal} {Phys. Rep.}\ }%
  \textbf{\bibinfo {volume} {417}},\ \bibinfo {pages} {211} (\bibinfo {year}
  {2005})%
  \bibAnnoteFile{NoStop}{Kramer}%
\bibitem{Florens1}%
  \BibitemOpen
  \bibfield{author}{%
  \bibinfo {author} {\bibfnamefont{T.}~\bibnamefont{Champel}}\ and\ \bibinfo
  {author} {\bibfnamefont{S.}~\bibnamefont{Florens}},\ }%
  \bibfield{journal}{%
  \Doi{10.1103/PhysRevB.82.045421}{\bibinfo {journal} {Phys. Rev. B}}\ }%
  \textbf{\bibinfo {volume} {82}},\ \bibinfo {pages} {045421} (\bibinfo {year}
  {2010})%
  \bibAnnoteFile{NoStop}{Florens1}%
\bibitem{Mirlin}%
  \BibitemOpen
  \bibfield{author}{%
  \bibinfo {author} {\bibfnamefont{A.}~\bibnamefont{Mirlin}}, \bibinfo {author}
  {\bibfnamefont{E.}~\bibnamefont{Altshuler}},\ and\ \bibinfo {author}
  {\bibfnamefont{P.}~\bibnamefont{Wolfle}},\ }%
  \bibfield{journal}{%
  \bibinfo {journal} {Ann. Phys.}\ }%
  \textbf{\bibinfo {volume} {5}},\ \bibinfo {pages} {281} (\bibinfo {year}
  {1996})%
  \bibAnnoteFile{NoStop}{Mirlin}%
\bibitem{Driftstates}%
  \BibitemOpen
  \bibfield{author}{%
  \bibinfo {author} {\bibfnamefont{M.}~\bibnamefont{Morgenstern}}, \bibinfo
  {author} {\bibfnamefont{J.}~\bibnamefont{Klijn}}, \bibinfo {author}
  {\bibfnamefont{C.}~\bibnamefont{Meyer}},\ and\ \bibinfo {author}
  {\bibfnamefont{R.}~\bibnamefont{Wiesendanger}},\ }%
  \bibfield{journal}{%
  \Doi{10.1103/PhysRevLett.90.056804}{\bibinfo {journal} {Phys. Rev. Lett.}}\
  }%
  \textbf{\bibinfo {volume} {90}},\ \bibinfo {pages} {056804} (\bibinfo {year}
  {2003})%
  \bibAnnoteFile{NoStop}{Driftstates}%
\bibitem{Evers}%
  \BibitemOpen
  \bibfield{author}{%
  \bibinfo {author} {\bibfnamefont{F.}~\bibnamefont{Evers}}\ and\ \bibinfo
  {author} {\bibfnamefont{A.~D.}\ \bibnamefont{Mirlin}},\ }%
  \bibfield{journal}{%
  \Doi{10.1103/RevModPhys.80.1355}{\bibinfo {journal} {Rev. Mod. Phys.}}\ }%
  \textbf{\bibinfo {volume} {80}},\ \bibinfo {pages} {1355} (\bibinfo {year}
  {2008})%
  \bibAnnoteFile{NoStop}{Evers}%
\bibitem{Slevin}%
  \BibitemOpen
  \bibfield{author}{%
  \bibinfo {author} {\bibfnamefont{K.}~\bibnamefont{Slevin}}\ and\ \bibinfo
  {author} {\bibfnamefont{T.}~\bibnamefont{Ohtsuki}},\ }%
  \bibfield{journal}{%
  \Doi{10.1103/PhysRevB.80.041304}{\bibinfo {journal} {Phys. Rev. B}}\ }%
  \textbf{\bibinfo {volume} {80}},\ \bibinfo {pages} {041304} (\bibinfo {year}
  {2009})%
  \bibAnnoteFile{NoStop}{Slevin}%
\bibitem{Roemer}%
  \BibitemOpen
  \bibfield{author}{%
  \bibinfo {author} {\bibfnamefont{P.}~\bibnamefont{Cain}}, \bibinfo {author}
  {\bibfnamefont{R.~A.}\ \bibnamefont{R\"omer}},\ and\ \bibinfo {author}
  {\bibfnamefont{M.~E.}\ \bibnamefont{Raikh}},\ }%
  \bibfield{journal}{%
  \Doi{10.1103/PhysRevB.67.075307}{\bibinfo {journal} {Phys. Rev. B}}\ }%
  \textbf{\bibinfo {volume} {67}},\ \bibinfo {pages} {075307} (\bibinfo {year}
  {2003})%
  \bibAnnoteFile{NoStop}{Roemer}%
\bibitem{Chalker}%
  \BibitemOpen
  \bibfield{author}{%
  \bibinfo {author} {\bibfnamefont{J.}~\bibnamefont{Chalker}}\ and\ \bibinfo
  {author} {\bibfnamefont{P.}~\bibnamefont{Coddington}},\ }%
  \bibfield{journal}{%
  \bibinfo {journal} {J. Phys. C}\ }%
  \textbf{\bibinfo {volume} {21}},\ \bibinfo {pages} {2665} (\bibinfo {year}
  {1988})%
  \bibAnnoteFile{NoStop}{Chalker}%
\bibitem{Aoki}%
  \BibitemOpen
  \bibfield{author}{%
  \bibinfo {author} {\bibfnamefont{H.}~\bibnamefont{Aoki}}\ and\ \bibinfo
  {author} {\bibfnamefont{T.}~\bibnamefont{Ando}},\ }%
  \bibfield{journal}{%
  \Doi{10.1103/PhysRevLett.54.831}{\bibinfo {journal} {Phys. Rev. Lett.}}\ }%
  \textbf{\bibinfo {volume} {54}},\ \bibinfo {pages} {831} (\bibinfo {year}
  {1985})%
  \bibAnnoteFile{NoStop}{Aoki}%
\bibitem{Evers2}%
  \BibitemOpen
  \bibfield{author}{%
  \bibinfo {author} {\bibfnamefont{F.}~\bibnamefont{Evers}}, \bibinfo {author}
  {\bibfnamefont{A.}~\bibnamefont{Mildenberger}},\ and\ \bibinfo {author}
  {\bibfnamefont{A.~D.}\ \bibnamefont{Mirlin}},\ }%
  \bibfield{journal}{%
  \Doi{10.1103/PhysRevB.64.241303}{\bibinfo {journal} {Phys. Rev. B}}\ }%
  \textbf{\bibinfo {volume} {64}},\ \bibinfo {pages} {241303} (\bibinfo {year}
  {2001})%
  \bibAnnoteFile{NoStop}{Evers2}%
\bibitem{Sohrmann}%
  \BibitemOpen
  \bibfield{author}{%
  \bibinfo {author} {\bibfnamefont{C.}~\bibnamefont{Sohrmann}}\ and\ \bibinfo
  {author} {\bibfnamefont{R.~A.}\ \bibnamefont{Roemer}},\ }%
  \bibfield{journal}{%
  \bibinfo {journal} {New J. Phys.}\ }%
  \textbf{\bibinfo {volume} {9}},\ \bibinfo {pages} {97} (\bibinfo {year}
  {2007})%
  \bibAnnoteFile{NoStop}{Sohrmann}%
\bibitem{Florens2}%
  \BibitemOpen
  \bibfield{author}{%
  \bibinfo {author} {\bibfnamefont{T.}~\bibnamefont{Champel}}\ and\ \bibinfo
  {author} {\bibfnamefont{S.}~\bibnamefont{Florens}},\ }%
  \bibfield{journal}{%
  \Doi{10.1103/PhysRevB.80.161311}{\bibinfo {journal} {Phys. Rev. B}}\ }%
  \textbf{\bibinfo {volume} {80}},\ \bibinfo {pages} {161311} (\bibinfo {year}
  {2009})%
  \bibAnnoteFile{NoStop}{Florens2}%
\bibitem{Menashe}%
  \BibitemOpen
  \bibfield{author}{%
  \bibinfo {author} {\bibfnamefont{D.}~\bibnamefont{Menashe}}, \bibinfo
  {author} {\bibfnamefont{O.}~\bibnamefont{Biham}}, \bibinfo {author}
  {\bibfnamefont{B.~D.}\ \bibnamefont{Laikhtman}},\ and\ \bibinfo {author}
  {\bibfnamefont{A.~L.}\ \bibnamefont{Efros}},\ }%
  \bibfield{journal}{%
  \Doi{10.1103/PhysRevB.64.115209}{\bibinfo {journal} {Phys. Rev. B}}\ }%
  \textbf{\bibinfo {volume} {64}},\ \bibinfo {pages} {115209} (\bibinfo {year}
  {2001})%
  \bibAnnoteFile{NoStop}{Menashe}%
\bibitem{Niimi}%
  \BibitemOpen
  \bibfield{author}{%
  \bibinfo {author} {\bibfnamefont{Y.}~\bibnamefont{Niimi}}, \bibinfo {author}
  {\bibfnamefont{H.}~\bibnamefont{Kambara}},\ and\ \bibinfo {author}
  {\bibfnamefont{H.}~\bibnamefont{Fukuyama}},\ }%
  \bibfield{journal}{%
  \bibinfo {journal} {Phys. Rev. Lett.}\ }%
  \textbf{\bibinfo {volume} {102}},\ \bibinfo {pages} {026803} (\bibinfo {year}
  {2009})%
  \bibAnnoteFile{NoStop}{Niimi}%
\bibitem{Miller}%
  \BibitemOpen
  \bibfield{author}{%
  \bibinfo {author} {\bibfnamefont{J.~B.}\ \bibnamefont{Miller}}, \bibinfo
  {author} {\bibfnamefont{D.~M.}\ \bibnamefont{Zumbuhl}}, \bibinfo {author}
  {\bibfnamefont{C.~M.}\ \bibnamefont{Marcus}}, \bibinfo {author}
  {\bibfnamefont{Y.~B.}\ \bibnamefont{Lyanda-Geller}}, \bibinfo {author}
  {\bibfnamefont{D.}~\bibnamefont{Goldhaber-Gordon}}, \bibinfo {author}
  {\bibfnamefont{K.}~\bibnamefont{Campman}},\ and\ \bibinfo {author}
  {\bibfnamefont{A.~C.}\ \bibnamefont{Gossard}},\ }%
  \bibfield{journal}{%
  \bibinfo {journal} {Phys. Rev. Lett.}\ }%
  \textbf{\bibinfo {volume} {90}},\ \bibinfo {pages} {076807} (\bibinfo {year}
  {2003})%
  \bibAnnoteFile{NoStop}{Miller}%
\bibitem{Stroscio2}%
  \BibitemOpen
  \bibfield{author}{%
  \bibinfo {author} {\bibfnamefont{D.~L.}\ \bibnamefont{Miller}}, \bibinfo
  {author} {\bibfnamefont{K.~D.}\ \bibnamefont{Kubista}}, \bibinfo {author}
  {\bibfnamefont{G.~M.}\ \bibnamefont{Rutter}}, \bibinfo {author}
  {\bibfnamefont{M.}~\bibnamefont{Ruan}}, \bibinfo {author}
  {\bibfnamefont{W.~A.}\ \bibnamefont{de~Heer}}, \bibinfo {author}
  {\bibfnamefont{P.~N.}\ \bibnamefont{First}},\ and\ \bibinfo {author}
  {\bibfnamefont{J.~A.}\ \bibnamefont{Stroscio}},\ }%
  \bibfield{journal}{%
  \bibinfo {journal} {Science}\ }%
  \textbf{\bibinfo {volume} {324}},\ \bibinfo {pages} {924} (\bibinfo {year}
  {2009})%
  \bibAnnoteFile{NoStop}{Stroscio2}%
\bibitem{GrapheneReview}%
  \BibitemOpen
  \bibfield{author}{%
  \bibinfo {author} {\bibfnamefont{M.}~\bibnamefont{Morgenstern}},\ }%
  \bibfield{journal}{%
  \bibinfo {journal} {phys. stat. sol. B}\ }%
  \textbf{\bibinfo {volume} {248}},\ \bibinfo {pages} {2423} (\bibinfo {year}
  {2011})%
  \bibAnnoteFile{NoStop}{GrapheneReview}%
\bibitem{Hashi2}%
  \BibitemOpen
  \bibfield{author}{%
  \bibinfo {author} {\bibfnamefont{K.}~\bibnamefont{Hashimoto}}, \bibinfo
  {author} {\bibfnamefont{T.}~\bibnamefont{Champel}}, \bibinfo {author}
  {\bibfnamefont{S.}~\bibnamefont{Florens}}, \bibinfo {author}
  {\bibfnamefont{C.}~\bibnamefont{Sohrmann}}, \bibinfo {author}
  {\bibfnamefont{J.}~\bibnamefont{Wiebe}}, \bibinfo {author}
  {\bibfnamefont{Y.}~\bibnamefont{Hirayama}}, \bibinfo {author}
  {\bibfnamefont{R.~A.}\ \bibnamefont{Roemer}}, \bibinfo {author}
  {\bibfnamefont{R.}~\bibnamefont{Wiesendanger}},\ and\ \bibinfo {author}
  {\bibfnamefont{M.}~\bibnamefont{M.}},\ }%
  \bibfield{journal}{%
  \bibinfo {journal} {ArXiV},\ \bibinfo {pages} {045421}}%
   (\bibinfo {year} {2011})%
  \bibAnnoteFile{NoStop}{Hashi2}%
\bibitem{Florens3}%
  \BibitemOpen
  \bibfield{author}{%
  \bibinfo {author} {\bibfnamefont{T.}~\bibnamefont{Champel}}\ and\ \bibinfo
  {author} {\bibfnamefont{S.}~\bibnamefont{Florens}},\ }%
  \bibfield{journal}{%
  \Doi{10.1103/PhysRevB.75.245326}{\bibinfo {journal} {Phys. Rev. B}}\ }%
  \textbf{\bibinfo {volume} {75}},\ \bibinfo {pages} {245326} (\bibinfo {year}
  {2007})%
  \bibAnnoteFile{NoStop}{Florens3}%
\bibitem{Kohmoto}%
  \BibitemOpen
  \bibfield{author}{%
  \bibinfo {author} {\bibfnamefont{M.}~\bibnamefont{Kohmoto}},\ }%
  \bibfield{journal}{%
  \bibinfo {journal} {Ann. Phys.}\ }%
  \textbf{\bibinfo {volume} {160}},\ \bibinfo {pages} {343} (\bibinfo {year}
  {1985})%
  \bibAnnoteFile{NoStop}{Kohmoto}%
\bibitem{Avron}%
  \BibitemOpen
  \bibfield{author}{%
  \bibinfo {author} {\bibfnamefont{J.~E.}\ \bibnamefont{Avron}}, \bibinfo
  {author} {\bibfnamefont{R.}~\bibnamefont{Seiler}},\ and\ \bibinfo {author}
  {\bibfnamefont{B.}~\bibnamefont{Simon}},\ }%
  \bibfield{journal}{%
  \Doi{10.1103/PhysRevLett.51.51}{\bibinfo {journal} {Phys. Rev. Lett.}}\ }%
  \textbf{\bibinfo {volume} {51}},\ \bibinfo {pages} {51} (\bibinfo {year}
  {1983})%
  \bibAnnoteFile{NoStop}{Avron}%
\bibitem{Laughlin}%
  \BibitemOpen
  \bibfield{author}{%
  \bibinfo {author} {\bibfnamefont{R.~B.}\ \bibnamefont{Laughlin}},\ }%
  \bibfield{journal}{%
  \Doi{10.1103/PhysRevB.23.5632}{\bibinfo {journal} {Phys. Rev. B}}\ }%
  \textbf{\bibinfo {volume} {23}},\ \bibinfo {pages} {5632} (\bibinfo {year}
  {1981})%
  \bibAnnoteFile{NoStop}{Laughlin}%
\bibitem{Halperin}%
  \BibitemOpen
  \bibfield{author}{%
  \bibinfo {author} {\bibfnamefont{B.~I.}\ \bibnamefont{Halperin}},\ }%
  \bibfield{journal}{%
  \Doi{10.1103/PhysRevB.25.2185}{\bibinfo {journal} {Phys. Rev. B}}\ }%
  \textbf{\bibinfo {volume} {25}},\ \bibinfo {pages} {2185} (\bibinfo {year}
  {1982})%
  \bibAnnoteFile{NoStop}{Halperin}%
\bibitem{McEuen}%
  \BibitemOpen
  \bibfield{author}{%
  \bibinfo {author} {\bibfnamefont{M.~T.}\ \bibnamefont{Woodside}}, \bibinfo
  {author} {\bibfnamefont{C.}~\bibnamefont{Vale}}, \bibinfo {author}
  {\bibfnamefont{P.~L.}\ \bibnamefont{McEuen}}, \bibinfo {author}
  {\bibfnamefont{C.}~\bibnamefont{Kadow}}, \bibinfo {author}
  {\bibfnamefont{K.~D.}\ \bibnamefont{Maranowski}},\ and\ \bibinfo {author}
  {\bibfnamefont{A.~C.}\ \bibnamefont{Gossard}},\ }%
  \bibfield{journal}{%
  \Doi{10.1103/PhysRevB.64.041310}{\bibinfo {journal} {Phys. Rev. B}}\ }%
  \textbf{\bibinfo {volume} {64}},\ \bibinfo {pages} {041310} (\bibinfo {year}
  {2001})%
  \bibAnnoteFile{NoStop}{McEuen}%
\bibitem{Weiss}%
  \BibitemOpen
  \bibfield{author}{%
  \bibinfo {author} {\bibfnamefont{E.}~\bibnamefont{Ahlswede}}, \bibinfo
  {author} {\bibfnamefont{J.}~\bibnamefont{Weis}}, \bibinfo {author}
  {\bibfnamefont{K.}~\bibnamefont{von Klitzing}},\ and\ \bibinfo {author}
  {\bibfnamefont{K.}~\bibnamefont{Eberl}},\ }%
  \bibfield{journal}{%
  \bibinfo {journal} {Physica E}\ }%
  \textbf{\bibinfo {volume} {12}},\ \bibinfo {pages} {165} (\bibinfo {year}
  {2002})%
  \bibAnnoteFile{NoStop}{Weiss}%
\bibitem{Ashoori}%
  \BibitemOpen
  \bibfield{author}{%
  \bibinfo {author} {\bibfnamefont{G.}~\bibnamefont{Finkelstein}}, \bibinfo
  {author} {\bibfnamefont{P.}~\bibnamefont{Glicofridis}}, \bibinfo {author}
  {\bibfnamefont{R.}~\bibnamefont{Ashoori}},\ and\ \bibinfo {author}
  {\bibfnamefont{M.}~\bibnamefont{Shayegan}},\ }%
  \bibfield{journal}{%
  \bibinfo {journal} {Science}\ }%
  \textbf{\bibinfo {volume} {289}},\ \bibinfo {pages} {90} (\bibinfo {year}
  {2000})%
  \bibAnnoteFile{NoStop}{Ashoori}%
\bibitem{Ito}%
  \BibitemOpen
  \bibfield{author}{%
  \bibinfo {author} {\bibfnamefont{H.}~\bibnamefont{Ito}}, \bibinfo {author}
  {\bibfnamefont{K.}~\bibnamefont{Furuya}}, \bibinfo {author}
  {\bibfnamefont{Y.}~\bibnamefont{Shibata}}, \bibinfo {author}
  {\bibfnamefont{S.}~\bibnamefont{Kashiwaya}}, \bibinfo {author}
  {\bibfnamefont{M.}~\bibnamefont{Yamaguchi}}, \bibinfo {author}
  {\bibfnamefont{T.}~\bibnamefont{Akazaki}}, \bibinfo {author}
  {\bibfnamefont{H.}~\bibnamefont{Tamura}}, \bibinfo {author}
  {\bibfnamefont{Y.}~\bibnamefont{Ootuka}},\ and\ \bibinfo {author}
  {\bibfnamefont{S.}~\bibnamefont{Nomura}},\ }%
  \bibfield{journal}{%
  \Doi{10.1103/PhysRevLett.107.256803}{\bibinfo {journal} {Phys. Rev. Lett.}}\
  }%
  \textbf{\bibinfo {volume} {107}},\ \bibinfo {pages} {256803} (\bibinfo {year}
  {2011})%
  \bibAnnoteFile{NoStop}{Ito}%
\bibitem{Yacoby2}%
  \BibitemOpen
  \bibfield{author}{%
  \bibinfo {author} {\bibfnamefont{A.}~\bibnamefont{Yacoby}}, \bibinfo {author}
  {\bibfnamefont{H.}~\bibnamefont{Hess}}, \bibinfo {author}
  {\bibfnamefont{T.}~\bibnamefont{Fulton}}, \bibinfo {author}
  {\bibfnamefont{L.}~\bibnamefont{Pfeiffer}},\ and\ \bibinfo {author}
  {\bibfnamefont{K.}~\bibnamefont{West}},\ }%
  \bibfield{journal}{%
  \bibinfo {journal} {Sol. St. Com.}\ }%
  \textbf{\bibinfo {volume} {111}},\ \bibinfo {pages} {1} (\bibinfo {year}
  {1999})%
  \bibAnnoteFile{NoStop}{Yacoby2}%
\bibitem{Ilani}%
  \BibitemOpen
  \bibinfo {author} {\bibfnamefont{S.}~\bibnamefont{Ilani}}, \bibinfo {author}
  {\bibfnamefont{J.}~\bibnamefont{Martin}}, \bibinfo {author}
  {\bibfnamefont{E.}~\bibnamefont{Teitelbaum}}, \bibinfo {author}
  {\bibfnamefont{J.}~\bibnamefont{Smet}}, \bibinfo {author}
  {\bibfnamefont{D.}~\bibnamefont{Mahalu}}, \bibinfo {author}
  {\bibfnamefont{V.}~\bibnamefont{Umansky}},\ and\ \bibinfo {author}
  {\bibfnamefont{A.}~\bibnamefont{Yacoby}}%
  \bibAnnoteFile{NoStop}{Ilani}%
\bibitem{Stroscio1}%
  \BibitemOpen
\bibfield{author}{%
    }%
  \bibfield{author}{%
  \bibinfo {author} {\bibfnamefont{Y.~J.}\ \bibnamefont{Song}}, \bibinfo
  {author} {\bibfnamefont{A.~F.}\ \bibnamefont{Otte}}, \bibinfo {author}
  {\bibfnamefont{Y.}~\bibnamefont{Kuk}}, \bibinfo {author}
  {\bibfnamefont{Y.}~\bibnamefont{Hu}}, \bibinfo {author}
  {\bibfnamefont{D.~B.}\ \bibnamefont{Torrance}}, \bibinfo {author}
  {\bibfnamefont{P.~N.}\ \bibnamefont{First}}, \bibinfo {author}
  {\bibfnamefont{W.~A.}\ \bibnamefont{de~Heer}}, \bibinfo {author}
  {\bibfnamefont{H.}~\bibnamefont{Min}}, \bibinfo {author}
  {\bibfnamefont{S.}~\bibnamefont{Adam}}, \bibinfo {author}
  {\bibfnamefont{M.~D.}\ \bibnamefont{Stiles}}, \bibinfo {author}
  {\bibfnamefont{A.~H.}\ \bibnamefont{MacDonald}},\ and\ \bibinfo {author}
  {\bibfnamefont{J.~A.}\ \bibnamefont{Stroscio}},\ }%
  \bibfield{journal}{%
  \Doi{10.1038/nature09330}{\bibinfo {journal} {Nature}}\ }%
  \textbf{\bibinfo {volume} {467}},\ \bibinfo {pages} {185} (\bibinfo {year}
  {2010})%
  \bibAnnoteFile{NoStop}{Stroscio1}%
\bibitem{Janak}%
  \BibitemOpen
  \bibfield{author}{%
  \bibinfo {author} {\bibfnamefont{J.~F.}\ \bibnamefont{Janak}},\ }%
  \bibfield{journal}{%
  \Doi{10.1103/PhysRev.178.1416}{\bibinfo {journal} {Phys. Rev.}}\ }%
  \textbf{\bibinfo {volume} {178}},\ \bibinfo {pages} {1416} (\bibinfo {year}
  {1969})%
  \bibAnnoteFile{NoStop}{Janak}%
\bibitem{Ando_exchange}%
  \BibitemOpen
  \bibfield{author}{%
  \bibinfo {author} {\bibfnamefont{T.}~\bibnamefont{Ando}}\ and\ \bibinfo
  {author} {\bibfnamefont{Y.}~\bibnamefont{Uemura}},\ }%
  \bibfield{journal}{%
  \bibinfo {journal} {J, Phys. Soc. Jpn.}\ }%
  \textbf{\bibinfo {volume} {37}},\ \bibinfo {pages} {1044} (\bibinfo {year}
  {1974})%
  \bibAnnoteFile{NoStop}{Ando_exchange}%
\bibitem{dial.nature}%
  \BibitemOpen
  \bibfield{author}{%
  \bibinfo {author} {\bibfnamefont{O.~E.}\ \bibnamefont{Dial}}, \bibinfo
  {author} {\bibfnamefont{R.~C.}\ \bibnamefont{Ashoori}}, \bibinfo {author}
  {\bibfnamefont{L.~N.}\ \bibnamefont{Pfeiffer}},\ and\ \bibinfo {author}
  {\bibfnamefont{K.~W.}\ \bibnamefont{West}},\ }%
  \bibfield{journal}{%
  \Doi{10.1038/nature05982}{\bibinfo {journal} {Nature}}\ }%
  \textbf{\bibinfo {volume} {448}},\ \bibinfo {pages} {176} (\bibinfo {year}
  {2007})%
  \bibAnnoteFile{NoStop}{dial.nature}%
\bibitem{Note1}%
  \BibitemOpen
  \bibinfo {note} {\protect \href
  {http://www.mathworks.com/help/toolbox/curvefit/}{MathWorks Curve Fitting
  Toolbox V2.1 User's Guide}}%
  \bibAnnoteFile{NoStop}{Note1}%
\bibitem{Bohm}%
  \BibitemOpen
  \bibfield{author}{%
  \bibinfo {author} {\bibfnamefont{D.}~\bibnamefont{Bohm}}\ and\ \bibinfo
  {author} {\bibfnamefont{D.}~\bibnamefont{Pines}},\ }%
  \bibfield{journal}{%
  \Doi{10.1103/PhysRev.92.609}{\bibinfo {journal} {Phys. Rev.}}\ }%
  \textbf{\bibinfo {volume} {92}},\ \bibinfo {pages} {609} (\bibinfo {year}
  {1953})%
  \bibAnnoteFile{NoStop}{Bohm}%
\bibitem{Gellmann}%
  \BibitemOpen
  \bibfield{author}{%
  \bibinfo {author} {\bibfnamefont{M.}~\bibnamefont{Gell-Mann}}\ and\ \bibinfo
  {author} {\bibfnamefont{K.~A.}\ \bibnamefont{Brueckner}},\ }%
  \bibfield{journal}{%
  \Doi{10.1103/PhysRev.106.364}{\bibinfo {journal} {Phys. Rev.}}\ }%
  \textbf{\bibinfo {volume} {106}},\ \bibinfo {pages} {364} (\bibinfo {year}
  {1957})%
  \bibAnnoteFile{NoStop}{Gellmann}%
\bibitem{MacDonald}%
  \BibitemOpen
  \bibfield{author}{%
  \bibinfo {author} {\bibfnamefont{A.~P.}\ \bibnamefont{Smith}}, \bibinfo
  {author} {\bibfnamefont{A.~H.}\ \bibnamefont{MacDonald}},\ and\ \bibinfo
  {author} {\bibfnamefont{G.}~\bibnamefont{Gumbs}},\ }%
  \bibfield{journal}{%
  \Doi{10.1103/PhysRevB.45.8829}{\bibinfo {journal} {Phys. Rev. B}}\ }%
  \textbf{\bibinfo {volume} {45}},\ \bibinfo {pages} {8829(R)} (\bibinfo {year}
  {1992})%
  \bibAnnoteFile{NoStop}{MacDonald}%
\bibitem{Pollak1970}%
  \BibitemOpen
  \bibfield{author}{%
  \bibinfo {author} {\bibfnamefont{M.}~\bibnamefont{Pollak}},\ }%
  \bibfield{journal}{%
  \Doi{10.1039/DF9705000013}{\bibinfo {journal} {Discuss. Faraday Soc.}}\ }%
  \textbf{\bibinfo {volume} {50}},\ \bibinfo {pages} {13} (\bibinfo {year}
  {1970})%
  \bibAnnoteFile{NoStop}{Pollak1970}%
\bibitem{Efros1975}%
  \BibitemOpen
  \bibfield{author}{%
  \bibinfo {author} {\bibfnamefont{A.~L.}\ \bibnamefont{Efros}}\ and\ \bibinfo
  {author} {\bibfnamefont{B.~I.}\ \bibnamefont{Shklovskii}},\ }%
  \bibfield{journal}{%
  \Doi{10.1088/0022-3719/8/4/003}{\bibinfo {journal} {J. Phys. C}}\ }%
  \textbf{\bibinfo {volume} {8}},\ \bibinfo {pages} {L49} (\bibinfo {year}
  {1975})%
  \bibAnnoteFile{NoStop}{Efros1975}%
\bibitem{Pikus1995}%
  \BibitemOpen
  \bibfield{author}{%
  \bibinfo {author} {\bibfnamefont{F.~G.}\ \bibnamefont{Pikus}}\ and\ \bibinfo
  {author} {\bibfnamefont{A.~L.}\ \bibnamefont{Efros}},\ }%
  \bibfield{journal}{%
  \Doi{10.1103/PhysRevB.51.16871}{\bibinfo {journal} {Phys. Rev. B}}\ }%
  \textbf{\bibinfo {volume} {51}},\ \bibinfo {pages} {16871} (\bibinfo {year}
  {1995})%
  \bibAnnoteFile{NoStop}{Pikus1995}%
\bibitem{Floehr}%
  \BibitemOpen
  \bibfield{author}{%
  \bibinfo {author} {\bibfnamefont{K.}~\bibnamefont{Floehr}}, \bibinfo {author}
  {\bibfnamefont{M.}~\bibnamefont{Liebmann}}, \bibinfo {author}
  {\bibfnamefont{K.}~\bibnamefont{Sladek}}, \bibinfo {author}
  {\bibfnamefont{H.~Y.}\ \bibnamefont{Guenel}}, \bibinfo {author}
  {\bibfnamefont{R.}~\bibnamefont{Frielinghaus}}, \bibinfo {author}
  {\bibfnamefont{F.}~\bibnamefont{Haas}}, \bibinfo {author}
  {\bibfnamefont{C.}~\bibnamefont{Meyer}}, \bibinfo {author}
  {\bibfnamefont{H.}~\bibnamefont{Hardtdegen}}, \bibinfo {author}
  {\bibfnamefont{T.}~\bibnamefont{Schaepers}}, \bibinfo {author}
  {\bibfnamefont{D.}~\bibnamefont{Gruetzmacher}},\ and\ \bibinfo {author}
  {\bibfnamefont{M.}~\bibnamefont{Morgenstern}},\ }%
  \bibfield{journal}{%
  \bibinfo {journal} {Rev. Sci. Instr.}\ }%
  \textbf{\bibinfo {volume} {82}},\ \bibinfo {pages} {113705} (\bibinfo {year}
  {2011})%
  \bibAnnoteFile{NoStop}{Floehr}%
\bibitem{Song}%
  \BibitemOpen
  \bibfield{author}{%
  \bibinfo {author} {\bibfnamefont{Y.~J.}\ \bibnamefont{Song}}, \bibinfo
  {author} {\bibfnamefont{A.~F.}\ \bibnamefont{Otte}}, \bibinfo {author}
  {\bibfnamefont{Y.}~\bibnamefont{Kuk}}, \bibinfo {author}
  {\bibfnamefont{Y.}~\bibnamefont{Hu}}, \bibinfo {author}
  {\bibfnamefont{D.~B.}\ \bibnamefont{Torrance}}, \bibinfo {author}
  {\bibfnamefont{P.~N.}\ \bibnamefont{First}}, \bibinfo {author}
  {\bibfnamefont{W.~A.}\ \bibnamefont{de~Heer}}, \bibinfo {author}
  {\bibfnamefont{H.}~\bibnamefont{Min}}, \bibinfo {author}
  {\bibfnamefont{S.}~\bibnamefont{Adam}}, \bibinfo {author}
  {\bibfnamefont{M.~D.}\ \bibnamefont{Stiles}}, \bibinfo {author}
  {\bibfnamefont{A.~H.}\ \bibnamefont{MacDonald}},\ and\ \bibinfo {author}
  {\bibfnamefont{J.~A.}\ \bibnamefont{Stroscio}},\ }%
  \bibfield{journal}{%
  \bibinfo {journal} {Nature}\ }%
  \textbf{\bibinfo {volume} {467}},\ \bibinfo {pages} {185} (\bibinfo {year}
  {2010})%
  \bibAnnoteFile{NoStop}{Song}%
\bibitem{AndradaeSilva1994}%
  \BibitemOpen
  \bibfield{author}{%
  \bibinfo {author} {\bibfnamefont{E.~A.}\ \bibnamefont{de~Andrada~e Silva}},
  \bibinfo {author} {\bibfnamefont{G.~C.}\ \bibnamefont{La~Rocca}},\ and\
  \bibinfo {author} {\bibfnamefont{F.}~\bibnamefont{Bassani}},\ }%
  \bibfield{journal}{%
  \Doi{10.1103/PhysRevB.50.8523}{\bibinfo {journal} {Phys. Rev. B}}\ }%
  \textbf{\bibinfo {volume} {50}},\ \bibinfo {pages} {8523} (\bibinfo {year}
  {1994})%
  \bibAnnoteFile{NoStop}{AndradaeSilva1994}%
\bibitem{Fu}%
  \BibitemOpen
  \bibfield{author}{%
  \bibinfo {author} {\bibfnamefont{L.}~\bibnamefont{Fu}}, \bibinfo {author}
  {\bibfnamefont{C.~L.}\ \bibnamefont{Kane}},\ and\ \bibinfo {author}
  {\bibfnamefont{E.~J.}\ \bibnamefont{Mele}},\ }%
  \bibfield{journal}{%
  \Doi{10.1103/PhysRevLett.98.106803}{\bibinfo {journal} {Phys. Rev. Lett.}}\
  }%
  \textbf{\bibinfo {volume} {98}},\ \bibinfo {pages} {106803} (\bibinfo {year}
  {2007})%
  \bibAnnoteFile{NoStop}{Fu}%
\bibitem{Moore}%
  \BibitemOpen
  \bibfield{author}{%
  \bibinfo {author} {\bibfnamefont{J.~E.}\ \bibnamefont{Moore}}\ and\ \bibinfo
  {author} {\bibfnamefont{L.}~\bibnamefont{Balents}},\ }%
  \bibfield{journal}{%
  \Doi{10.1103/PhysRevB.75.121306}{\bibinfo {journal} {Phys. Rev. B}}\ }%
  \textbf{\bibinfo {volume} {75}},\ \bibinfo {pages} {121306} (\bibinfo {year}
  {2007})%
  \bibAnnoteFile{NoStop}{Moore}%
\bibitem{Murakami}%
  \BibitemOpen
  \bibfield{author}{%
  \bibinfo {author} {\bibfnamefont{S.}~\bibnamefont{Murakami}},\ }%
  \bibfield{journal}{%
  \bibinfo {journal} {New J. Phys.}\ }%
  \textbf{\bibinfo {volume} {9}},\ \bibinfo {pages} {356} (\bibinfo {year}
  {2007})%
  \bibAnnoteFile{NoStop}{Murakami}%
\bibitem{Hsieh}%
  \BibitemOpen
  \bibfield{author}{%
  \bibinfo {author} {\bibfnamefont{D.}~\bibnamefont{Hsieh}}, \bibinfo {author}
  {\bibfnamefont{D.}~\bibnamefont{Qian}}, \bibinfo {author}
  {\bibfnamefont{L.}~\bibnamefont{Wray}}, \bibinfo {author}
  {\bibfnamefont{Y.}~\bibnamefont{Xia}}, \bibinfo {author}
  {\bibfnamefont{Y.~S.}\ \bibnamefont{Hor}}, \bibinfo {author}
  {\bibfnamefont{R.~J.}\ \bibnamefont{Cava}},\ and\ \bibinfo {author}
  {\bibfnamefont{M.~Z.}\ \bibnamefont{Hasan}},\ }%
  \bibfield{journal}{%
  \bibinfo {journal} {Nature}\ }%
  \textbf{\bibinfo {volume} {452}},\ \bibinfo {pages} {970} (\bibinfo {year}
  {2008})%
  \bibAnnoteFile{NoStop}{Hsieh}%
\bibitem{Zhang}%
  \BibitemOpen
  \bibfield{author}{%
  \bibinfo {author} {\bibfnamefont{H.}~\bibnamefont{Zhang}}, \bibinfo {author}
  {\bibfnamefont{C.-X.}\ \bibnamefont{Liu}}, \bibinfo {author}
  {\bibfnamefont{X.-L.}\ \bibnamefont{Qi}}, \bibinfo {author}
  {\bibfnamefont{X.}~\bibnamefont{Dai}}, \bibinfo {author}
  {\bibfnamefont{Z.}~\bibnamefont{Fang}},\ and\ \bibinfo {author}
  {\bibfnamefont{S.-C.}\ \bibnamefont{Zhang}},\ }%
  \bibfield{journal}{%
  \bibinfo {journal} {Nature Phys.}\ }%
  \textbf{\bibinfo {volume} {5}},\ \bibinfo {pages} {438} (\bibinfo {year}
  {2009})%
  \bibAnnoteFile{NoStop}{Zhang}%
\bibitem{Winkler2}%
  \BibitemOpen
  \bibfield{author}{%
  \bibinfo {author} {\bibfnamefont{R.}~\bibnamefont{Winkler}},\ }%
  \bibfield{journal}{%
  \bibinfo {journal} {Physica E}\ }%
  \textbf{\bibinfo {volume} {22}},\ \bibinfo {pages} {450} (\bibinfo {year}
  {2004})%
  \bibAnnoteFile{NoStop}{Winkler2}%
\bibitem{Hofmann}%
  \BibitemOpen
  \bibfield{author}{%
  \bibinfo {author} {\bibfnamefont{Y.~M.}\ \bibnamefont{Koroteev}}, \bibinfo
  {author} {\bibfnamefont{G.}~\bibnamefont{Bihlmayer}}, \bibinfo {author}
  {\bibfnamefont{J.~E.}\ \bibnamefont{Gayone}}, \bibinfo {author}
  {\bibfnamefont{E.~V.}\ \bibnamefont{Chulkov}}, \bibinfo {author}
  {\bibfnamefont{S.}~\bibnamefont{Bl\"ugel}}, \bibinfo {author}
  {\bibfnamefont{P.~M.}\ \bibnamefont{Echenique}},\ and\ \bibinfo {author}
  {\bibfnamefont{P.}~\bibnamefont{Hofmann}},\ }%
  \bibfield{journal}{%
  \Doi{10.1103/PhysRevLett.93.046403}{\bibinfo {journal} {Phys. Rev. Lett.}}\
  }%
  \textbf{\bibinfo {volume} {93}},\ \bibinfo {pages} {046403} (\bibinfo {year}
  {2004})%
  \bibAnnoteFile{NoStop}{Hofmann}%
\bibitem{Ast}%
  \BibitemOpen
  \bibfield{author}{%
  \bibinfo {author} {\bibfnamefont{C.~R.}\ \bibnamefont{Ast}}, \bibinfo
  {author} {\bibfnamefont{J.}~\bibnamefont{Henk}}, \bibinfo {author}
  {\bibfnamefont{A.}~\bibnamefont{Ernst}}, \bibinfo {author}
  {\bibfnamefont{L.}~\bibnamefont{Moreschini}}, \bibinfo {author}
  {\bibfnamefont{M.~C.}\ \bibnamefont{Falub}}, \bibinfo {author}
  {\bibfnamefont{D.}~\bibnamefont{Pacile}}, \bibinfo {author}
  {\bibfnamefont{P.}~\bibnamefont{Bruno}}, \bibinfo {author}
  {\bibfnamefont{K.}~\bibnamefont{Kern}},\ and\ \bibinfo {author}
  {\bibfnamefont{M.}~\bibnamefont{Grioni}},\ }%
  \bibfield{journal}{%
  \bibinfo {journal} {Phys. Rev. Lett.}\ }%
  \textbf{\bibinfo {volume} {98}},\ \bibinfo {pages} {186807} (\bibinfo {year}
  {2007})%
  \bibAnnoteFile{NoStop}{Ast}%
\bibitem{LaShell}%
  \BibitemOpen
  \bibfield{author}{%
  \bibinfo {author} {\bibfnamefont{S.}~\bibnamefont{LaShell}}, \bibinfo
  {author} {\bibfnamefont{B.~A.}\ \bibnamefont{McDougall}},\ and\ \bibinfo
  {author} {\bibfnamefont{E.}~\bibnamefont{Jensen}},\ }%
  \bibfield{journal}{%
  \bibinfo {journal} {Phys. Rev. Lett.}\ }%
  \textbf{\bibinfo {volume} {77}},\ \bibinfo {pages} {3419} (\bibinfo {year}
  {1996})%
  \bibAnnoteFile{NoStop}{LaShell}%
\bibitem{Hoesch}%
  \BibitemOpen
  \bibfield{author}{%
  \bibinfo {author} {\bibfnamefont{M.}~\bibnamefont{Hoesch}}, \bibinfo {author}
  {\bibfnamefont{M.}~\bibnamefont{Muntwiler}}, \bibinfo {author}
  {\bibfnamefont{V.~N.}\ \bibnamefont{Petrov}}, \bibinfo {author}
  {\bibfnamefont{M.}~\bibnamefont{Hengsberger}}, \bibinfo {author}
  {\bibfnamefont{L.}~\bibnamefont{Patthey}}, \bibinfo {author}
  {\bibfnamefont{M.}~\bibnamefont{Shi}}, \bibinfo {author}
  {\bibfnamefont{M.}~\bibnamefont{Falub}}, \bibinfo {author}
  {\bibfnamefont{T.}~\bibnamefont{Greber}},\ and\ \bibinfo {author}
  {\bibfnamefont{J.}~\bibnamefont{Osterwalder}},\ }%
  \bibfield{journal}{%
  \bibinfo {journal} {Phys. Rev. B}\ }%
  \textbf{\bibinfo {volume} {69}},\ \bibinfo {pages} {241401} (\bibinfo {year}
  {2004})%
  \bibAnnoteFile{NoStop}{Hoesch}%
\bibitem{Dedkov}%
  \BibitemOpen
  \bibfield{author}{%
  \bibinfo {author} {\bibfnamefont{Y.~S.}\ \bibnamefont{Dedkov}}, \bibinfo
  {author} {\bibfnamefont{M.}~\bibnamefont{Fonin}}, \bibinfo {author}
  {\bibfnamefont{U.}~\bibnamefont{R\"udiger}},\ and\ \bibinfo {author}
  {\bibfnamefont{C.}~\bibnamefont{Laubschat}},\ }%
  \bibfield{journal}{%
  \bibinfo {journal} {Phys. Rev. Lett.}\ }%
  \textbf{\bibinfo {volume} {100}},\ \bibinfo {pages} {107602} (\bibinfo {year}
  {2008})%
  \bibAnnoteFile{NoStop}{Dedkov}%
\bibitem{gierz_silicon_2009}%
  \BibitemOpen
  \bibfield{author}{%
  \bibinfo {author} {\bibfnamefont{I.}~\bibnamefont{Gierz}}, \bibinfo {author}
  {\bibfnamefont{T.}~\bibnamefont{Suzuki}}, \bibinfo {author}
  {\bibfnamefont{E.}~\bibnamefont{Frantzeskakis}}, \bibinfo {author}
  {\bibfnamefont{S.}~\bibnamefont{Pons}}, \bibinfo {author}
  {\bibfnamefont{S.}~\bibnamefont{Ostanin}}, \bibinfo {author}
  {\bibfnamefont{A.}~\bibnamefont{Ernst}}, \bibinfo {author}
  {\bibfnamefont{J.}~\bibnamefont{Henk}}, \bibinfo {author}
  {\bibfnamefont{M.}~\bibnamefont{Grioni}}, \bibinfo {author}
  {\bibfnamefont{K.}~\bibnamefont{Kern}},\ and\ \bibinfo {author}
  {\bibfnamefont{C.~R.}\ \bibnamefont{Ast}},\ }%
  \bibfield{journal}{%
  \bibinfo {journal} {Phys. Rev. Lett.}\ }%
  \textbf{\bibinfo {volume} {103}},\ \bibinfo {pages} {046803} (\bibinfo {year}
  {2009})%
  \bibAnnoteFile{NoStop}{gierz_silicon_2009}%
\bibitem{Hoepfner}%
  \BibitemOpen
  \bibfield{author}{%
  \bibinfo {author} {\bibfnamefont{P.}~\bibnamefont{H\"opfner}}, \bibinfo
  {author} {\bibfnamefont{J.}~\bibnamefont{Sch\"afer}}, \bibinfo {author}
  {\bibfnamefont{A.}~\bibnamefont{Fleszar}}, \bibinfo {author}
  {\bibfnamefont{S.}~\bibnamefont{Meyer}}, \bibinfo {author}
  {\bibfnamefont{C.}~\bibnamefont{Blumenstein}}, \bibinfo {author}
  {\bibfnamefont{T.}~\bibnamefont{Schramm}}, \bibinfo {author}
  {\bibfnamefont{M.}~\bibnamefont{He\ss{}mann}}, \bibinfo {author}
  {\bibfnamefont{X.}~\bibnamefont{Cui}}, \bibinfo {author}
  {\bibfnamefont{L.}~\bibnamefont{Patthey}}, \bibinfo {author}
  {\bibfnamefont{W.}~\bibnamefont{Hanke}},\ and\ \bibinfo {author}
  {\bibfnamefont{R.}~\bibnamefont{Claessen}},\ }%
  \bibfield{journal}{%
  \bibinfo {journal} {Phys. Rev. B}\ }%
  \textbf{\bibinfo {volume} {83}},\ \bibinfo {pages} {235435} (\bibinfo {year}
  {2011})%
  \bibAnnoteFile{NoStop}{Hoepfner}%
\bibitem{Frantze}%
  \BibitemOpen
  \bibfield{author}{%
  \bibinfo {author} {\bibfnamefont{E.}~\bibnamefont{Frantzeskakis}}, \bibinfo
  {author} {\bibfnamefont{S.}~\bibnamefont{Pons}},\ and\ \bibinfo {author}
  {\bibfnamefont{M.}~\bibnamefont{Grioni}},\ }%
  \bibfield{journal}{%
  \bibinfo {journal} {Phys. Rev. B}\ }%
  \textbf{\bibinfo {volume} {82}},\ \bibinfo {pages} {085440} (\bibinfo {year}
  {2010})%
  \bibAnnoteFile{NoStop}{Frantze}%
\bibitem{Ohtsubo}%
  \BibitemOpen
  \bibfield{author}{%
  \bibinfo {author} {\bibfnamefont{Y.}~\bibnamefont{Ohtsubo}}, \bibinfo
  {author} {\bibfnamefont{S.}~\bibnamefont{Hatta}}, \bibinfo {author}
  {\bibfnamefont{K.}~\bibnamefont{Yaji}}, \bibinfo {author}
  {\bibfnamefont{H.}~\bibnamefont{Okuyama}}, \bibinfo {author}
  {\bibfnamefont{K.}~\bibnamefont{Miyamoto}}, \bibinfo {author}
  {\bibfnamefont{T.}~\bibnamefont{Okuda}}, \bibinfo {author}
  {\bibfnamefont{A.}~\bibnamefont{Kimura}}, \bibinfo {author}
  {\bibfnamefont{H.}~\bibnamefont{Namatame}}, \bibinfo {author}
  {\bibfnamefont{M.}~\bibnamefont{Taniguchi}},\ and\ \bibinfo {author}
  {\bibfnamefont{T.}~\bibnamefont{Aruga}},\ }%
  \bibfield{journal}{%
  \bibinfo {journal} {Phys. Rev. B}\ }%
  \textbf{\bibinfo {volume} {82}},\ \bibinfo {pages} {201307} (\bibinfo {year}
  {2010})%
  \bibAnnoteFile{NoStop}{Ohtsubo}%
\bibitem{Yaji}%
  \BibitemOpen
  \bibfield{author}{%
  \bibinfo {author} {\bibfnamefont{K.}~\bibnamefont{Yaji}}, \bibinfo {author}
  {\bibfnamefont{Y.}~\bibnamefont{Ohtsubo}}, \bibinfo {author}
  {\bibfnamefont{S.}~\bibnamefont{Hatta}}, \bibinfo {author}
  {\bibfnamefont{H.}~\bibnamefont{Okuyama}}, \bibinfo {author}
  {\bibfnamefont{K.}~\bibnamefont{Miyamoto}}, \bibinfo {author}
  {\bibfnamefont{T.}~\bibnamefont{Okuda}}, \bibinfo {author}
  {\bibfnamefont{A.}~\bibnamefont{Kimura}}, \bibinfo {author}
  {\bibfnamefont{H.}~\bibnamefont{Namatame}}, \bibinfo {author}
  {\bibfnamefont{M.}~\bibnamefont{Taniguchi}},\ and\ \bibinfo {author}
  {\bibfnamefont{T.}~\bibnamefont{Aruga}},\ }%
  \bibfield{journal}{%
  \bibinfo {journal} {Nat Commun}\ }%
  \textbf{\bibinfo {volume} {1}},\ \bibinfo {pages} {17} (\bibinfo {year}
  {2010})%
  \bibAnnoteFile{NoStop}{Yaji}%
\bibitem{Seiler}%
  \BibitemOpen
  \bibfield{author}{%
  \bibinfo {author} {\bibfnamefont{D.~G.}\ \bibnamefont{Seiler}}, \bibinfo
  {author} {\bibfnamefont{B.~D.}\ \bibnamefont{Bajaj}},\ and\ \bibinfo {author}
  {\bibfnamefont{A.~E.}\ \bibnamefont{Stephens}},\ }%
  \bibfield{journal}{%
  \Doi{10.1103/PhysRevB.16.2822}{\bibinfo {journal} {Phys. Rev. B}}\ }%
  \textbf{\bibinfo {volume} {16}},\ \bibinfo {pages} {2822} (\bibinfo {year}
  {1977})%
  \bibAnnoteFile{NoStop}{Seiler}%
\bibitem{Nitta2}%
  \BibitemOpen
  \bibfield{author}{%
  \bibinfo {author} {\bibfnamefont{T.}~\bibnamefont{Koga}}, \bibinfo {author}
  {\bibfnamefont{J.}~\bibnamefont{Nitta}}, \bibinfo {author}
  {\bibfnamefont{T.}~\bibnamefont{Akazaki}},\ and\ \bibinfo {author}
  {\bibfnamefont{H.}~\bibnamefont{Takayanagi}},\ }%
  \bibfield{journal}{%
  \bibinfo {journal} {Phys. Rev. Lett.}\ }%
  \textbf{\bibinfo {volume} {89}},\ \bibinfo {pages} {046801} (\bibinfo {year}
  {2002})%
  \bibAnnoteFile{NoStop}{Nitta2}%
\bibitem{PhysRevB.61.15588}%
  \BibitemOpen
  \bibfield{author}{%
  \bibinfo {author} {\bibfnamefont{T.}~\bibnamefont{Matsuyama}}, \bibinfo
  {author} {\bibfnamefont{R.}~\bibnamefont{K\"ursten}}, \bibinfo {author}
  {\bibfnamefont{C.}~\bibnamefont{Mei\ss{}ner}},\ and\ \bibinfo {author}
  {\bibfnamefont{U.}~\bibnamefont{Merkt}},\ }%
  \bibfield{journal}{%
  \Doi{10.1103/PhysRevB.61.15588}{\bibinfo {journal} {Phys. Rev. B}}\ }%
  \textbf{\bibinfo {volume} {61}},\ \bibinfo {pages} {15588} (\bibinfo {year}
  {2000})%
  \bibAnnoteFile{NoStop}{PhysRevB.61.15588}%
\bibitem{Winkler.PhysRevB.48.8918}%
  \BibitemOpen
  \bibfield{author}{%
  \bibinfo {author} {\bibfnamefont{R.}~\bibnamefont{Winkler}}\ and\ \bibinfo
  {author} {\bibfnamefont{U.}~\bibnamefont{R\"ossler}},\ }%
  \bibfield{journal}{%
  \Doi{10.1103/PhysRevB.48.8918}{\bibinfo {journal} {Phys. Rev. B}}\ }%
  \textbf{\bibinfo {volume} {48}},\ \bibinfo {pages} {8918} (\bibinfo {year}
  {1993})%
  \bibAnnoteFile{NoStop}{Winkler.PhysRevB.48.8918}%
\bibitem{Pettersen}%
  \BibitemOpen
  \bibfield{author}{%
  \bibinfo {author} {\bibfnamefont{L.}~\bibnamefont{Petersen}}\ and\ \bibinfo
  {author} {\bibfnamefont{P.}~\bibnamefont{Hedegard}},\ }%
  \bibfield{journal}{%
  \bibinfo {journal} {Surf. Sci.}\ }%
  \textbf{\bibinfo {volume} {459}},\ \bibinfo {pages} {49} (\bibinfo {year}
  {2000})%
  \bibAnnoteFile{NoStop}{Pettersen}%
\bibitem{Walls}%
  \BibitemOpen
  \bibfield{author}{%
  \bibinfo {author} {\bibfnamefont{J.~E.~D.}\ \bibnamefont{Walls}}\ and\
  \bibinfo {author} {\bibfnamefont{E.~J.}\ \bibnamefont{Heller}},\ }%
  \bibfield{journal}{%
  \bibinfo {journal} {Nano Lett.}\ }%
  \textbf{\bibinfo {volume} {7}},\ \bibinfo {pages} {3377} (\bibinfo {year}
  {2007})%
  \bibAnnoteFile{NoStop}{Walls}%
\bibitem{Bluegel}%
  \BibitemOpen
  \bibfield{author}{%
  \bibinfo {author} {\bibfnamefont{S.}~\bibnamefont{Lounis}}, \bibinfo {author}
  {\bibfnamefont{A.}~\bibnamefont{Bringer}},\ and\ \bibinfo {author}
  {\bibfnamefont{S.}~\bibnamefont{Bl\"ugel}},\ }%
  \bibfield{journal}{%
  \bibinfo {journal} {ArXiV},\ \bibinfo {pages} {1204.0999}}%
   (\bibinfo {year} {2012})%
  \bibAnnoteFile{NoStop}{Bluegel}%
\bibitem{Ast2}%
  \BibitemOpen
  \bibfield{author}{%
  \bibinfo {author} {\bibfnamefont{C.~R.}\ \bibnamefont{Ast}}, \bibinfo
  {author} {\bibfnamefont{G.}~\bibnamefont{Wittich}}, \bibinfo {author}
  {\bibfnamefont{P.}~\bibnamefont{Wahl}}, \bibinfo {author}
  {\bibfnamefont{R.}~\bibnamefont{Vogelgesang}}, \bibinfo {author}
  {\bibfnamefont{D.}~\bibnamefont{Pacile}}, \bibinfo {author}
  {\bibfnamefont{M.~C.}\ \bibnamefont{Falub}}, \bibinfo {author}
  {\bibfnamefont{L.}~\bibnamefont{Moreschini}}, \bibinfo {author}
  {\bibfnamefont{M.}~\bibnamefont{Papagno}}, \bibinfo {author}
  {\bibfnamefont{M.}~\bibnamefont{Grioni}},\ and\ \bibinfo {author}
  {\bibfnamefont{K.}~\bibnamefont{Kern}},\ }%
  \bibfield{journal}{%
  \bibinfo {journal} {Phys. Rev. B}\ }%
  \textbf{\bibinfo {volume} {75}},\ \bibinfo {pages} {201401} (\bibinfo {year}
  {2007})%
  \bibAnnoteFile{NoStop}{Ast2}%
\bibitem{Pascual}%
  \BibitemOpen
  \bibfield{author}{%
  \bibinfo {author} {\bibfnamefont{J.~I.}\ \bibnamefont{Pascual}}, \bibinfo
  {author} {\bibfnamefont{G.}~\bibnamefont{Bihlmayer}}, \bibinfo {author}
  {\bibfnamefont{Y.~M.}\ \bibnamefont{Koroteev}}, \bibinfo {author}
  {\bibfnamefont{H.~P.}\ \bibnamefont{Rust}}, \bibinfo {author}
  {\bibfnamefont{G.}~\bibnamefont{Ceballos}}, \bibinfo {author}
  {\bibfnamefont{M.}~\bibnamefont{Hansmann}}, \bibinfo {author}
  {\bibfnamefont{K.}~\bibnamefont{Horn}}, \bibinfo {author}
  {\bibfnamefont{E.~V.}\ \bibnamefont{Chulkov}}, \bibinfo {author}
  {\bibfnamefont{S.}~\bibnamefont{Bl\"ugel}}, \bibinfo {author}
  {\bibfnamefont{P.~M.}\ \bibnamefont{Echenique}},\ and\ \bibinfo {author}
  {\bibfnamefont{P.}~\bibnamefont{Hofmann}},\ }%
  \bibfield{journal}{%
  \bibinfo {journal} {Phys. Rev. Lett.}\ }%
  \textbf{\bibinfo {volume} {93}},\ \bibinfo {pages} {196802} (\bibinfo {year}
  {2004})%
  \bibAnnoteFile{NoStop}{Pascual}%
\bibitem{Yazdani}%
  \BibitemOpen
  \bibfield{author}{%
  \bibinfo {author} {\bibfnamefont{P.}~\bibnamefont{Roushan}}, \bibinfo
  {author} {\bibfnamefont{J.}~\bibnamefont{Seo}}, \bibinfo {author}
  {\bibfnamefont{C.~V.}\ \bibnamefont{Parker}}, \bibinfo {author}
  {\bibfnamefont{Y.~S.}\ \bibnamefont{Hor}}, \bibinfo {author}
  {\bibfnamefont{D.}~\bibnamefont{Hsieh}}, \bibinfo {author}
  {\bibfnamefont{D.}~\bibnamefont{Qian}}, \bibinfo {author}
  {\bibfnamefont{A.}~\bibnamefont{Richardella}}, \bibinfo {author}
  {\bibfnamefont{M.~Z.}\ \bibnamefont{Hasan}}, \bibinfo {author}
  {\bibfnamefont{R.~J.}\ \bibnamefont{Cava}},\ and\ \bibinfo {author}
  {\bibfnamefont{A.}~\bibnamefont{Yazdani}},\ }%
  \bibfield{journal}{%
  \bibinfo {journal} {Nature}\ }%
  \textbf{\bibinfo {volume} {460}},\ \bibinfo {pages} {1106} (\bibinfo {year}
  {2009})%
  \bibAnnoteFile{NoStop}{Yazdani}%
\bibitem{Sherman2}%
  \BibitemOpen
  \bibfield{author}{%
  \bibinfo {author} {\bibfnamefont{M.~M.}\ \bibnamefont{Glazov}}\ and\ \bibinfo
  {author} {\bibfnamefont{E.~Y.}\ \bibnamefont{Sherman}},\ }%
  \bibfield{journal}{%
  \Doi{10.1103/PhysRevB.71.241312}{\bibinfo {journal} {Phys. Rev. B}}\ }%
  \textbf{\bibinfo {volume} {71}},\ \bibinfo {pages} {241312} (\bibinfo {year}
  {2005})%
  \bibAnnoteFile{NoStop}{Sherman2}%
\bibitem{sherman3}%
  \BibitemOpen
  \bibfield{author}{%
  \bibinfo {author} {\bibfnamefont{E.~Y.}\ \bibnamefont{Sherman}},\ }%
  \bibfield{journal}{%
  \Doi{10.1103/PhysRevB.67.161303}{\bibinfo {journal} {Phys. Rev. B}}\ }%
  \textbf{\bibinfo {volume} {67}},\ \bibinfo {pages} {161303} (\bibinfo {year}
  {2003})%
  \bibAnnoteFile{NoStop}{sherman3}%
\bibitem{Sherman4}%
  \BibitemOpen
  \bibfield{author}{%
  \bibinfo {author} {\bibfnamefont{E.}~\bibnamefont{Sherman}},\ }%
  \bibfield{journal}{%
  \bibinfo {journal} {Appl. Phys. Lett.}\ }%
  \textbf{\bibinfo {volume} {82}},\ \bibinfo {pages} {209} (\bibinfo {year}
  {2003})%
  \bibAnnoteFile{NoStop}{Sherman4}%
\bibitem{Dyakonov}%
  \BibitemOpen
  \bibfield{author}{%
  \bibinfo {author} {\bibfnamefont{M.}~\bibnamefont{Dyakonov}}\ and\ \bibinfo
  {author} {\bibfnamefont{V.}~\bibnamefont{Perel}},\ }%
  \bibfield{journal}{%
  \bibinfo {journal} {Phys. Lett. A}\ }%
  \textbf{\bibinfo {volume} {A 35}},\ \bibinfo {pages} {459} (\bibinfo {year}
  {1971})%
  \bibAnnoteFile{NoStop}{Dyakonov}%
\bibitem{Dietl}%
  \BibitemOpen
  \bibfield{author}{%
  \bibinfo {author} {\bibfnamefont{T.}~\bibnamefont{Dietl}}, \bibinfo {author}
  {\bibfnamefont{H.}~\bibnamefont{Ohno}}, \bibinfo {author}
  {\bibfnamefont{F.}~\bibnamefont{Matsukura}}, \bibinfo {author}
  {\bibfnamefont{J.}~\bibnamefont{Cibert}},\ and\ \bibinfo {author}
  {\bibfnamefont{D.}~\bibnamefont{Ferrand}},\ }%
  \bibfield{journal}{%
  \bibinfo {journal} {Science}\ }%
  \textbf{\bibinfo {volume} {287}},\ \bibinfo {pages} {1019} (\bibinfo {year}
  {2000})%
  \bibAnnoteFile{NoStop}{Dietl}%
\bibitem{Volnianska}%
  \BibitemOpen
  \bibfield{author}{%
  \bibinfo {author} {\bibfnamefont{O.}~\bibnamefont{Volnianska}}\ and\ \bibinfo
  {author} {\bibfnamefont{P.}~\bibnamefont{Boguslawski}},\ }%
  \bibfield{journal}{%
  \bibinfo {journal} {J. Phys: Cond. Matt.}\ }%
  \textbf{\bibinfo {volume} {22}},\ \bibinfo {pages} {073202} (\bibinfo {year}
  {2010})%
  \bibAnnoteFile{NoStop}{Volnianska}%
\bibitem{AndoK}%
  \BibitemOpen
  \bibfield{author}{%
  \bibinfo {author} {\bibfnamefont{K.}~\bibnamefont{Ando}},\ }%
  \bibfield{journal}{%
  \bibinfo {journal} {Science}\ }%
  \textbf{\bibinfo {volume} {312}},\ \bibinfo {pages} {1883} (\bibinfo {year}
  {2006})%
  \bibAnnoteFile{NoStop}{AndoK}%
\bibitem{Koenraad1}%
  \BibitemOpen
  \bibfield{author}{%
  \bibinfo {author} {\bibfnamefont{A.~M.}\ \bibnamefont{Yakunin}}, \bibinfo
  {author} {\bibfnamefont{A.~Y.}\ \bibnamefont{Silov}}, \bibinfo {author}
  {\bibfnamefont{P.~M.}\ \bibnamefont{Koenraad}}, \bibinfo {author}
  {\bibfnamefont{J.~H.}\ \bibnamefont{Wolter}}, \bibinfo {author}
  {\bibfnamefont{W.}~\bibnamefont{Van~Roy}}, \bibinfo {author}
  {\bibfnamefont{J.}~\bibnamefont{De~Boeck}}, \bibinfo {author}
  {\bibfnamefont{J.-M.}\ \bibnamefont{Tang}},\ and\ \bibinfo {author}
  {\bibfnamefont{M.~E.}\ \bibnamefont{Flatt\'e}},\ }%
  \bibfield{journal}{%
  \Doi{10.1103/PhysRevLett.92.216806}{\bibinfo {journal} {Phys. Rev. Lett.}}\
  }%
  \textbf{\bibinfo {volume} {92}},\ \bibinfo {pages} {216806} (\bibinfo {year}
  {2004})%
  \bibAnnoteFile{NoStop}{Koenraad1}%
\bibitem{Koenraad2}%
  \BibitemOpen
  \bibfield{author}{%
  \bibinfo {author} {\bibfnamefont{A.~M.}\ \bibnamefont{Yakunin}}, \bibinfo
  {author} {\bibfnamefont{A.~Y.}\ \bibnamefont{Silov}}, \bibinfo {author}
  {\bibfnamefont{P.~M.}\ \bibnamefont{Koenraad}}, \bibinfo {author}
  {\bibfnamefont{J.-M.}\ \bibnamefont{Tang}}, \bibinfo {author}
  {\bibfnamefont{M.~E.}\ \bibnamefont{Flatt\'e}}, \bibinfo {author}
  {\bibfnamefont{W.~V.}\ \bibnamefont{Roy}}, \bibinfo {author}
  {\bibfnamefont{J.~D.}\ \bibnamefont{Boeck}},\ and\ \bibinfo {author}
  {\bibfnamefont{J.~H.}\ \bibnamefont{Wolter}},\ }%
  \bibfield{journal}{%
  \Doi{10.1103/PhysRevLett.95.256402}{\bibinfo {journal} {Phys. Rev. Lett.}}\
  }%
  \textbf{\bibinfo {volume} {95}},\ \bibinfo {pages} {256402} (\bibinfo {year}
  {2005})%
  \bibAnnoteFile{NoStop}{Koenraad2}%
\bibitem{MeierF}%
  \BibitemOpen
  \bibfield{author}{%
  \bibinfo {author} {\bibfnamefont{F.}~\bibnamefont{Marczinowski}}, \bibinfo
  {author} {\bibfnamefont{J.}~\bibnamefont{Wiebe}}, \bibinfo {author}
  {\bibfnamefont{J.-M.}\ \bibnamefont{Tang}}, \bibinfo {author}
  {\bibfnamefont{M.~E.}\ \bibnamefont{Flatt\'e}}, \bibinfo {author}
  {\bibfnamefont{F.}~\bibnamefont{Meier}}, \bibinfo {author}
  {\bibfnamefont{M.}~\bibnamefont{Morgenstern}},\ and\ \bibinfo {author}
  {\bibfnamefont{R.}~\bibnamefont{Wiesendanger}},\ }%
  \bibfield{journal}{%
  \Doi{10.1103/PhysRevLett.99.157202}{\bibinfo {journal} {Phys. Rev. Lett.}}\
  }%
  \textbf{\bibinfo {volume} {99}},\ \bibinfo {pages} {157202} (\bibinfo {year}
  {2007})%
  \bibAnnoteFile{NoStop}{MeierF}%
\bibitem{Koenraad3}%
  \BibitemOpen
  \bibfield{author}{%
  \bibinfo {author} {\bibfnamefont{C.}~\bibnamefont{\ifmmode~\mbox{\c{C}}\else
  \c{C}\fi{}elebi}}, \bibinfo {author} {\bibfnamefont{J.~K.}\
  \bibnamefont{Garleff}}, \bibinfo {author} {\bibfnamefont{A.~Y.}\
  \bibnamefont{Silov}}, \bibinfo {author} {\bibfnamefont{A.~M.}\
  \bibnamefont{Yakunin}}, \bibinfo {author} {\bibfnamefont{P.~M.}\
  \bibnamefont{Koenraad}}, \bibinfo {author}
  {\bibfnamefont{W.}~\bibnamefont{Van~Roy}}, \bibinfo {author}
  {\bibfnamefont{J.-M.}\ \bibnamefont{Tang}},\ and\ \bibinfo {author}
  {\bibfnamefont{M.~E.}\ \bibnamefont{Flatt\'e}},\ }%
  \bibfield{journal}{%
  \Doi{10.1103/PhysRevLett.104.086404}{\bibinfo {journal} {Phys. Rev. Lett.}}\
  }%
  \textbf{\bibinfo {volume} {104}},\ \bibinfo {pages} {086404} (\bibinfo {year}
  {2010})%
  \bibAnnoteFile{NoStop}{Koenraad3}%
\bibitem{Yazdani2}%
  \BibitemOpen
  \bibfield{author}{%
  \bibinfo {author} {\bibfnamefont{D.}~\bibnamefont{Kitchen}}, \bibinfo
  {author} {\bibfnamefont{A.}~\bibnamefont{Richardella}}, \bibinfo {author}
  {\bibfnamefont{J.-M.}\ \bibnamefont{Tang}}, \bibinfo {author}
  {\bibfnamefont{M.~E.}\ \bibnamefont{Flatte}},\ and\ \bibinfo {author}
  {\bibfnamefont{A.}~\bibnamefont{Yazdani}},\ }%
  \bibfield{journal}{%
  \bibinfo {journal} {Nature}\ }%
  \textbf{\bibinfo {volume} {442}},\ \bibinfo {pages} {436} (\bibinfo {year}
  {2006})%
  \bibAnnoteFile{NoStop}{Yazdani2}%
\bibitem{Yazdani3}%
  \BibitemOpen
  \bibfield{author}{%
  \bibinfo {author} {\bibfnamefont{A.}~\bibnamefont{Richardella}}, \bibinfo
  {author} {\bibfnamefont{P.}~\bibnamefont{Roushan}}, \bibinfo {author}
  {\bibfnamefont{S.}~\bibnamefont{Mack}}, \bibinfo {author}
  {\bibfnamefont{B.}~\bibnamefont{Zhou}}, \bibinfo {author}
  {\bibfnamefont{D.~A.}\ \bibnamefont{Huse}}, \bibinfo {author}
  {\bibfnamefont{D.~D.}\ \bibnamefont{Awschalom}},\ and\ \bibinfo {author}
  {\bibfnamefont{A.}~\bibnamefont{Yazdani}},\ }%
  \bibfield{journal}{%
  \bibinfo {journal} {Science}\ }%
  \textbf{\bibinfo {volume} {327}},\ \bibinfo {pages} {665} (\bibinfo {year}
  {2010})%
  \bibAnnoteFile{NoStop}{Yazdani3}%
\bibitem{Dederichs}%
  \BibitemOpen
  \bibfield{author}{%
  \bibinfo {author} {\bibfnamefont{H.}~\bibnamefont{Katayama-Yoshida}},
  \bibinfo {author} {\bibfnamefont{K.}~\bibnamefont{Sato}}, \bibinfo {author}
  {\bibfnamefont{T.}~\bibnamefont{Fukushima}}, \bibinfo {author}
  {\bibfnamefont{M.}~\bibnamefont{Toyoda}}, \bibinfo {author}
  {\bibfnamefont{H.}~\bibnamefont{Kizaki}}, \bibinfo {author}
  {\bibfnamefont{V.~A.}\ \bibnamefont{Dinh}},\ and\ \bibinfo {author}
  {\bibfnamefont{P.~H.}\ \bibnamefont{Dederichs}},\ }%
  \bibfield{journal}{%
  \bibinfo {journal} {phys. stat. sol. A}\ }%
  \textbf{\bibinfo {volume} {204}},\ \bibinfo {pages} {15} (\bibinfo {year}
  {2007})%
  \bibAnnoteFile{NoStop}{Dederichs}%
\end{thebibliography}%

\end{document}